\documentclass[5p,times,twocolumn]{elsarticle}

\usepackage[]{natbib}
\usepackage{graphicx}        
\usepackage{subcaption}      
\usepackage{multirow}        
\usepackage{booktabs}        
\usepackage{amsmath,amssymb,amsfonts} 
\usepackage{siunitx}         
\usepackage{algorithm}       
\usepackage{algpseudocode}   
\usepackage{hyperref}        

\usepackage{cleveref}        
\usepackage{xcolor}          
\usepackage{float}           
\usepackage{caption}         
\usepackage{enumitem}        
\usepackage{geometry}        
\usepackage{mathrsfs}        
\usepackage{pgfplots}        
\usepackage[title]{appendix} 
\usepackage{overpic}
\pgfplotsset{compat=1.18}

\hypersetup{
    hidelinks,               
    colorlinks=false,        
    linkcolor=black,         
    citecolor=black,         
    urlcolor=black           
}

\newcommand{\etal}{\textit{et al.}}

\usepackage[version=4]{mhchem}

\usepackage{microtype}

\usepackage{placeins}

\usepackage{soul}
\usepackage{lipsum}

\begin{document}

\begin{frontmatter}

\title{\Large \textbf{Accelerated Multi-Objective Alloy Discovery through Efficient Bayesian Methods: Application to the FCC Alloy Space}}

\author[inst1]{Trevor Hastings\corref{cor1}}
\ead{trevorhastings@tamu.edu}
\author[inst1]{Mrinalini Mulukutla}
\author[inst1]{Danial Khatamsaz}
\author[inst1]{Daniel Salas}
\author[inst1]{Wenle Xu}
\author[inst1]{Daniel Lewis}
\author[inst1]{Nicole Person}
\author[inst1]{Matthew Skokan}
\author[inst1]{Braden Miller}
\author[inst1]{James Paramore}
\author[inst1,inst2]{Brady Butler}
\author[inst3]{Douglas Allaire}
\author[inst1]{Vahid Attari}
\author[inst1]{Ibrahim Karaman}
\author[inst1]{George Pharr}
\author[inst1]{Ankit Srivastava}
\author[inst1,inst3]{Raymundo Arr\'{o}yave\corref{cor2}}
\ead{rarroyave@tamu.edu}

\cortext[cor1]{Corresponding Author}
\cortext[cor2]{Corresponding Author}

\address[inst1]{Texas A\&M University, Department of Materials Science and Engineering, College Station, TX, 77843-3003, USA}
\address[inst2]{DEVCOM Army Research Laboratory South at Texas A\&M University, College Station, TX 77843-3003, USA}
\address[inst3]{Texas A\&M University, Department of Mechanical Engineering, College Station, TX, 77843, USA}

\begin{abstract}
{This study introduces BIRDSHOT, an integrated Bayesian materials discovery framework designed to efficiently explore complex compositional spaces while optimizing multiple material properties. We applied this framework to the CoCrFeNiVAl FCC high entropy alloy (HEA) system, targeting three key performance objectives: ultimate tensile strength/yield strength ratio, hardness, and strain rate sensitivity. The experimental campaign employed an integrated cyber-physical approach that combined vacuum arc melting (VAM) for alloy synthesis with advanced mechanical testing, including tensile and high-strain-rate nanoindentation testing. By incorporating batch Bayesian optimization schemes that allowed the parallel exploration of the alloy space, we completed five iterative design-make-test-learn loops, identifying a non-trivial three-objective Pareto set in a high-dimensional alloy space. Notably, this was achieved by exploring only 0.15\% of the feasible design space, representing a significant acceleration in discovery rate relative to traditional methods. This work demonstrates the capability of BIRDSHOT to navigate complex, multi-objective optimization challenges and highlights its potential for broader application in accelerating materials discovery.}
\end{abstract}

\begin{keyword}
Bayesian Optimization \sep Materials Discovery \sep High-Throughput \sep Multi-Objective Optimization \sep Machine Learning
\end{keyword}

\end{frontmatter}

\section{Introduction}
\label{intro}

\subsection{Motivation and State of the Art}
Accelerating the materials development cycle is crucial to addressing pressing global challenges that require groundbreaking technologies. These technologies, spanning sectors such as clean energy, efficient transportation, and resilient infrastructure, typically demand materials with properties far beyond the capabilities of those currently available. For instance, structural alloys face a persistent tradeoff between strength and ductility---a challenge that becomes even more complex when additional performance metrics are required\cite{vela2023high}.

Unfortunately, though informed by scientific understanding, traditional approaches to materials development often rely on sequential experimentation and (mostly) incremental adjustments, making it difficult to efficiently navigate vast and complex materials design spaces. This stepwise, ad-hoc approach inherently limits the speed at which new materials can be discovered and optimized. The slow pace of materials innovation often results in significant delays between the initial conceptualization of a potentially transformative technology and its realization, which depends on the discovery and deployment of the optimal enabling material(s) \cite{maine2006commercializing,national2011materials}.

These limitations have spurred the emergence of more systematic approaches to accelerated materials development. For instance, two decades ago Integrated Computational Materials Engineering (ICME) \cite{national2008integrated} emerged as a strategy to expedite the materials development cycle through the integration of experiments and simulation to create quantitative process-structure-property-performance (PSPP) relationships\cite{arroyave2019systems}. These PSPP linkages help establish causal, quantitative connections between processing conditions and performance and, when inverted, provide a pathway to materials optimization. While, in theory, ICME-based approaches enable the algorithmic optimization of materials, they often encounter significant challenges\cite{panchal2013key}, including the difficulty of seamlessly integrating diverse simulation tools across multiple length scales, the substantial uncertainties present in model and parameters\cite{honarmandi2020uncertainty}, and the high computational costs associated with predictive multi-scale simulations, which preclude the in-line integration of ICME methods within materials development efforts.

Unlike ICME, which integrates feedback loops between experiments and simulations to identify locally optimal process-material combinations, high-throughput (HTP) approaches prioritize the rapid exploration of vast materials design spaces in “open-loop” settings \cite{potyrailo2011combinatorial,jain2011high}. Computational and experimental HTP methods are designed to screen vast materials spaces through combinatorial approaches rapidly, generating extensive datasets that can, in turn, be screened to identify materials with promising properties\cite{rajan2008combinatorial}. However, these methods lack mechanisms for iterative refinement and often struggle with high-dimensional spaces. This focus on broad exploration can result in inefficient resource use, with significant effort devoted to synthesizing, characterizing, testing, and simulating materials in suboptimal regions of the design space. Moreover, HTP approaches often face inherent bottlenecks; while synthesis may be efficiently scaled, characterization techniques frequently lag behind, limiting overall throughput. As a result, the efficiency of HTP workflows is dictated by the slowest step in the pipeline, ultimately constraining the pace and effectiveness of discovery campaigns.  

Machine learning (ML) approaches have increasingly been adopted in materials discovery, complementing existing frameworks like ICME and HTP methods. ML models such as support vector machines, random forests, and active learning strategies have been employed for single-objective optimization, enabling rapid computational screening of candidates \cite{yang2022machine, liu2022accelerated, wen2019machine, rao2022machine, giles2022machine}. However, these methods often lack iterative feedback and focus primarily on computational predictions, with experimental validation deferred to later stages. This limitation is especially pronounced in multi-objective optimization, where experimental efforts are limited \cite{del2020assessing, jablonka2021bias, gopakumar2018multi}. For bulk materials, high synthesis, characterization, and testing costs further delay experimental validation, reducing the overall efficiency of ML-driven discovery.

ICME, HTP, and ML-based methods hold great promise but face distinct challenges. ICME is a powerful approach to generating quantitative PSPP relationships, but struggles with integrating multi-scale simulations, leading to inefficiencies and high computational costs. HTP methods focus on exploring broad design spaces but often lack feedback mechanisms, making them resource-intensive while lacking a path toward systematic refinement. ML approaches offer powerful data-driven tools for property prediction and design optimization but frequently delay experimental validation, particularly for bulk materials where synthesis and testing are expensive. 

Recent advances in Bayesian Optimization (BO) have addressed challenges in materials discovery by employing probabilistic approaches to iteratively select the most informative experiments or simulations based on prior data \cite{frazier2018tutorial,arroyave2022perspective}. For example, Khatamsaz \etal \cite{khatamsaz2023bayesian} applied entropy-based BO for efficient, constrained multi-objective optimization of high-dimensional high-entropy alloy (HEA) spaces, identifying feasible regions before performing detailed optimization. Couperthwaite \etal \cite{couperthwaite2020materials} introduced a batch BO framework that integrates multiple information sources, leveraging parallel high-throughput (HTP) queries to optimize materials design. Vela \etal \cite{vela2023data} combined predictions from physics-based models with sparse literature data using Bayesian methods, enabling robust property predictions for a Bayesian discovery campaign. Acemi \etal \cite{acemi2024multi} explored high-dimensional refractory HEA (RHEA) spaces under multiple objectives and constraints through iterative experiments, incorporating a formal Bayesian step in the campaign’s final stages. Recently, Paramore \etal~\cite{paramore2024two} demonstrated the "two-shot" optimization of a very high-dimensional space, although in this case the objective space was relatively straightforward and the optimization process involved a single Bayesian iteration.

\subsection{Batch-wise Improvement in Reduced materials Design Space using a Holistic Optimization Technique}

\begin{figure}[hbt]
        \centering
        \includegraphics[width=0.9\columnwidth]{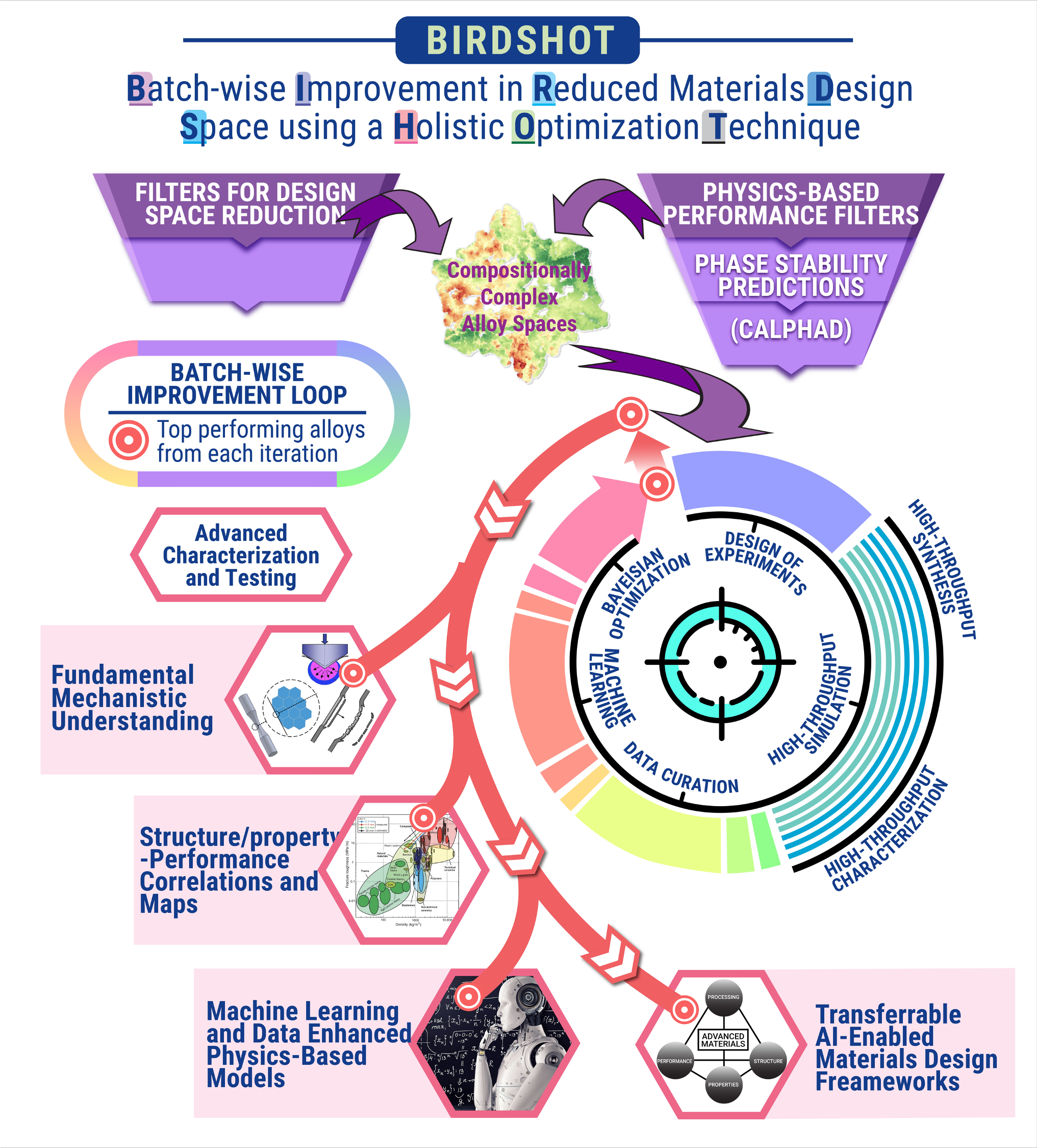}
        \caption{An overview of the BIRDSHOT framework.}
        \label{fig:birdshot} 
    \end{figure}

Building on the recent advancements described above, this work introduces BIRDSHOT, a multi-objective optimization framework that integrates and improves upon ICME, HTP, and ML-based approaches to materials development. BIRDSHOT—Batch-wise Improvement in Reduced Design Space using a Holistic Optimization Technique—accelerates materials discovery through an iterative, closed-loop cycle combining computational tools and experimental protocols. It leverages advanced strategies to address existing limitations by: (i) integrating simulations, physics-based models, and machine learning (ML) with phase stability analysis and intelligent search algorithms to efficiently identify feasible regions for optimization \cite{abu2018efficient,galvan2017constraint}; (ii) fusing computational and experimental data to build robust predictive ML models \cite{khatamsaz2021efficiently,ghoreishi2018multi}; (iii) employing Bayesian Optimization (BO) \cite{arroyave2022perspective} for globally informed, iterative exploration; (iv) utilizing batch adaptations of BO (Batch BO) \cite{couperthwaite2020materials} to evaluate multiple candidates simultaneously; and (v) addressing multiple objectives and constraints to comprehensively tackle complex materials optimization problems \cite{khatamsaz2022multi}. BIRDSHOT is then demonstrated by an iterative, multi-objective optimization campaign over a high-dimensional FCC HEA space.


The main elements of BIRDSHOT are depicted in \autoref{fig:birdshot}: (1) \textit{Filtering}: Utilizing state-of-the-art (SOA) ML-augmented search algorithms and leveraging mechanistic models, the design space is initially narrowed to focus on high-value regions. (2) \textit{Design of Experiments (DOE)}: Novel clustering methods are employed to ensure broad and representative coverage of the feasible design space, maximizing the effectiveness of exploratory experiments. (3) \textit{High-Throughput (HTP) Simulation}: Preliminary stages rely on rapid simulations to predict key properties, helping refine the experimental focus. (4) \textit{HTP Synthesis and Characterization}: The framework integrates advanced synthesis techniques with comprehensive characterization protocols to assess critical properties, including phase stability, mechanical performance, and environmental resilience. (5) \textit{Data Curation and Augmentation}: This involves building and refining feature-rich datasets through systematic experimental validation, enhancing existing models and databases. (6) \textit{Machine Learning Models}: ML models are further refined to efficiently predict material properties, using classifiers and regressors trained on both simulated and experimental data. (7) \textit{Adaptive BO Approaches}: These methods actively adapt the ML models and exploration paths, incorporating multiple information sources, and allowing simultaneous selection of multiple candidates through Batch BO. The ability to make parallel decisions is critical for efficient exploration under practical time and resource constraints.

\subsection{Multi-Objective Optimization of FCC HEA Spaces}

In this study, we used BIRDSHOT to discover new alloys within the system \(\ce{Al_{x1}V_{x2}Cr_{x3}Fe_{x4}Co_{x5}Ni_{x6}}\), where \(x_i\) ranges from 0 to 0.95 in increments of 0.05. The goal was to optimize three key performance metrics: (1) ultimate strength/yield strength ratio, (2) hardness, and (3) strain-rate sensitivity. These properties are critical for dynamic loading applications and are exceedingly challenging to predict computationally. To efficiently navigate this design space, BIRDSHOT applied four constraints: (i) FCC phase stability above \(700\;^\circ \text{C}\), (ii) a solidification range below \(100\;\text{K}\), (iii) thermal conductivity exceeding \(5\;\text{W}/(\text{m} \cdot \text{K})\), and (iv) density under \(8.5\;\text{g}/\text{cm}^{3}\). These constraints ensured the selected compositions met mechanical performance goals and practical requirements, including phase stability, manufacturability, and thermal properties. Applying these filters reduced the design space to approximately 53,000 candidate compositions.

\begin{figure}[hbt]
        \centering
        \includegraphics[width=1.0\columnwidth]{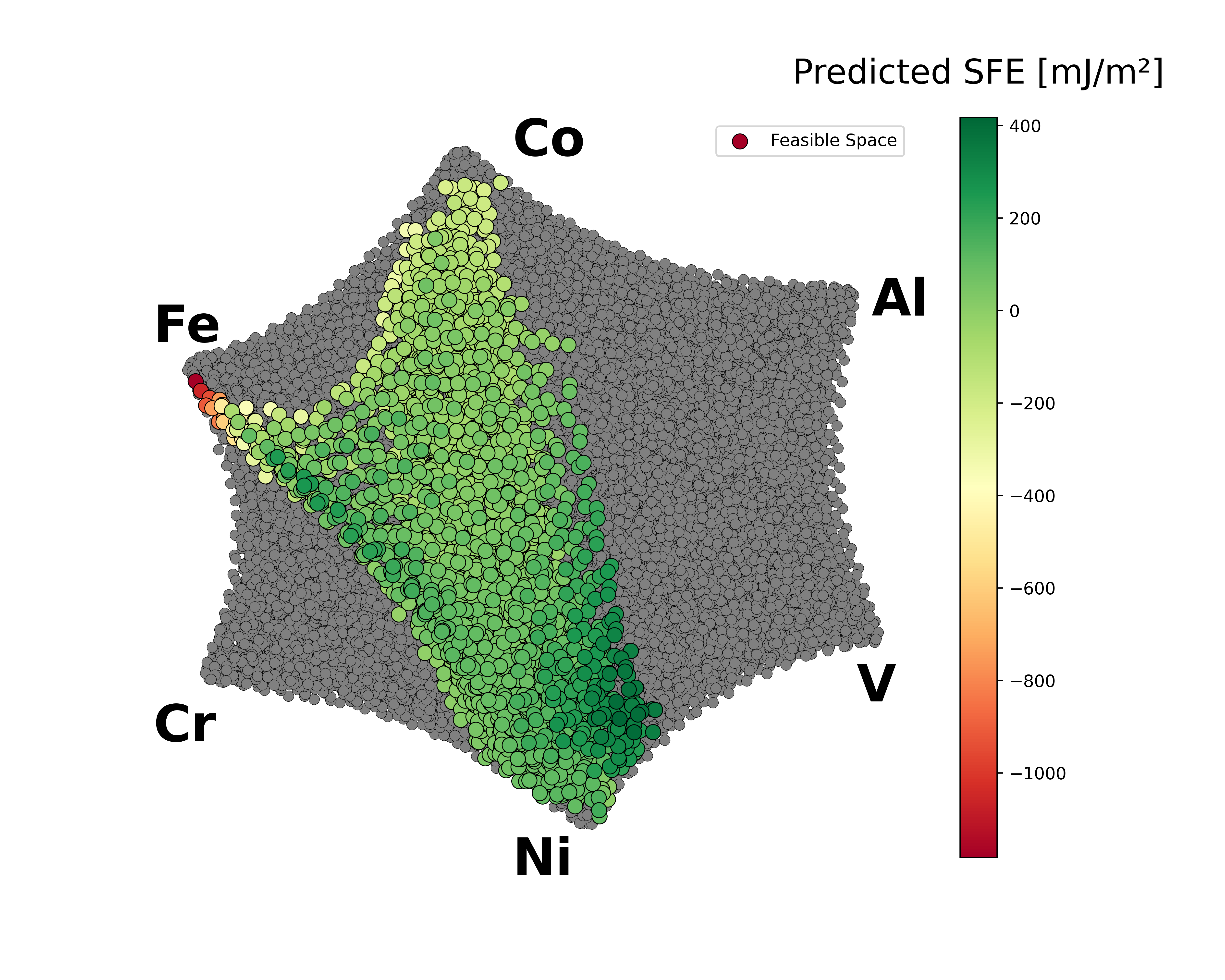}
        \caption{Visualization of the DFT-predicted stacking-fault energy (SFE) across a UMAP embedding of the CoCrFeNiVAl composition space. The data points are color-coded according to their SFE values, with darker red representing lower SFE values and green indicating higher SFE values. The plot highlights the distribution of SFE within the composition space, with the feasible space filtered according to constraints designed for the problem. Reproduced with permission from Khan\etal~\cite{khan2022towards}.}
        \label{fig:sfe} 
    \end{figure}

The design of this discovery campaign was further informed by the hypothesis that differences in the stacking fault energy (SFE) can lead to distinct deformation mechanisms, significantly impacting the mechanical behavior of the alloys. In face-centered cubic (FCC) materials, SFE plays a crucial role in determining whether deformation occurs primarily through dislocation slip, twinning, or phase transformation. Alloys with low SFE are more prone to twinning-induced plasticity (TWIP) and transformation-induced plasticity (TRIP), which enhance ductility and strain hardening. Conversely, higher SFE values favor slip-dominated deformation.

To explore these effects, the design space was divided into two categories based on SFE: alloys with low SFE (below $50\;\text{mJ}/\text{m}^{2}$) and those with high SFE (above $50\;\text{mJ}/\text{m}^{2}$). This division allowed for targeted optimization within each category, focusing on how SFE could modulate mechanical responses under varying loading conditions. SFE predictions across the alloy space were done using a machine learning (ML) model trained on data derived from density functional theory (DFT) calculations, developed by Khan \etal~\cite{khan2022towards}. This ML model enabled efficient and accurate screening of the vast composition space without requiring computationally expensive direct simulations---see \autoref{fig:sfe}. 

BIRDSHOT’s Bayesian optimization (BO) loop was run separately for these two subspaces, allowing for parallel exploration. SFE estimates were generated using a machine learning model trained on DFT-derived data, enabling accurate classification across the SFE groups. The overall exploration involved multiple iterative loops, incorporating advanced experimental setups. Alloys were synthesized via Vacuum Arc Melting (VAM) and then comprehensively characterized using Scanning Electron Microscopy (SEM), Electron Backscatter Diffraction (EBSD), and X-ray Diffraction (XRD). Mechanical properties were assessed through tensile testing and high strain-rate nanoindentation.

\section{Methods}   
\label{s3_methods}

\begin{figure*}[htb]
    \centering
    \includegraphics[width=1.0\linewidth]{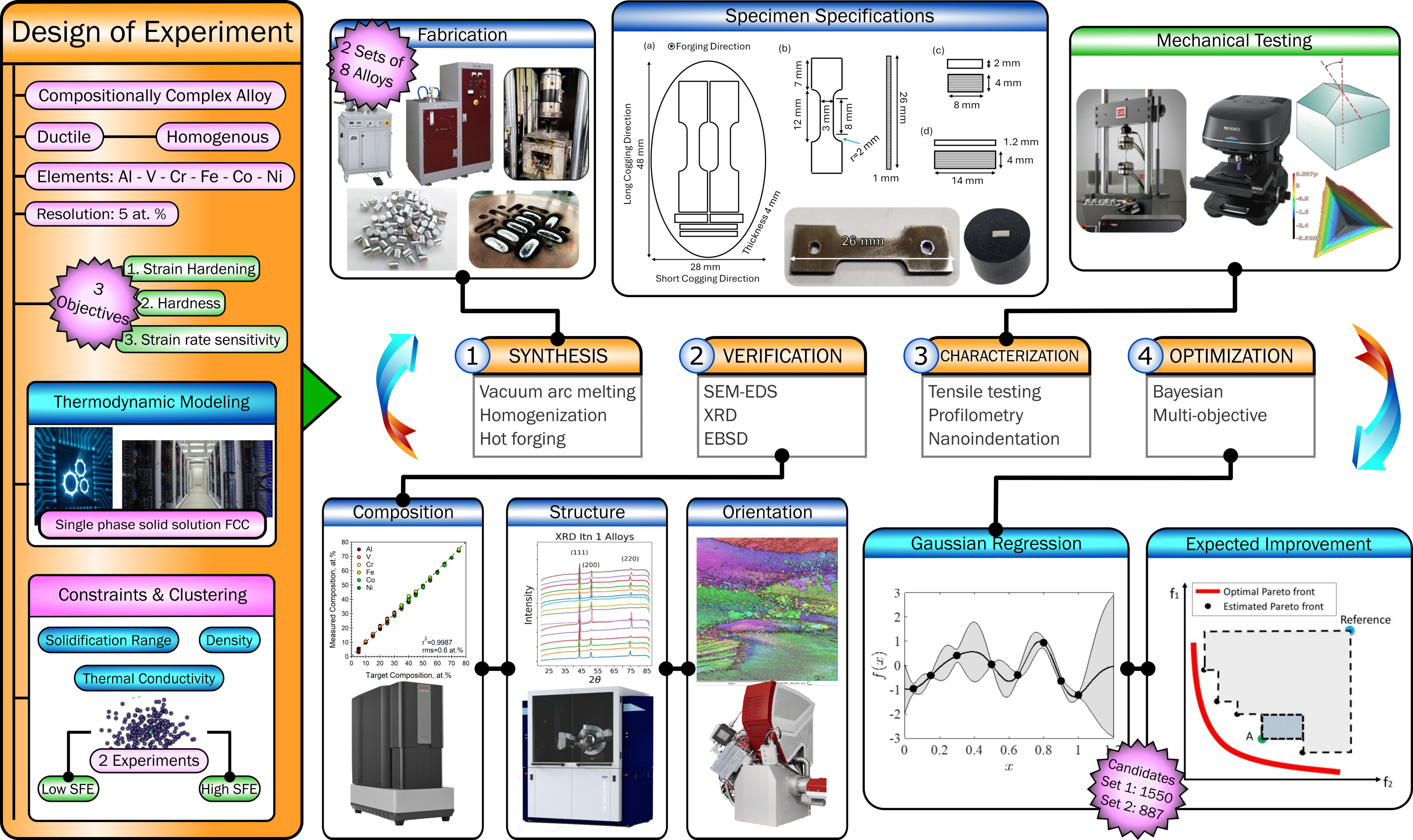}
    \caption{The BIRDSHOT Materials Discovery framework for the complex \ce{Al-V-Cr-Fe-Co-Ni} alloy system. The project scope is outlined and reduced to a small pool of testable candidates. Materials are synthesized in batches, their compositions are verified, and their properties of interest are characterized. After acquiring experimental data, all iterations of data are provided to a Batch BO algorithm to make suggestions for the next batch of materials. These suggestions are then cross-checked for their expected equilibrium phases, and the process is repeated.}
    \label{fig:schematic} 
\end{figure*}

One of the primary objectives of BIRDSHOT is to accelerate the discovery of the Pareto set—the set of optimal solutions in a multi-objective optimization problem where improving one objective requires compromising another—in a high-dimensional alloy space by iteratively refining each step of a collaborative workflow. This approach emphasizes both efficiency and accuracy while advancing scientific understanding. A key challenge lies in managing complex workflows, handling large data volumes generated during processing and characterization, and systematically analyzing this information to address specific application targets. \autoref{fig:schematic} provides a schematic overview of the integrated computational and experimental approaches used in this work, highlighting the identification of candidate alloys, adherence to compositions and constraints, and evaluation of key performance metrics.

\vspace{-0.4em}\subsection{Design}\vspace{-0.2em}

\subsubsection{Filtering}

The \(\ce{Al-V-Cr-Fe-Co-Ni}\) system, along with all its possible subsystems from two to six elements, was selected based on the well-studied \(\ce{Cr-Fe-Co-Ni}\) system\cite{cantor2021multicomponent}. The compositional space was defined with a resolution of 5 at.\%, resulting in a design space comprising $\sim$53,000 compositions. Aluminum was included to further control stacking fault energy (SFE)\cite{khan2022towards}, while vanadium was added due to its role in enhancing mechanical properties, as observed in V-Co-Ni ternary alloys, which are known for their high ultimate tensile strengths \cite{sohn2019ultrastrong}. 

To guide the alloy design process, CALPHAD (CALculation of PHAse Diagrams) predictions were employed using Thermo-Calc with the TCHEA v5.1 to predict the phase stability and thermo-physical properties across the  
\(\ce{Al-V-Cr-Fe-Co-Ni}\) alloy space. Modern CALPHAD methods have demonstrated success in accurately predicting single-phase solid solutions in complex alloys by considering higher-order mixing parameters \cite{li2022superior, senkov2015accelerated, zhang2014understanding, gao2015design, wang2012effects, liu2022formation, tsai2014high, gao2016senary}. Alternative approaches for predicting solid solution stability---detailed in the \emph{Supplementary Information} (\textbf{SI})---proved inadequate for this high-dimensional problem. The TCHEA database, tailored for multi-principal element alloys (MPEAs), offers robust thermodynamic models to understand the behavior of these complex systems. The key design constraints included phase stability, solidification range, density, and thermal conductivity, all critical for ensuring manufacturability and reliable performance under realistic conditions. 

The goal was to maintain a high volume fraction of the face-centered cubic (FCC) structure (\(\sim100\%\)) above \(700^{\circ} \mathrm{C}\), ensuring stability upon cooling to minimize phase transformations during service. A single-phase FCC structure simplifies processing and enhances predictability in manufacturing and service conditions. Phase fraction predictions from the TCHEA v5.1 database were used to screen compositions, with phase stability confirmed through XRD analysis of homogenized samples to address potential inaccuracies in equilibrium-based predictions. A narrow solidification range (\(<100\,K\)) was targeted to reduce solidification defects and improve manufacturability, with liquidus and solidus temperatures predicted using Thermo-Calc's Property Model Calculator. Thermal conductivity above \(5\,W/(m\cdot K)\) at room temperature ensured efficient heat dissipation, minimizing overheating and thermal fatigue risks. Density was limited to \(<8.5\,g/cm^{3}\) to support weight-sensitive applications, with predictions based on atomic weights and phase fractions of the constituents.

\subsubsection{Candidate Selection}

The initial compositional space comprised 53,124 possible alloy combinations. Applying four constraints—FCC phase stability \(> 700^{\circ} \mathrm{C}\), solidification range \(< 100 \,K\), thermal conductivity \(> 5\,W/(m\cdot K)\), and density \(< 8.5\,g/cm^{3}\)---refined the design space to 2,437 feasible candidates. As described above, these candidates were further examined for potential deformation and strengthening mechanisms based on their predicted stacking fault energies (SFE). SFE values were calculated using a machine learning model trained on density functional theory (DFT) data \cite{khan2022towards}. A Gaussian Process Regressor (GPR) trained on 498 DFT-calculated compositions, with a Matern kernel, achieved \(R^2 = 0.939\) using 320 medoids. GPR and Support Vector Regressors (SVR) produced similar root mean squared errors (RMSE): \(22.6 \,mJ/m^{2}\) and \(24.8 \,mJ/m^{2}\), respectively. The GPR model was used to classify the alloys into two subsets: 1,550 with "low" SFE (\(< 50 \,mJ/m^{2}\)) and 887 with "high" SFE (\(> 50 \,mJ/m^{2}\)).

A key challenge in alloy design is minimizing bias in the selection of initial samples for exploration. Traditional design of experiments (DOE) methods, including random sampling, often fail to provide comprehensive coverage of the design space, leading to inefficiencies. To address this, we used a space-filling approach with a K-Medoids clustering algorithm \cite{kaur2014k}, which systematically partitions the compositional dataset into clusters and selects a representative medoid point from each. This ensures unbiased and diverse sampling, avoiding over-representation of specific regions and guaranteeing that selected points correspond to real, feasible compositions. As demonstrated in prior work by the authors \cite{vela2023high,couperthwaite2020materials,khan2022towards}, this method enables effective exploration of the design space. For both the low and high SFE groups, 8 well-distributed compositions were selected, providing a strong foundation for subsequent testing and optimization.

\subsubsection{Curation}\vspace{-0.2em}

Data curation is an essential step in any optimization framework, as all optimization methods fundamentally assume that both the user and models operate perfectly—a presumption that will inevitably be incorrect to some degree. In this study, \textit{valid} materials were defined by their compliance with single-phase solid solution constraints; however, despite employing robust models, certain compositions may not meet these criteria. When such designs were identified, they were excluded from the optimization scheme. A complete dataset of this information for all alloys is available in the \textbf{SI}, and the extent of materials that failed to meet project constraints is presented in \autoref{s4_results}.

\subsection{Synthesis}\vspace{-0.2em}

\begin{figure}[hbt]
    \centering
    \includegraphics[width=0.85\columnwidth]{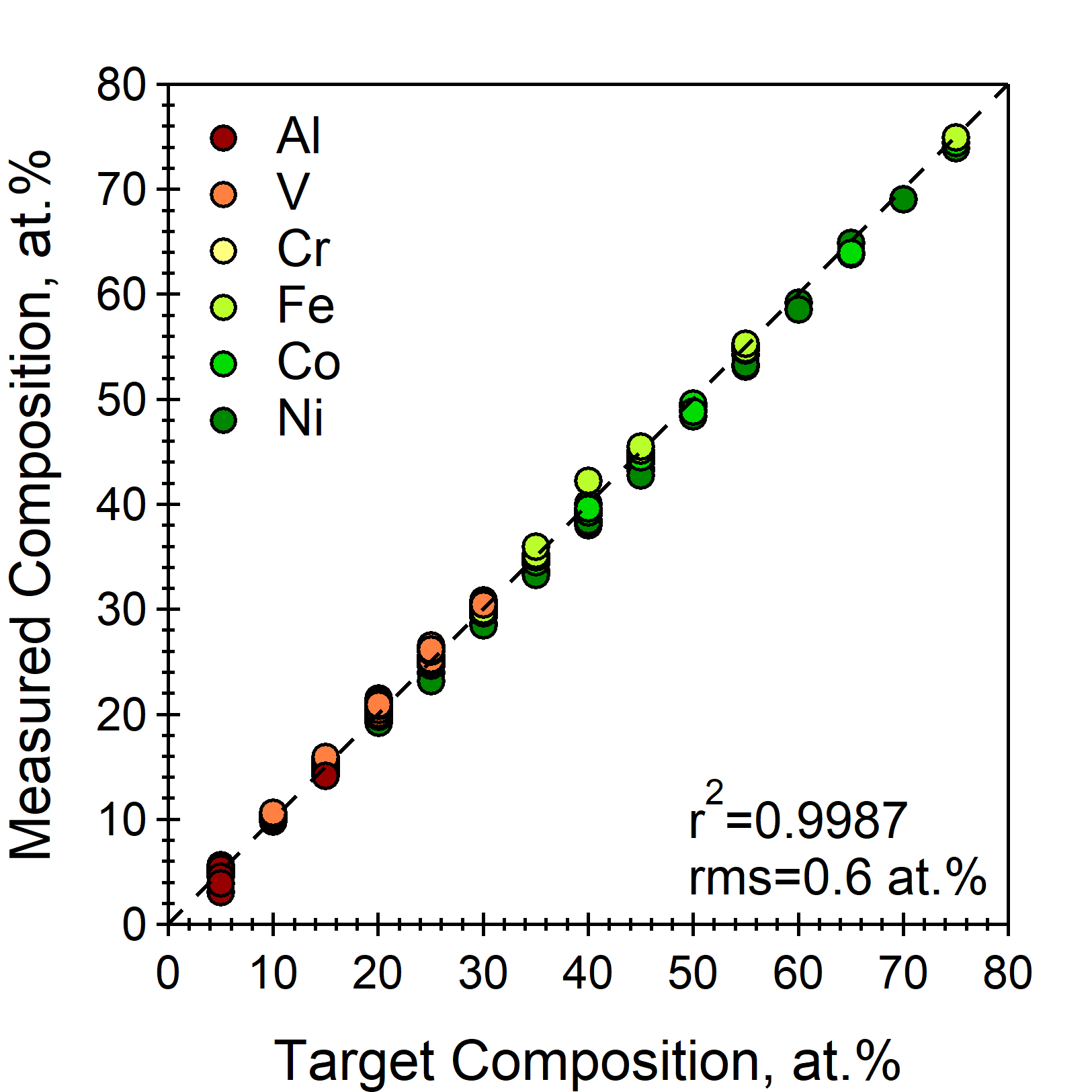}
    \caption{Comparison between target and realized chemistries in the samples generated in this project.}
    \label{fig:EDS}
\end{figure}

All designed alloys were synthesized from high-purity elements ($>$99.9 wt.\%) using an Edmund Bühler AM 200 Vacuum Arc Melter (VAM) to produce 30 g ingots. The fabrication process involved initially melting all elements in copper crucibles under a high-purity Ar atmosphere using arc melting. To maximize chemical homogeneity, each ingot was flipped and remelted seven times. Throughout the fabrication process, weight loss was recorded to monitor material integrity. Samples that exhibited weight losses above a certain threshold were discarded and re-synthesized. After melting, the ingots were solution heat-treated at 1,150$^{\circ}$C for 24 hours using a Centorr high-temperature furnace (LF Series, Model 22). The atmosphere within the furnace's chamber was purged with Ar gas at least three times down to $5\cdot10^{-5}\;torr$, and the thermal treatment was performed under a controlled, very low-pressure Ar atmosphere ($<10^{-2}\;torr$). Finally, the material was slowly cooled in the furnace.

Next, the ingots were wrapped in a protective stainless steel foil, including small pieces of titanium sponge as a sacrificial material. They were soaked for 30 minutes at 1150 °C, hot-forged in a single step in a press pre-heated to 400 °C, yielding an approximate 65\% thickness reduction, and then slowly cooled in air. The stainless steel foil was subsequently removed. For the as-forged ingots shown in \autoref{fig:schematic}, one can distinguish between long and short rolling directions perpendicular to the forging loading axis, as the original length and width of the samples are approximately 39 mm and 15 mm, respectively.

Next, sample profiles were cut via wire electrical discharge machining (wire-cut EDM). These samples included:

\begin{itemize}
    \setlength{\itemsep}{0pt} 
    \item Four dog bone-like miniaturized flat tension samples (a variation of SS-3 tension specimens \cite{klueh1985miniature,gussev2017sub}) with a total length of $26\;mm$, a gauge length of $8\;mm$, and a typical cross-section of $3\cdot1\;mm^2$, with the loading direction parallel to the long rolling direction (see \autoref{fig:schematic}), used for quasistatic tension tests.
    \item Two samples with a thickness of $2\;mm$ and a face of $8\cdot4\;mm^2$ normal to the long rolling direction (see \autoref{fig:schematic}), used for nanoindentation experiments.
    \item Two samples with a thickness of $1.2\;mm$ and a face of $14\cdot4\;mm^2$ normal to the long rolling direction (see \autoref{fig:schematic}, employed for SEM/EBSD, chemical (EDS), XRD, and Vickers hardness analyses.    
\end{itemize}

Finally, the tension samples were rough-polished to ensure parallel faces in the grip section, and pinholes were drilled. The $8\cdot4$ and $14\cdot4$ flat samples were then prepared to achieve the high-quality surface required for microscopy techniques and nanoindentation.

\subsection{Verification}\vspace{-0.2em}

\subsubsection{Metallography}

Samples were polished for nanoindentation and general microscopy using a Buehler AutoMet 250 autopolisher. A Buehler Vibromet 2 vibratory polisher was used specifically for electron backscatter diffraction (EBSD) but not for nanoindentation, as the surfaces it produces are unsuitable for that application.

\subsubsection{Chemical, Microstructural, and Phase Analyses}

Compositional measurements were completed using a Tescan FERA-3 Model GMH Focused Ion Beam Microscope equipped with an energy-dispersive X-ray spectroscopy detector (EDS) at a voltage of 15 kV. EDS point measurements were acquired for 50 seconds each across three separate sections, with five points per section. The overall ingot composition was determined as the average of these 15 data points, paying particular attention to possible compositional differences between sections to detect any chemical inhomogeneities. The ingots were found to be homogeneous, and the average variance between the measured and target element contents, shown in \autoref{fig:EDS}, was 0.6\%, with 90\% of the cases deviating less than 1\%. Atomic fractions of each fabricated alloy are included in the \textbf{SI}.

X-ray diffraction (XRD) data were obtained at room temperature using a Bruker D8 Discover diffractometer equipped with a Cu K-$\alpha$ X-ray source and a collimator with a $1 \, \text{mm}$ aperture. Diffraction spectra were obtained using a Vantec 500 area detector, covering a diffraction angle between 21 and 85 degrees. The spectra were evaluated using MAUD software.

The microstructure resulting from the thermomechanical processing of the ingots was characterized using a Phenom XL scanning electron microscope (SEM). Backscattered electron (BSE) images were taken of each sample at a working distance of 6 mm from the undeformed alloys. EBSD of the samples was performed in the same SEM, using the Oxford Symmetry CMOS-based EBSD detector at a voltage of 20 kV and a current of 20 nA. Crystallographic orientation maps were obtained on a surface normal to the cold rolling direction, with an area of $1 \times 1 \, \text{mm}$ and a step size of 1 $\mu$m. The \textbf{SI} includes a collage of representative microscopy images.

\subsection{Characterization}\vspace{-0.2em}

\subsubsection{Quasistatic Tensile Deformation}

Isothermal tensile tests were conducted using an MTS hydraulic test system with a 2500 lbf (11.12 kN) load cell, operating at a strain rate of \(5 \times 10^{-4} \, \text{s}^{-1}\), and an MTS extensometer directly attached to the gauge section of the specimens. Each test captured the elastic and plastic deformation regimes up to fracture. True stress-strain (\(\sigma\)-\(\epsilon\)) curves were generated for each alloy to determine the true stress at ultimate tensile stress (UTS) and yield stress (YS) at a 0.2\% offset, allowing for calculation of the strain hardening ratio (unitless). A sample stress-strain curve for a representative alloy, indicating data extraction for modulus, yield, and ultimate tensile strengths, is provided in the \textbf{SI}.
    
\subsubsection{Nanoindentation}

\begin{figure}[hbt]
    \centering
    \includegraphics[width=0.99\columnwidth]{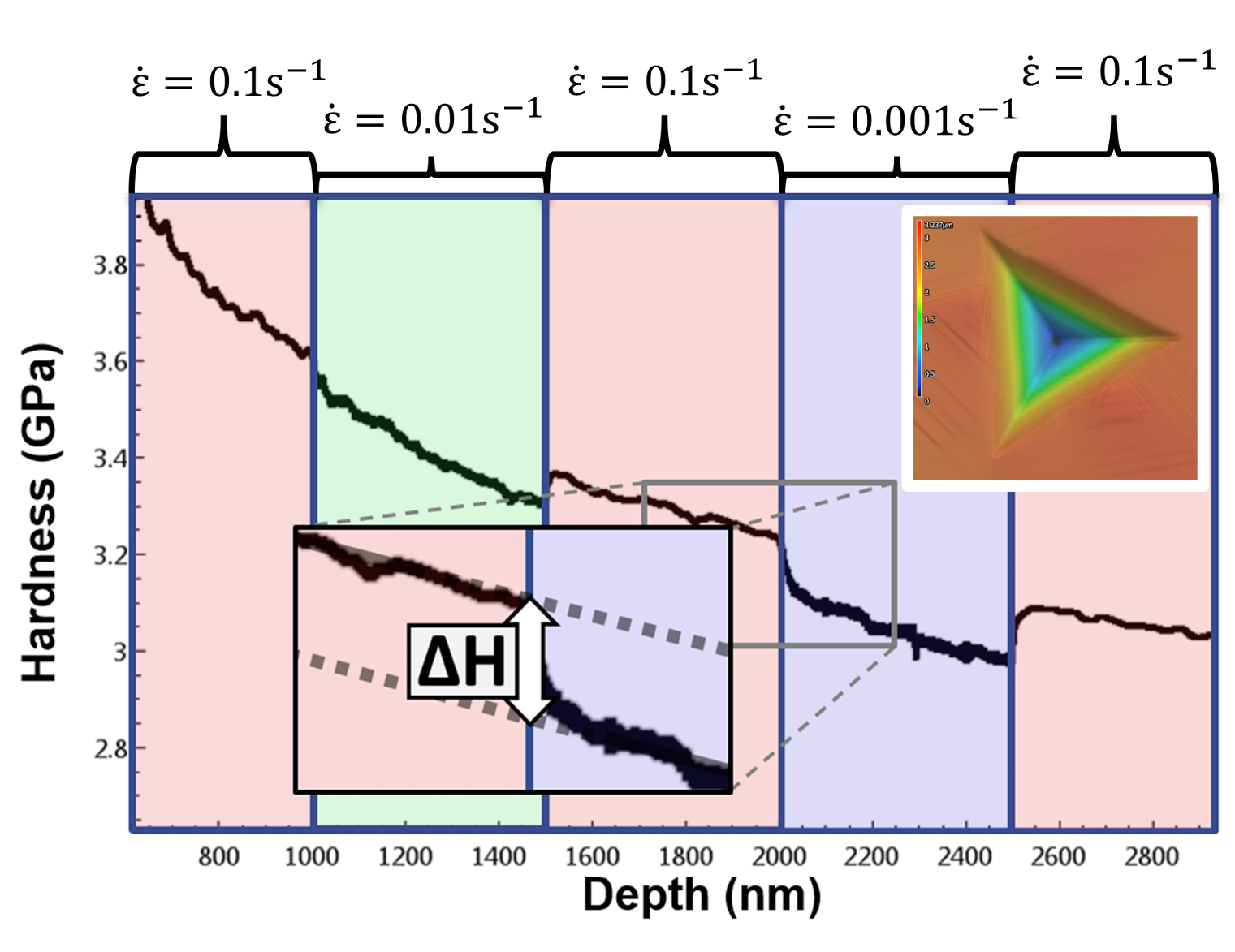}
    \caption{Nanoindentation tests for a representative alloy. SRJT data and optically measured indent (inset). Pile-up is visible outside the indent periphery in regions with positive depth values.}
    \label{fig:NI_SRJT}
\end{figure}

Each sample was indented using a KLA Instruments iMicro nanoindenter equipped with a Berkovich diamond tip. Hardness ($H$) and modulus ($E_r$) were calculated based on the methods of Oliver and Pharr \cite{oliver1992improved}, using \autoref{hardness} and \autoref{reduced_modulus}, respectively.

\begin{equation}
\begin{aligned}
\label{hardness}
H &= \frac{P}{A_c}
\end{aligned}
\end{equation}

\begin{equation}
\begin{aligned}
\label{reduced_modulus}
E_r &= \frac{\sqrt{\pi}}{2} \frac{S}{\sqrt{A_c}}
\end{aligned}
\end{equation}

While these methods account for elastic contributions to the indentation depth and corresponding area, they do not consider the additional contact area caused by plastic deformation, where material tends to pile-up around the indenter tip. To account for this pile-up, we defined the ratio between the projected area of the indent including pile-up ($A_{c_{\text{pile-up}}}$) and the projected area of the indent excluding pile-up, defined as the area below the sample surface plane ($A_{c_{\text{beneath-surface}}}$) as shown in \autoref{pileup_ratio}:

\begin{equation}
\begin{aligned}
\label{pileup_ratio}
\xi &= \frac{A_{c_{\text{pile-up}}}}{A_{c_{\text{beneath-surface}}}}
\end{aligned}
\end{equation}

These ratios were measured for every indent using optical profilometry with a Keyence VKX1000 3D Surface Profiler. Corrected hardness and corrected reduced modulus were then calculated using \autoref{corrected_hardness} and \autoref{corrected_reduced_modulus}, respectively.

\begin{equation}
\begin{aligned}
\label{corrected_hardness}
H_{\text{corrected}} &= \frac{P}{A_c} \cdot \frac{A_{c_{\text{beneath-surface}}}}{A_{c_{\text{pile-up}}}} = \frac{H}{\xi}
\end{aligned}
\end{equation}

\begin{equation}
\begin{aligned}
\label{corrected_reduced_modulus}
E_{r_{\text{corrected}}} &= \frac{\sqrt{\pi}}{2} \frac{S}{\sqrt{A_c}} \cdot \sqrt{\frac{A_{c_{\text{beneath-surface}}}}{A_{c_{\text{pile-up}}}}} = \frac{E_r}{\sqrt{\xi}}
\end{aligned}
\end{equation}

The strain rate sensitivity exponent ($m$) was determined based on the power-law relationship, using the strain rate jump test (SRJT) methods of Maier-Kiener and Durst \cite{maier2017advanced}.

\begin{equation}
\begin{aligned}
\label{strain_rate_hardening_exp}
m &= \frac{\partial \ln H}{\partial \ln \dot{\epsilon}} 
\approx \frac{\ln H_2 - \ln H_1}{\ln \dot{\epsilon}_2 - \ln \dot{\epsilon}_1}
\end{aligned}
\end{equation}

Strain rates were controlled by adjusting the loading rate as described by Lucas and Oliver \cite{lucas1999indentation}, where the strain rate can be approximated from the loading rate, as shown in \autoref{strain_rate}.

\begin{equation}
\begin{aligned}
\label{strain_rate}
\dot{\epsilon} &= \frac{\dot{h}}{h} = \frac{1}{2} \left( \frac{\dot{P}}{P} - \frac{\dot{H}}{H} \right) \approx \frac{1}{2} \frac{\dot{P}}{P}
\end{aligned}
\end{equation}

Nanoindentation data for a representative alloy is presented in \autoref{fig:NI_SRJT}. Indents were performed to a depth of 3000 nm, with strain rate jumps at depths of 1000, 1500, 2000, and 2500 nm. The strain rate began at 0.2 s\(^{-1}\) and sequentially changed to 0.02, 0.2, 0.002, and back to 0.2 s\(^{-1}\). Linear fits were applied to the hardness data before and after each strain rate jump, excluding hardness data during the transition period to account for settling time. The fitted data were extrapolated to depths nearest the rate jumps, and the change in hardness with respect to strain rate was used to calculate the strain rate sensitivity exponent for each jump.

\subsection{Optimization}\vspace{-0.2em}

\begin{figure*}[h!]
    \centering
    \includegraphics[width=0.90\textwidth]{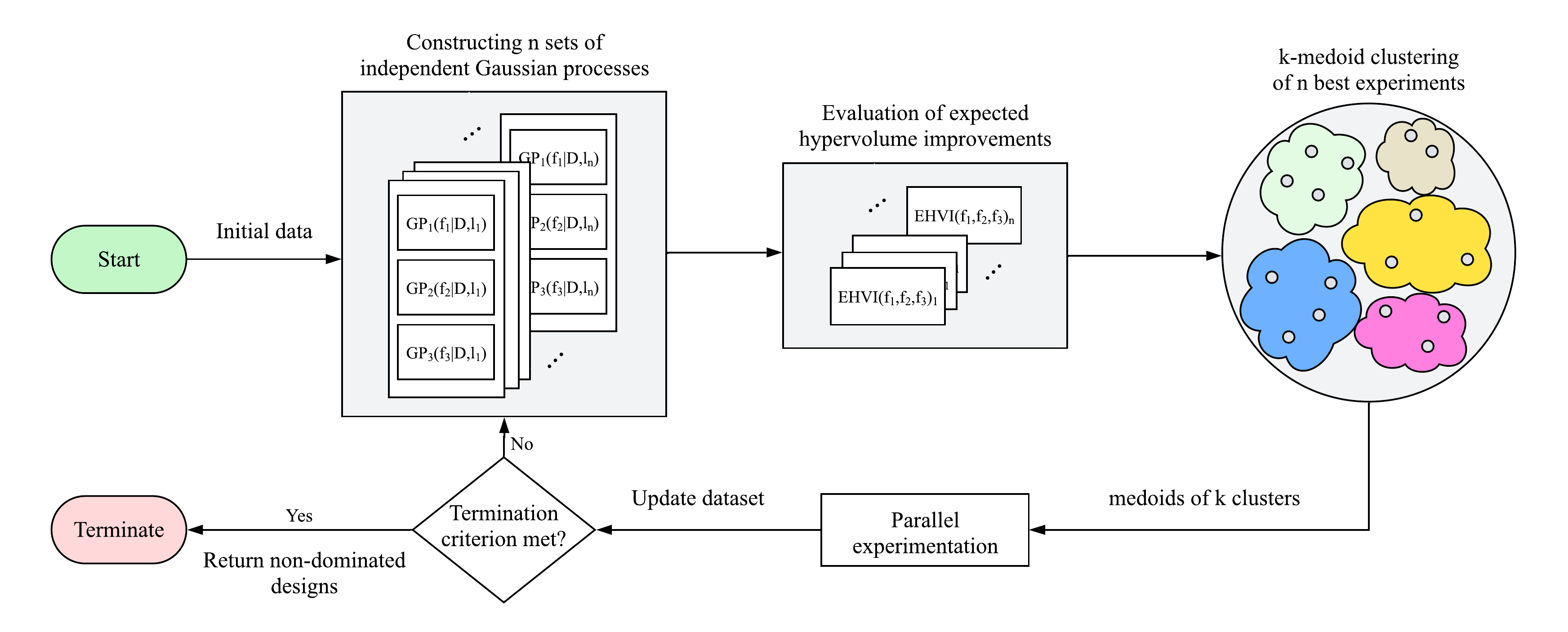}
    \caption{A step-by-step representation of the closed-loop 3-objective batch Bayesian optimization framework. Initial data, generated randomly as the first design stage, is used to construct $n$ sets of Gaussian process models with different length scales. Expected hypervolume improvements are calculated for each candidate, and a k-medoid clustering approach groups $n$ suggestions into the appropriate batch size. The cluster medoids are then proposed for parallel experimentation. The loop repeats until a termination criterion is met.}
    \label{fig:flowchart}
\end{figure*}

Bayesian Optimization (BO)\cite{frazier2018tutorial} is an effective framework for black-box optimization, particularly suited to applications where objective functions are expensive, lack analytical form, and do not provide gradients. BO operates over a predefined design space, representing the range of potential solutions under exploration. A central element of BO is the surrogate model, often a Gaussian Process (GP)\cite{williams1995gaussian,deringer2021gaussian}, which makes probabilistic predictions across this design space and is iteratively updated with new data through Bayesian updating. The acquisition function\cite{wilson2018maximizing} then leverages the surrogate model’s predictions to balance exploration (sampling lesser-known regions) and exploitation (refining areas with high potential), guiding each experiment to maximize the overall utility---measured in terms of improved performance or knowledge gain---while optimizing resource optimization. Further enhancements to BO may include batch querying\cite{couperthwaite2020materials}, allowing for simultaneous evaluation of multiple design points to speed convergence and multi-source fusion\cite{ghoreishi2018multi}, where data from various sources (e.g., experimental and simulated data) refines the surrogate model’s accuracy.

\subsubsection{Gaussian Process Regression}

Bayesian experimental design uses surrogate models to approximate costly black-box objective functions. Ideally, such models can be used to provide computationally efficient predictions, guiding decisions on the most informative experiments to run. Gaussian Process regression (GP) is a popular choice for this purpose and is widely used across various domains \cite{williams1995gaussian,williams2006gaussian}. A GP is a probabilistic model that uses distance-based correlations between data points to predict outputs---with quantified uncertainty---in unobserved regions. GPs are relatively easy to update through Bayesian updating, conditioned with new data ,and have other mathematical properties that make them especially suitable surrogate models for BO.

Assume a set of $N$ observed data points, denoted by $\{\mathbf{X}_{N}, \mathbf{y}_{N}\}$, where $\mathbf{X}_{N} = (\mathbf{x}_{1}, \ldots, \mathbf{x}_{N})$ and $\mathbf{y}_{N} = \left(f(\mathbf{x}_{1}), \ldots, f(\mathbf{x}_{N})\right)$. The GP provides a probabilistic prediction at an unobserved location \(\mathbf{x}\), modeled as a normal distribution:

\begin{equation}
\label{GP11}
f_{\textrm{GP}}(\mathbf{x}) \mid \mathbf{X}_{N}, \mathbf{y}_{N} \sim \mathcal{N}\left(\mu(\mathbf{x}),\sigma_{\textrm{GP}}^2(\mathbf{x})\right)
\end{equation}

with mean and variance given by:

\begin{equation}
\begin{aligned}
\label{meancov}
\mu(\mathbf{x}) &= K(\mathbf{X}_{N},\mathbf{x})^{\textrm{T}}[K(\mathbf{X}_{N},\mathbf{X}_{N})+\sigma^2_{n}I]^{-1} \mathbf{y}_{N}\\
\sigma_{\textrm{GP}}^2 (\mathbf{x})  &=  k(\mathbf{x},\mathbf{x}) - K(\mathbf{X}_{N},\mathbf{x})^{\textrm{T}} [K(\mathbf{X}_{N},\mathbf{X}_{N})+\sigma^2_{n}I]^{-1}K(\mathbf{X}_{N},\mathbf{x})
\end{aligned}
\end{equation}

where \( k \) is a kernel function that calculates correlations. \( K(\mathbf{X}_{N},\mathbf{X}_{N}) \) is an \(N \times N\) matrix with entries \(k(\mathbf{x}_{m}, \mathbf{x}_{n})\), and \( K(\mathbf{X}_{N}, \mathbf{x})\) is an \(N \times 1\) vector with entries \(k(\mathbf{x}_{m}, \mathbf{x})\). Experimental errors are accounted for with \(\sigma^2_{n}\).

A common choice for the kernel function is the squared exponential:

\begin{equation}
\label{eq:5}
k(\mathbf{x},\mathbf{x'}) = \sigma_s^2 \exp\left(- \sum_{h = 1}^{d} \frac{(x_h - x'_h)^2}{2 l_h^2}\right)
\end{equation}

where \( d \) is the dimensionality of the input space, \(\sigma_s^2\) is the signal variance as a measure of model uncertainty, and \( l_h \) for \( h = 1, 2, \ldots, d \) are the characteristic length-scales, determining correlation strength between data points across each dimension. Better performance can be achieved if these kernel functions are constructed with physics priors\cite{vela2023data,khatamsaz2023physics}.

\subsubsection{Multi-Objective Optimization}

A multi-objective optimization problem, typically defined in terms of minimization, is expressed as:

\begin{equation}
\label{eq:problem}
    \text{minimize} \,\, \{f_1(\mathbf{x}), \ldots, f_n(\mathbf{x})\}, \quad \mathbf{x} \in \mathcal{X}
\end{equation}

where $f_1(\mathbf{x}), \ldots, f_n(\mathbf{x})$ are the objective functions, and $\mathcal{X}$ is the feasible design space. In multi-objective settings, objectives often conflict, meaning a single solution that optimally satisfies all objectives may not exist. Instead, solutions are evaluated based on non-dominance, leading to a set of optimal solutions known as the Pareto front. Optimal solutions, $\mathbf{y}$, for a multi-objective problem with $n$ objectives, are denoted as $\mathbf{y} \prec \mathbf{y}'$ and are defined by:

\begin{equation}
\begin{aligned}
   \{\mathbf{y} : \mathbf{y} & = (y_1, y_2, \ldots, y_n), \; y_i \leq y_i' \quad \forall \; i \in \{1, 2, \ldots, n\}, \\ 
   & \exists \; j \in \{1, 2, \ldots, n\} : y_j < y'_j\}
\end{aligned}{}
\end{equation}

where $\mathbf{y}' = (y_1', y_2', \ldots, y_n')$ represents any possible objective output. The set of all non-dominated solutions, $\mathbf{y} \in \mathcal{Y}$, where $\mathcal{Y}$ is the objective space, is known as the Pareto front.

Various techniques exist for approximating or discovering the Pareto front, including the weighted sum approach \cite{marler2010weighted}, adaptive weighted sum \cite{kim2005adaptive}, normal boundary intersection \cite{das1998normal}, and hypervolume indicator methods \cite{beume2009s, emmerich2011hypervolume, bradstreet2010fast, fonseca2006improved, russo2013quick, yang2007novel, zitzler1999multiobjective}. Among these, hypervolume-based methods are particularly well-suited for Bayesian optimization due to their computational efficiency and compatibility with the iterative, stochastic nature of Bayesian design. In this context, the hypervolume represents the space enclosed by a fixed reference point and the approximated Pareto front in the objective space, providing a comprehensive metric for evaluating improvement during optimization. The goal is to maximize hypervolume, progressively advancing the approximated Pareto front toward the true Pareto front.

The selection of the next experiment is guided by the 'Expected HyperVolume Improvement' (EHVI), a criterion that is mathematically analogous to the Expected Improvement (EI) used in single-objective optimization. A key feature of EHVI is its ability to function as a scalar metric for experimental utility in a multi-objective setting. Regardless of the number of objectives being optimized, EHVI identifies the single best design point (in this case, an alloy) to query at any given time. This property enables EHVI to guide an optimal sequence of experiments, ensuring maximum expected gain at each stage of the experimental campaign. 

Following Ref. \cite{zhao2018fast}, we use a recursive approach for exact hypervolume and EHVI computations. As detailed in \cite{feliot2017bayesian}, EHVI is formulated as:

\begin{equation}
    \mathbb{E}[\textrm{HVI}({\mathbf{y}})] = \int_U \mathbb{P}(\mathbf{y} \prec {\mathbf{y}'}) d{{\mathbf{y}'}}
    \label{Expected}
\end{equation}

where $\mathbb{P}(\mathbf{y} \prec \mathbf{y}')$ is the probability that $\mathbf{y}'$ dominates $\mathbf{y}$, and $U$ represents the dominated hypervolume, bounded by the reference point.

The GP posterior given data is a random variable identified as $y_i \sim \mathcal{N}(\mu_i, \sigma_i^2)$ for $i \leq m$, where $\mu_i$ and $\sigma_i^2$ are the mean and variance of the $i^{th}$ objective. Independent GPs are built for each objective function. The probability that a new solution can dominate the current Pareto front is given by:

\begin{equation}
    \mathbb{P}(\mathbf{y} \prec \mathbf{y}') = \prod_{i=1}^m \Phi \left( \frac{y_i' - \mu_i}{\sigma_i} \right)
    \label{prob}
\end{equation}

where $\Phi$ is the cumulative distribution function (CDF) of the standard normal random variable. For more details on the closed-form expression of (\autoref{Expected}), see Refs. \cite{zhao2018fast, feliot2017bayesian, while2011fast, jaszkiewicz2018improved, khatamsaz2020efficient}.

\begin{figure*}[ht!]
    \centering
    \includegraphics[width=0.95\textwidth]{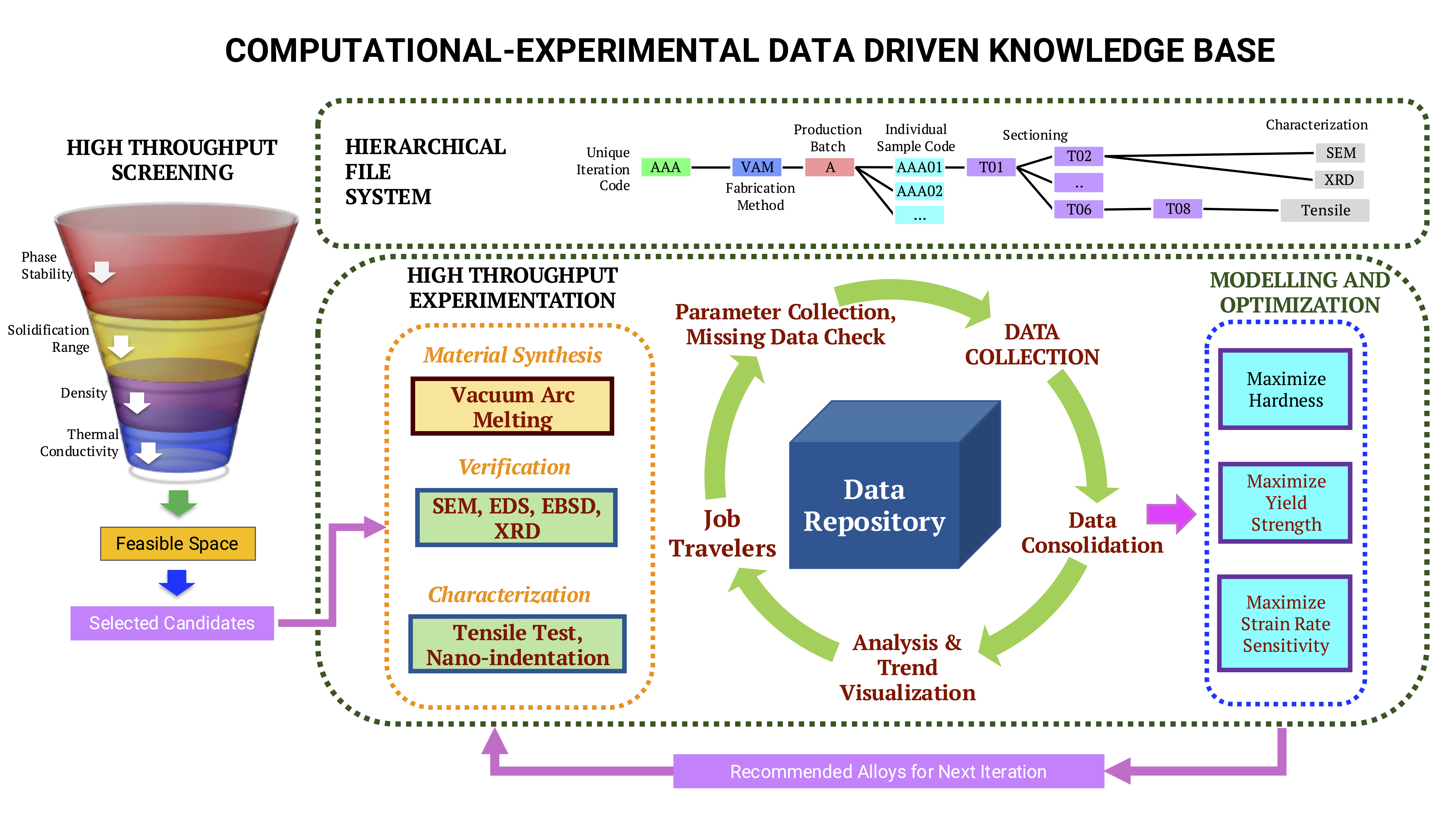}
    \caption{Schematic illustration of the data management workflow for accelerated materials discovery. The figure shows the initial high-throughput screening process from alloy design to candidate selection, followed by the sample naming convention and structured folder hierarchy that reflect various steps involved in the process. A centralized data management system with job travelers dedicated to each workflow step ensures traceability and data integrity. Consolidated performance data feeds into the Bayesian optimization framework, guiding the iterative selection of the next batch of alloy candidates.}
    \label{fig:data_management}
\end{figure*}

\subsubsection{Batch Bayesian Optimization}
\label{batch_bo}

In Gaussian Process (GP) models, a kernel function defines relationships between data points and computes the mean and standard deviation of predictions at unobserved locations. Length scales in the kernel determine correlation strength based on distances in each dimension, making their optimal selection essential for accurately modeling the objective function. However, Bayesian optimization (BO) often starts with sparse data, making it risky to adjust GP hyperparameters, such as those defining the kernel, due to limited information. Common methods for tuning length scales, including maximum likelihood estimation \cite{rasmussen2005gaussian} and cross-validation, are prone to local optima in this regime \cite{joy2020batch, couperthwaite2020materials}, hindering convergence and reducing optimization efficiency.

To address these inefficiencies, BIRDSHOT uses batch Bayesian optimization (BBO) to construct multiple GPs with randomly generated length scales \cite{joy2020batch}. Each GP provides a different interpretation of the objective function, capturing varying hypotheses about data point correlations and enabling diverse decisions for the next optimal design query:
\[
\textbf{x}_{1:n} = \arg\max_{\mathbf{x} \in \chi}  \textrm{EHVI}^{N}(\textbf{x} | \textrm{GP}(\mathbb{D}_N,\theta_{1:n}))
\]
Here, \( n \) GPs are constructed with distinct hyperparameter sets \( \theta \), based on data \( \mathbb{D} \). Depending on the batch size (parallel experimentation capacity), suggestions are clustered using the K-Medoids technique. The medoids, representing promising regions, are proposed as the next experiments. This batch approach accelerates learning by accommodating a wide range of smoothness scenarios for the objective function, bypassing the need for precise length scale tuning.

\subsubsection{Implementation}

To initialize the Gaussian Processes (GPs), Latin hypercube sampling is applied within the bounds \([0.05, 0.95]^6\), generating 1000 six-dimensional samples. For each objective, 1000 independent GPs are instantiated using a Squared Exponential kernel with fixed length scales across iterations. The kernel scaling factor, representing uncertainty, assumes a 20\% achievable increase for each objective. A 95\% confidence interval (\(\sim2\sigma\)) is applied, with scaling factors defined as \((\frac{1}{2}(1.2 \cdot y_{max}))^2\), where \(y_{max}\) is the maximum property value for each objective at each iteration. Strain rate sensitivity values are scaled by a factor of 100 to balance their influence in hypervolume and EHVI calculations.

For experiment selection, candidate alloy objectives are predicted by the GPs to calculate EHVI and identify optimal candidates. This process is repeated 1000 times, each iteration using a different GP set. From the pool of 1000 top candidates, the batch size for the next iteration is reduced to 8 using \textit{k-medoids}, the clustering algorithm also used for initial alloy selection.

Duplicate entries in the candidate pool reflect increased selection weight, as \textit{k-medoids} prioritizes frequently selected options. This indicates certain alloys are more likely to enhance hypervolume based on predicted properties, independent of the GP posterior's shape. These candidates are thus selected more often. A step-by-step illustration of this framework is shown in \autoref{fig:flowchart}. Additional fine-tuning options for the algorithm, such as kernel functions, acquisition functions, and batch sizes, are beyond the scope of this work. Example Bayesian code for this project is available in our \underline{GitHub Repository}.

\subsection{Data Management}\vspace{-0.2em}

Effective data management is critical in high-throughput materials discovery, particularly in interdisciplinary collaborations involving large datasets and complex workflows. Data systems must handle significant data volumes from diverse sources while ensuring consistency, accessibility, and traceability throughout the research cycle. In our framework, robust data management was key to accelerating material development.

A central component of our approach was a cloud-based data storage and management platform, implemented via Google Drive, serving as a centralized repository for all data generated during synthesis, characterization, and analysis. A standardized file structure enabled easy navigation and real-time collaboration, ensuring team members had immediate access to updated data. A custom sample naming convention and tracking system were developed to encode critical information, such as iteration, composition, and processing method, while sample travelers recorded metadata like synthesis conditions and test results. This ensured traceability and consistency at every project stage. 

Automation streamlined routine tasks, including metadata entry, sample traveler generation, and file organization, reducing human error and expediting data handling. Custom routines were also developed to analyze raw experimental data, ensuring consistency in processing and enabling the extraction of key material properties and performance metrics.

The workflow, shown schematically in \autoref{fig:data_management}, illustrates the high-throughput screening process, the progression of selected candidates, the structured folder system for characterization stages, and the sample travelers used at each workflow step. At the core is the data management system, which consolidates data into the Bayesian optimization framework, guiding the selection of the next alloy batch for exploration. Further details, including cloud storage protocols, naming conventions, automation tools, and analysis strategies, are extensively documented in a dedicated publication \cite{mulukutla2024illustrating}.

\section{Hypervolume Improvement}
\label{s3b_hypervolume}

\begin{figure}[ht]
    \centering
    \includegraphics[width=0.9\columnwidth]{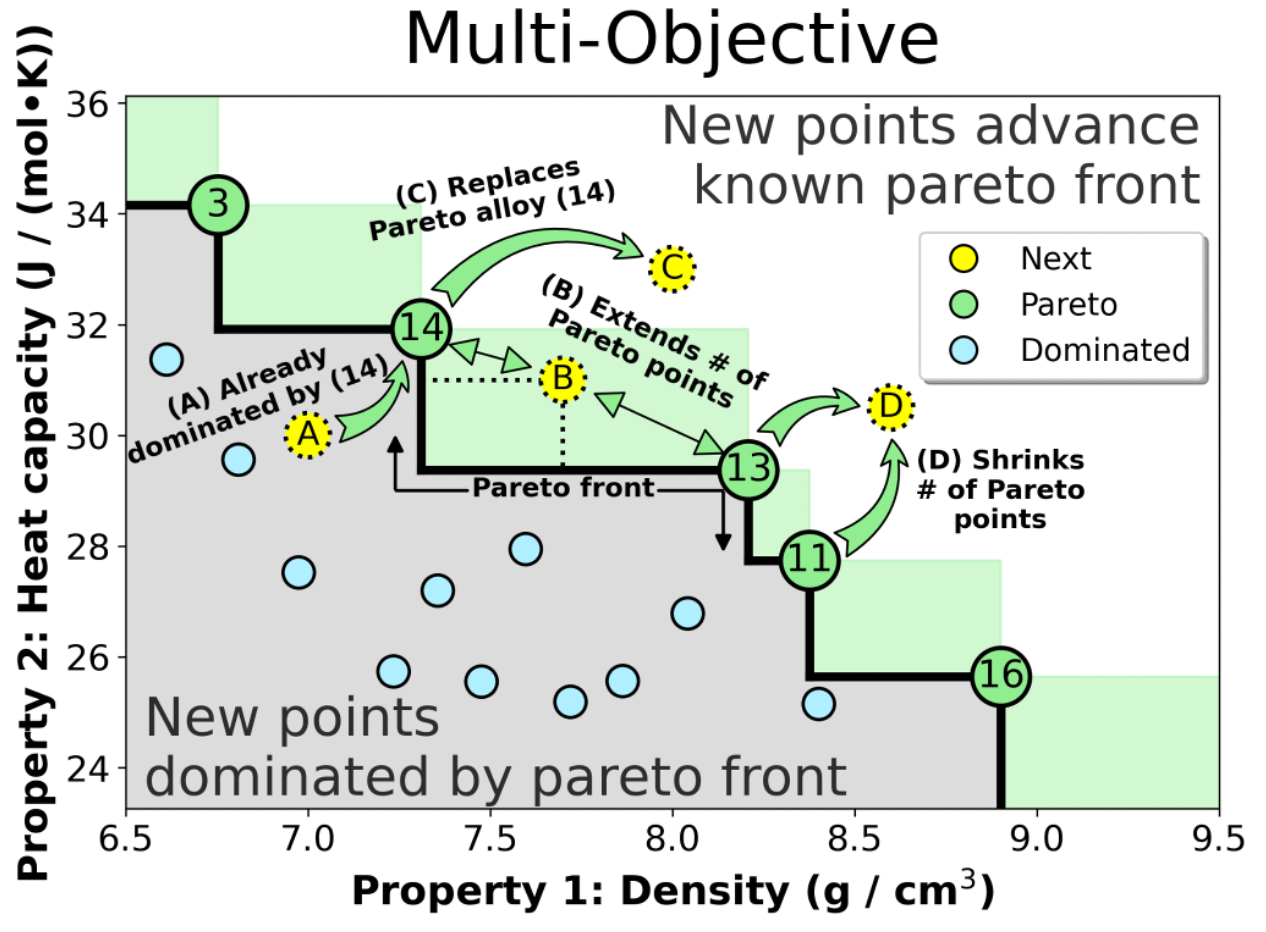}
    \caption{Visualization of data in a multi-objective optimization problem where the objectives are maximized. When analyzing objectives individually, the optimal data point is the one with the maximum value. When objectives are equally important, an array of “best possible” data points, or the Pareto front, emerges, representing trade-offs between objectives. The gray region represents the hypervolume, bounded by the origin and axes.}
    \label{fig:pareto_tutorial}
\end{figure}

The success of a multi-objective optimization problem can be evaluated by quantifying the progression of the \emph{hypervolume (HV)}, as defined in \autoref{s3_methods}. As a scalar metric, \emph{Expected Hypervolume Improvement (EHVI)} is an effective utility function for selecting optimal querying sequences, regardless of the number of material properties optimized simultaneously. \autoref{fig:pareto_tutorial} illustrates the HV of an arbitrary dataset and its Pareto Front, using an example set of materials with simulated CALPHAD-derived thermophysical properties.

In single-objective optimization, each new data point either sets a new maximum or does not. In contrast, multi-objective optimization presents four possible outcomes for each new data point, illustrated in \autoref{fig:pareto_tutorial}: (A) the point is dominated by at least one existing Pareto point (gray region); (B) it lies on the Pareto front, expanding the set of non-dominated points (green region); (C) it replaces a single Pareto point (white region), similar to achieving a new maximum in single-objective optimization; or (D) it dominates multiple Pareto points, pushing the Pareto front further into the white region.

Although this example considers only two objectives, the gray region in \autoref{fig:pareto_tutorial} represents the hypervolume, a key metric for assessing progress in high-dimensional spaces. In three dimensions, HV appears as a meshed surface enveloping data points. However, visualizing HV beyond three dimensions becomes less intuitive, as flattening the space can obscure the relationship between points and the Pareto front. Instead, HV is tracked iteratively, showing cumulative increases as optimization progresses.

\begin{figure}[hbt]
    \centering
    \includegraphics[width=0.99\columnwidth]{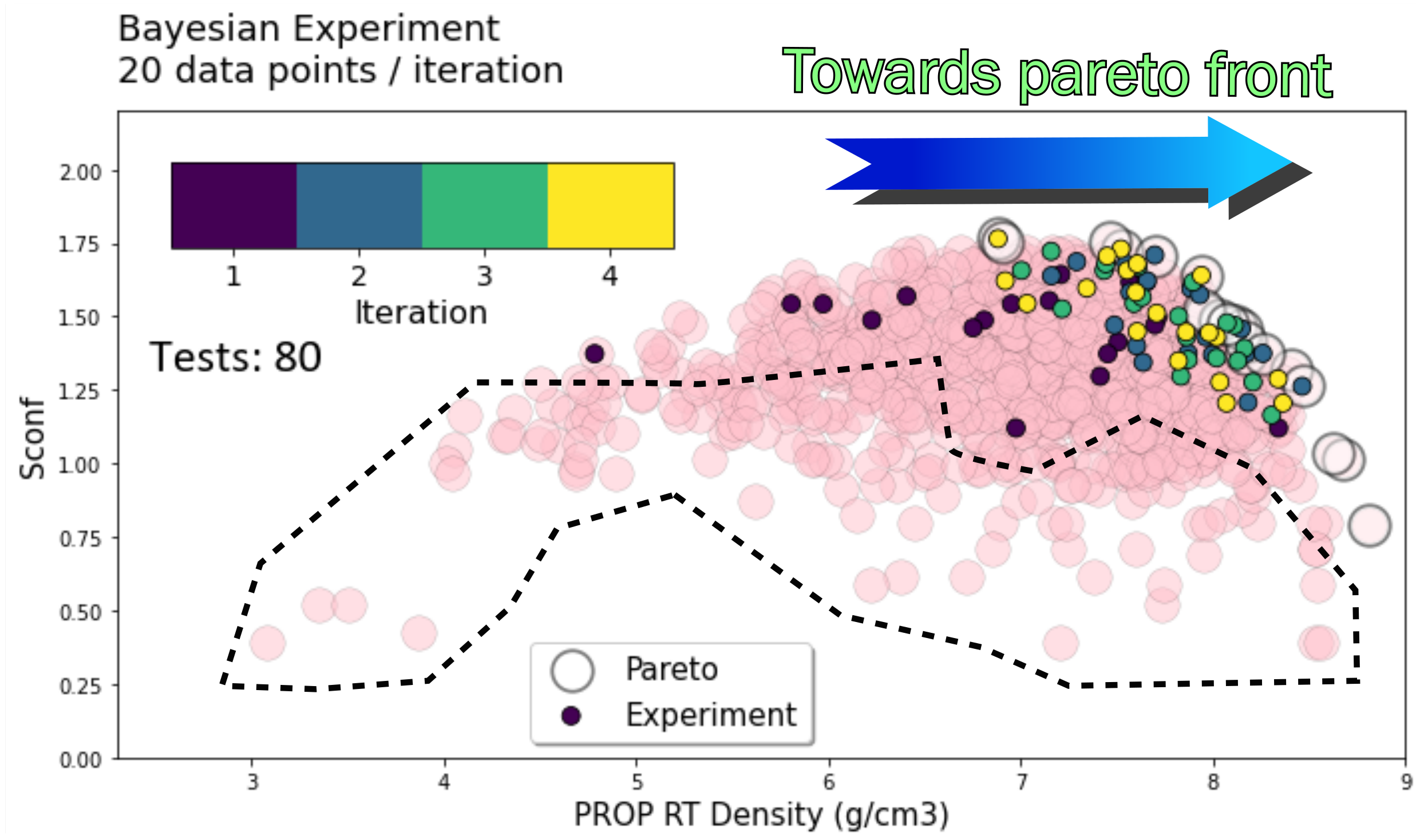}
    \caption{Example optimization simulation maximizing room temperature (RT) properties derived from thermodynamic calculations (Thermo-Calc). Among 1,000 candidates, 21 represent the ground truth Pareto points.  After an initial test of 20 alloys, Bayesian optimization selected the most promising candidates in each 20-alloy iteration, quickly converging toward the Pareto front. The dotted line marks a region containing alloys with suboptimal properties, which the algorithm avoided testing.}
    \label{fig:toy_optimization}
\end{figure}

The previous example illustrates HV improvement in a simplified two-objective scenario. However, practical optimization often involves more complex design spaces. To demonstrate this, we apply Bayesian optimization to maximize Density and Configurational Entropy in the \ce{Al-V-Cr-Fe-Co-Ni} alloy system, using high-confidence CALPHAD calculations as ground truth. In \autoref{fig:toy_optimization} , 979 points are dominated by 21 Pareto-optimal points; any subset containing all 21 achieves the maximum possible HV.

The simulation starts with an initial dataset of 20 points selected via k-medoids, representing material characterization in a real-world scenario. These black points in \autoref{fig:toy_optimization} yield an initial HV below the maximum since the probability of randomly selecting all Pareto points at once is approximately \(\frac{1}{\binom{1000}{21}}\), or about \(1\) in \(10^{44}\), due to the large number of possible combinations. Bayesian optimization, performed in batches of 20, finds over half of the Pareto points by the fourth iteration. As long as at least one data point advances the Pareto front each iteration (points in categories B, C, or D from \autoref{fig:pareto_tutorial}), HV increases progressively, ideally converging toward the maximum HV as efficiently as possible.

\autoref{fig:toy_optimization} represents an idealized scenario where all candidate properties are known. In real-world optimization, however, properties are only revealed after testing. Bayesian optimization addresses this challenge by leveraging all available data, including suboptimal points, to guide the search toward the Pareto front. As shown in \autoref{fig:toy_optimization} , this approach identifies the desired results within just 80 tests—compared to the ~500 tests that random sampling would require on average. Randomly finding half of the Pareto-optimal set within just 80 tests is highly unlikely, with a probability of approximately 0.3\% or lower, as estimated using the Coupon Collector Problem. This probability model estimates the number of random trials needed to collect all unique items from a finite set, demonstrating why random selection is inefficient compared to guided approaches like Bayesian optimization \cite{boneh1997coupon}.

As described in \autoref{s3_methods}, batch-mode Bayesian Optimization (BO) with an ensemble of Gaussian Processes (GPs) is crucial for accelerating Pareto front discovery, particularly in high-throughput workflows. Unlike traditional BO, which selects one candidate at a time, batch BO evaluates multiple candidates simultaneously, enabling faster exploration of the design space. To prevent overconfidence in sparse data, our approach uses multiple GPs with different kernel length scales. This variation allows the model to capture diverse spatial patterns, ensuring that candidate selection remains balanced and robust even when data are limited.

\begin{figure}[hbt]
    \centering
    \includegraphics[width=0.9\columnwidth]{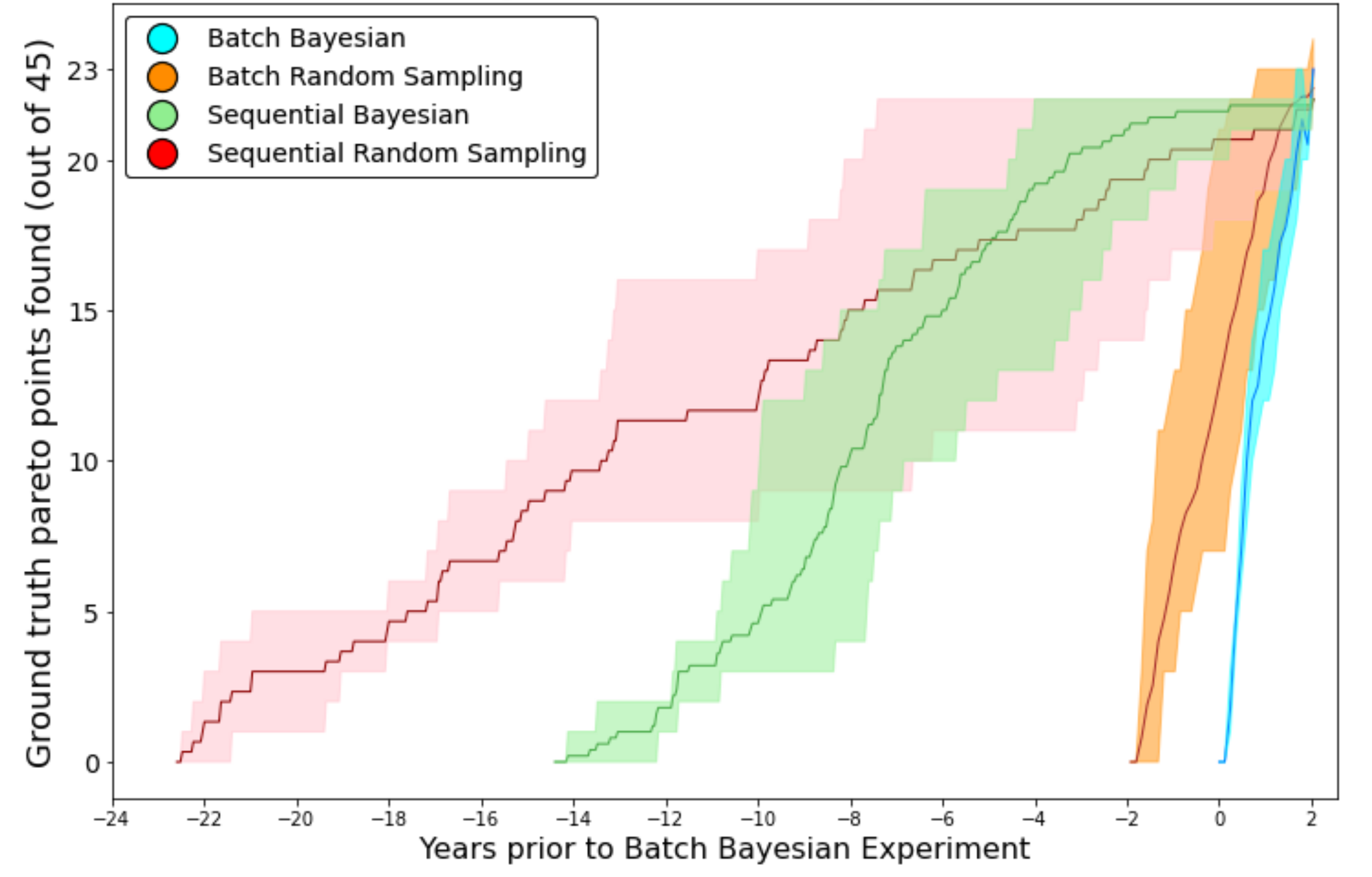}
    \caption{Performance comparison in a three-objective Bayesian optimization problem with 1,000 design choices (alloys). Time estimates for different optimization types: batch and sequential BO vs. random sampling in parallel and sequential modes. Batch experiments evaluate 20 candidates per iteration over six weeks, while sequential experiments evaluate one candidate per iteration over two weeks, all synchronized to conclude simultaneously.}
    \label{fig:optimization_timeline}
\end{figure}

\autoref{fig:optimization_timeline} compares four methods: batch BO, sequential BO, parallel random sampling, and sequential random sampling (sequential BO was modeled with different sets of identical lengthscales in each dimension). Batch BO advanced the Pareto front faster and more efficiently than the other methods. Each batch iteration takes six weeks—compared to two weeks for sequential experiments—yet testing multiple candidates at once significantly accelerates the search. Unlike sequential methods, which often get stuck in local maxima, batch BO maintains steady progress by exploring diverse regions of the design space. Our comparison shows that finding all 45 points in the Pareto set would take approximately 22 years using a random sequential approach—significantly longer than with batch BO. Additionally, batch BO is more consistent, exhibiting less variation between runs compared to other methods, regardless of the initial data used.

\section{Results and Discussion}
\label{s4_results}

\begin{figure}[hbt]
    \centering
    \includegraphics[width=0.99\linewidth]{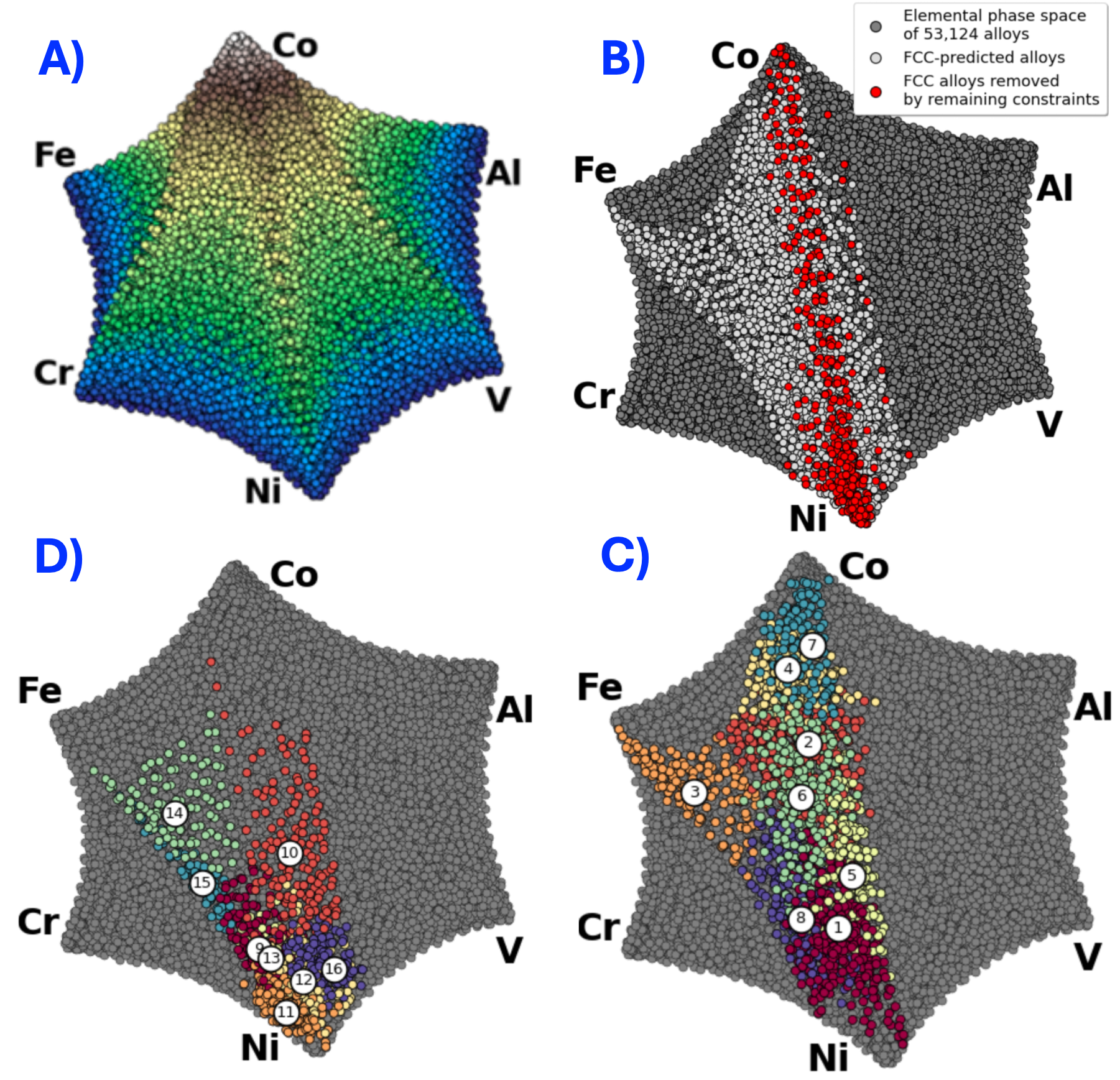}
    \caption{UMAP representations of the compositional space after filtering and selection steps for the first iteration of alloys. Clock-wise: \textbf{A}: Composition space ordered based on the atomic fraction of Co for each alloy. \textbf{B}: The full six-element compositional space, filtered FCC-predicted compositions, and compositions removed by additional property constraints. \textbf{C}: Alloys predicted to have low SFE (center) or \textbf{D}:high SFE (right). Colors correspond to the eight clusters obtained using k-medoids, with numbers marking their medoids, the alloys produced in the first iteration.}
\label{fig:filtering_selection} 
    
\end{figure}

\vspace{-0.4em}\subsection{Initial Alloy Selection and Design Space Mapping}\vspace{-0.2em}

High-dimensional alloy discovery problems present a key challenge: representing their design spaces in a generalized, interpretable manner. In this case, input spaces comprised of an integer number of elements can be mapped to the plane. \autoref{fig:filtering_selection} (left) provides a flattened schematic of the \ce{Al-V-Cr-Fe-Co-Ni} design space for the two parallel experiments in this study (gray data points). Later, we will show how hypervolume (HV) effectively represents all outputs.

\autoref{fig:filtering_selection} (left) highlights the predominantly \ce{Fe-Co-Ni}--based compositions that satisfy the constraints outlined in \autoref{s3_methods}. Panels C and D \autoref{fig:filtering_selection} show input spaces for two specific experiments, where compositions with higher predicted stacking fault energy (SFE) tend to be Ni-based alloys. These figures are generated using the Uniform Manifold Approximation and Projection (UMAP) for Dimension Reduction, a method detailed in the \textbf{SI}. Notably, metals with vertices further apart can still be present in small proportions in many materials. 

Using the clustering algorithm defined in \autoref{s3_methods}, the initial alloy set features compositions with distinct characteristics, providing a diverse range of starting materials for optimization. While cluster overlaps in \autoref{fig:filtering_selection} are visually apparent due to dimensional flattening, k-medoids clustering ensures a broad distribution of initial compositions. This variety is essential for Bayesian Batch Optimization (BBO), minimizing the risk of converging prematurely on local maxima and maximizing information gain across the design space. For high-resolution design spaces (e.g., finer atomic fraction steps than 5\% or more than six constituents), a method to calculate the number of unique partitions based on elements and atomic fraction resolution is presented in the \textbf{SI}. Code for this calculation is available on this project’s GitHub Repository.

\vspace{-0.4em}\subsection{Characterization and Testing}\vspace{-0.2em}

\begin{figure}[htb]
    \centering
    \includegraphics[width=0.8\columnwidth]{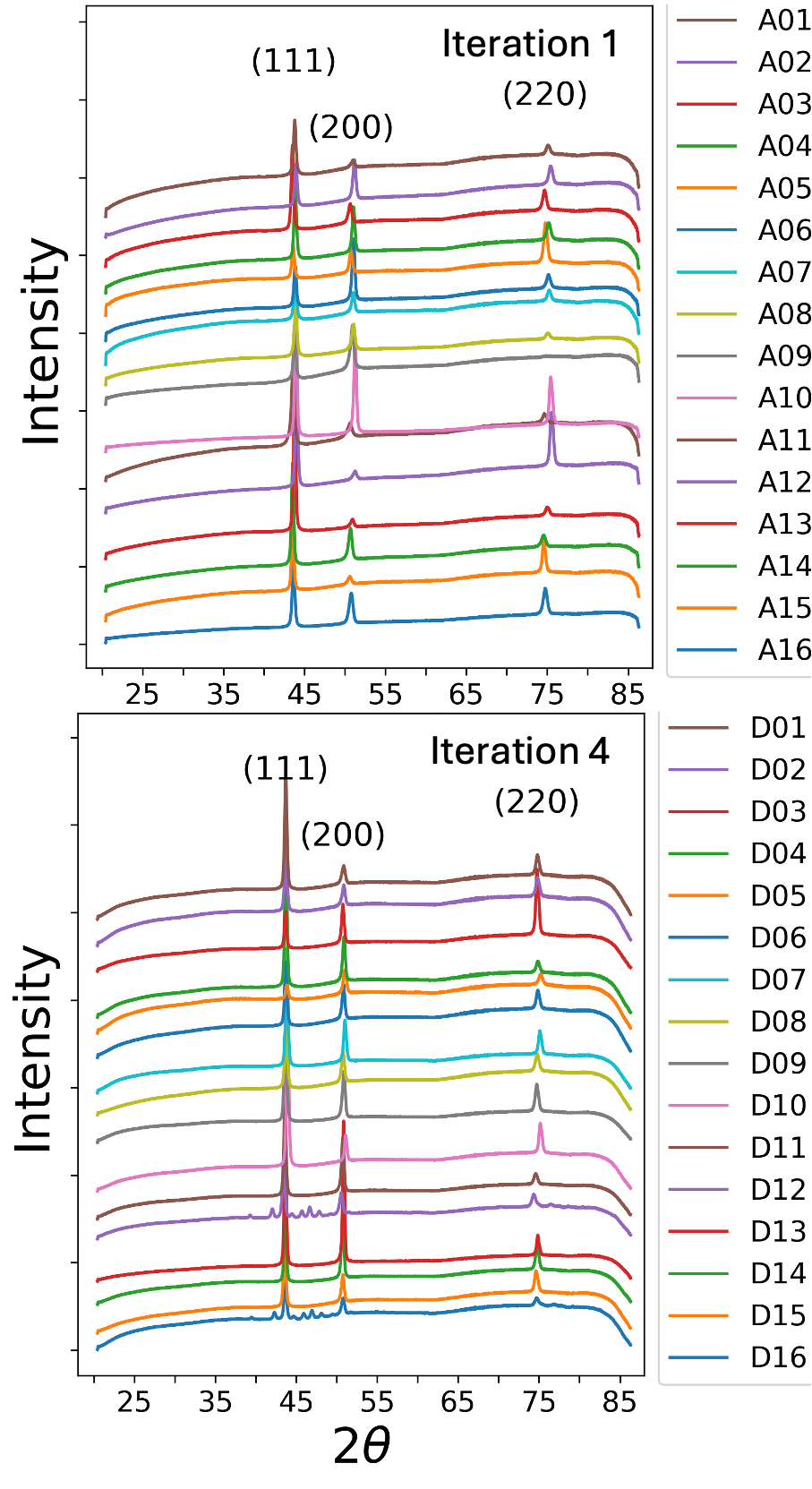}
    \caption{ (top) X-ray diffraction patterns for alloys from the first iteration. (bottom) X-ray diffraction patterns for alloys in the fourth iteration, showing non-FCC peaks in multiple alloys. Intensities within each iteration are offset for clarity. Expected positions for FCC (111), (200), and (220) peaks are labeled.}
    \label{fig:xrd}
\end{figure}


Phase stability of candidate alloys, evaluated via X-ray Diffraction (XRD) and shown in \autoref{fig:xrd} (top), confirmed the intended face-centered cubic (FCC) structure for most samples, ensuring phase homogeneity in alignment with the target design. However, some exceptions were observed: one alloy (Sample B02 from the second iteration) exhibited a single-phase body-centered cubic (BCC) structure, two alloys showed ordering peaks compatible with the $L1_{2}$ phase (an ordered FCC variant), indicating partial ordering or undetected nano-sized precipitates, and four samples revealed high fractions of Sigma (P4$_2$/mnm) secondary phases. \autoref{fig:xrd} (bottom) presents these Sigma phases in alloys from the fourth iteration. Complete XRD data for all alloys is available in the \textbf{SI}. We would like to note that not all secondary precipitates resulted in suboptimal properties, as the alloys with incipient formation of $L1_{2}$ precipitates exhibited significant strengthening.

Secondary phases primarily appeared in alloys with over 30\% vanadium (V) and chromium (Cr) content (added together). Complex inter-element interactions destabilized the FCC structure predicted by CALPHAD, particularly when elements like V and Al promoted BCC or intermetallic formations. The emergence of the $FCC-L1_{2}$ phase, often stabilized by Ni and Al, suggests localized atomic ordering affecting phase stability. V, due to its atomic size and electronic properties, exhibits a strong tendency to stabilize BCC structures. Notably, while CALPHAD anticipated non-FCC phases for alloys high in certain elements, a high-Fe alloy exhibited unexpected BCC phases, underscoring the complexity of phase behavior in compositionally complex alloys.

\begin{figure}[htb]
    \centering
    \includegraphics[width=0.9\columnwidth]{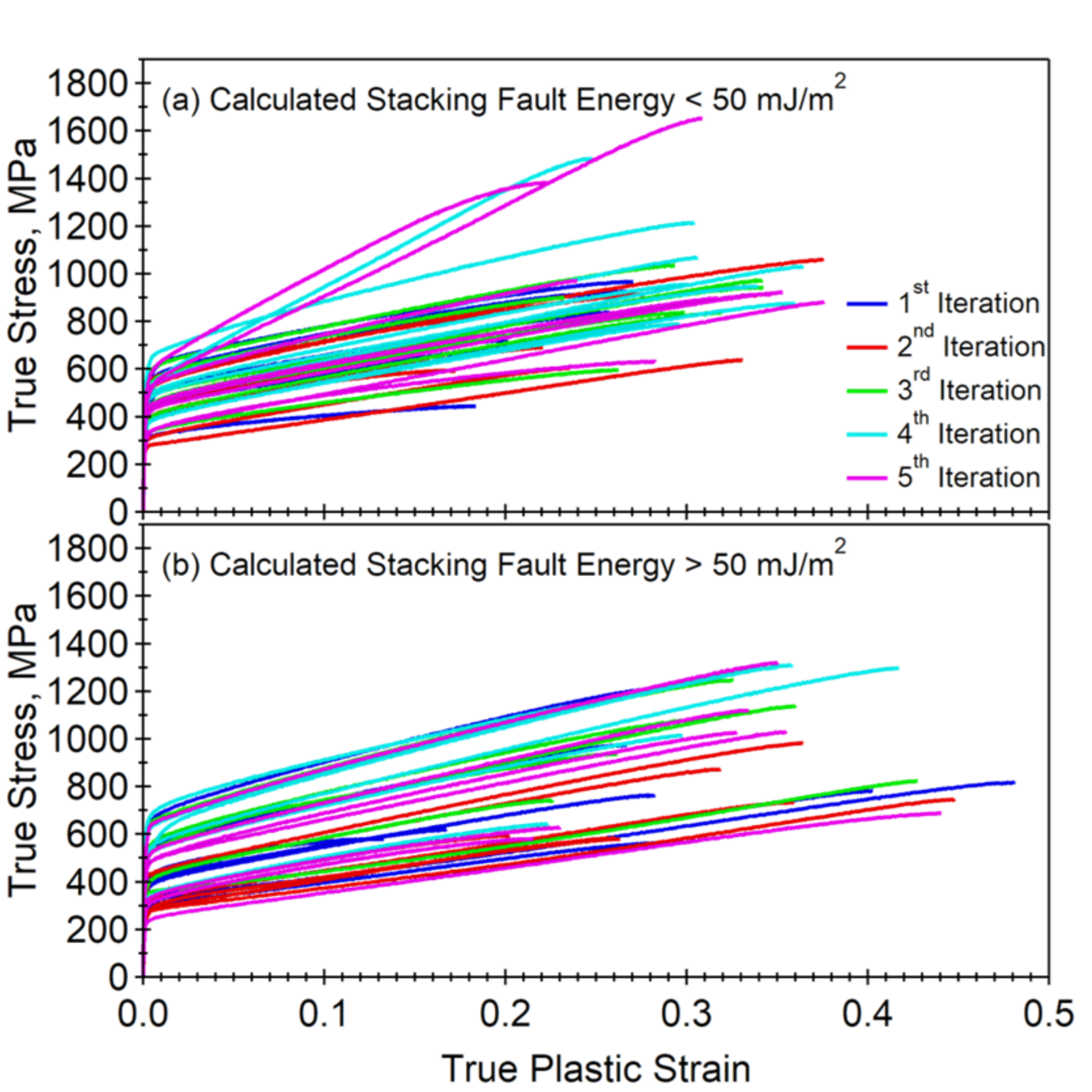}
    \caption{(left) True stress-true strain curves for all alloys. (center) X-ray diffraction patterns for alloys from the first iteration. (right) X-ray diffraction patterns for alloys in the fourth iteration, showing non-FCC peaks in multiple alloys. Intensities within each iteration are offset for clarity. Expected positions for FCC (111), (200), and (220) peaks are labeled.}
    \label{fig:tension}
\end{figure}

The strain rate sensitivity exponent ($m$) calculated using \autoref{strain_rate_hardening_exp} for all tested samples was consistently low, which is characteristic of FCC metals due to their high dislocation mobility and the availability of multiple slip systems. A notable exception was sample B02, which was determined to be BCC by XRD (see \textbf{SI}) and deviated significantly from this trend, displaying a markedly higher strain rate sensitivity. This behavior can be attributed to the characteristic deformation mechanisms in BCC structures, such as the increased sensitivity of screw dislocation movement to strain rate. \autoref{fig:tension} shows true stress versus true plastic strain curves, highlighting each alloy’s tensile performance. Progressive increases in true stress at fracture across iterations indicate that each design cycle effectively enhanced tensile strength without compromising ductility, critical for high-performance structural applications. In the following section, a detailed analysis of the effectiveness of the BO framework will be discussed.

In high-throughput projects, while XRD and tensile data are readily aggregated, microstructural data is less easily visualized in bulk. Although plotting the 80 microstructures characterized here would be impractical, selected SEM images are provided in the \textbf{SI}. These micrographs verified the absence of pores post-synthesis via vacuum arc melting and, in some cases, confirmed secondary phases like Sigma particles.

\vspace{-0.4em}\subsection{Progress of Bayesian Discovery Campaign}\vspace{-0.2em}

    \begin{figure*}[htb]
        \centering
        \includegraphics[width=0.9\textwidth]{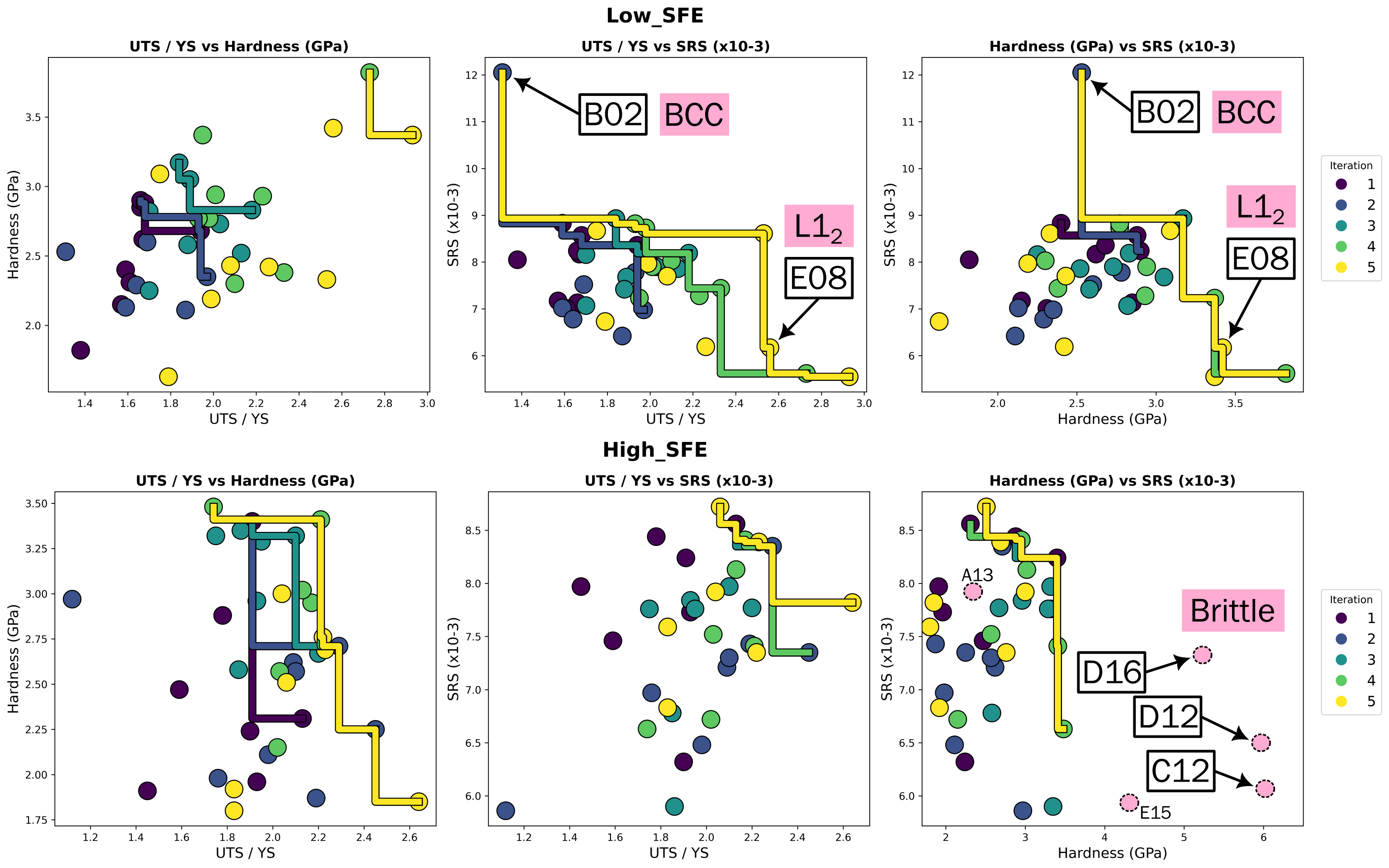}
        \caption{The Pareto (nondominated) sets for each BBO experiment, sliced to 2 of the 3 dimensions for each combination of material properties. The Pareto fronts are marked by lines, cumulatively, for each iteration of data; they can be seen increasing in each case as material properties are maximized. Data points which were not ductile single phase solid solution FCC are annotated with the alloy (white bounding box) and its constraint violation (pink), if they would have contributed the current Pareto front. For alloys that were brittle, since they do not have a yield strength, their approximate position is annotated in pink on the Hardness-SRS property slice.}
        \label{fig:2d_property_pareto_slices} 
    \end{figure*}

In this discovery campaign, aimed at discovering a 3-objective Pareto front, the progression of the batch Bayesian optimization (BBO) process can be directly visualized. \textcolor{black}{\autoref{fig:2d_property_pareto_slices} visually details the progression of these material properties across iterations using 2D slices of the three objectives. Pareto sets for the given slices are shown with lines corresponding to each iteration cumulatively. The slice of hardness and strain hardening for the low-SFE experiment in particular shows a significant improvement over 5 iterations.}

Reducing the multidimensional complexity of the optimization by visualizing material properties individually can be misleading. In this campaign, no \emph{a priori} assumption was made about the likelihood of improvement in any of the target properties. Given the focus on hypervolume improvement, the BIRDSHOT framework is designed to optimize all objectives simultaneously. Thus, improvement in individual properties is not guaranteed in every iteration as the HV could potentially be improved by merely improving the performance along one or two dimensions of the objective space. Thus, focusing on a single property can give the impression that an optimization has stalled, as that property may not be improving while the others continue to do so. In this specific case, the improvement along the strain rate sensitivity (SRS) direction is not as significant as is the improvement in hardness and UTS/$\sigma_y$. \textcolor{black}{\autoref{fig:2d_property_pareto_slices} shows this along the four 2D slices that included SRS: the low-SFE experiment is dominated by an alloy that does not meet the phase constraint, and the high-SFE experiment has little difference in the highest sensitivity values from iteration 1 to iteration 5.} The modest improvement in strain rate sensitivity (SRS) suggests that the alloy space under investigation exhibits limited strain rate dependence. This behavior is typical for materials with a strong solid solution matrix of FCC structures, where deformation mechanisms like dislocation glide dominate, leading to uniform behavior across moderate strain rates.

    \begin{figure*}[htb]
        \centering
        \includegraphics[width=0.9\textwidth]{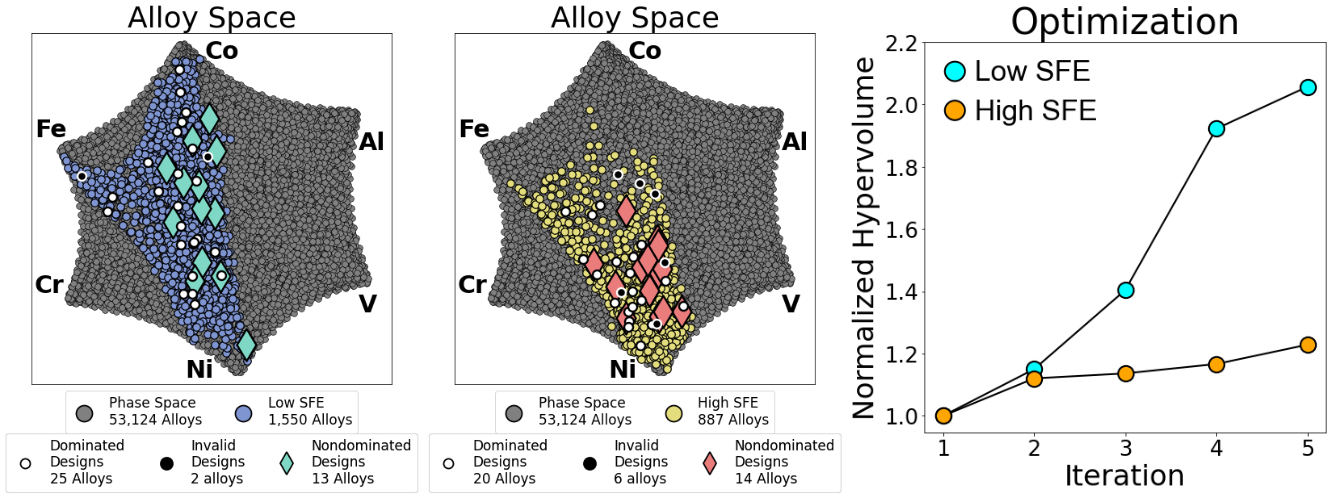}
        \caption{(left, center) The Pareto (nondominated) sets for each BBO experiment. Designs that did not meet project constraints despite predictions are in black, and were not included in optimization. (right) Hypervolumes of the data at each iteration, and hypervolumes predicted by data from the previous iteration using EHVI.}
        \label{fig:pareto_sets_umap} 
    \end{figure*}

\textcolor{black}{The phase predictions made by CALPHAD are of course not perfect. On occasion, a material discovered would be non-FCC or fail to have a valid strain hardening value given its brittle deformation. These are shown in pink in  \autoref{fig:2d_property_pareto_slices}. These alloys were excluded from the iterative optimization, as they likely would have resulted in additional similar predictions that violate the project's design constraints. Overall, five iterations of optimized alloys, their strain hardening ratio ranged from $1.38$ to $2.93$, and their hardness ranged from $1.63\;GPa$ to $3.82\;GPa$.} Individual property values for each material, along with summary results of hardness and strain rate sensitivity, are provided in the \textbf{SI}. 

\textcolor{black}{\autoref{fig:2d_property_pareto_slices} provides a convenient at-a-glance view of physical material properties that can be directly compared to other authors. These are, however, only approximations of the actual optimization, since they have been sliced for visual clarity. True progression of a multi-objective problem must be described by a metric which incorporates all objectives simultaneously, such as hypervolume (HV).} The HV representing \{Strain hardening---hardness---strain rate sensitivity\} can be seen in \autoref{fig:pareto_sets_umap} (right). As only the comparison of HV matters, it has been normalized to the value of the first iteration independently for each experiment. The flattened input spaces from \autoref{fig:filtering_selection} are shown in \autoref{fig:pareto_sets_umap} (left), (middle). They provide a qualitative visual aid for what types of alloys produced the Pareto set in this system. \textcolor{black}{Alloys which did not meet project design constraints are in black.}

Since it is inherently impossible---within a finite number of observation---to conclusively determine a black-box function's complete, ground-truth Pareto front, establishing a definitive “stopping indicator” is impractical. Rather than targeting a specific numerical threshold, this prototype demonstrates achievable progress over a set number of iterations, such as those feasible within a fiscal year. As shown in \autoref{fig:pareto_sets_umap} (right), the hypervolume consistently increases with each iteration in both experiments, a trend unlikely to be replicated by a non-optimized method. Additionally, \autoref{fig:pareto_sets_umap} (left) highlights regions within the design space that do not contribute to the Pareto set. In the high SFE experiment, compositions rich in iron or cobalt, though meeting design constraints, were outperformed by their nickel-rich counterparts.

\textcolor{black}{The relatively quick convergence observed in the high SFE experiment suggests that its initial medoids were closer to the true Pareto front than those in the low SFE case. Furthermore, even if new designs do not markedly increase hypervolume, they still have the capacity to increase the area of the Pareto surface; this indeed occurs significantly: the Pareto lines in \autoref{fig:2d_property_pareto_slices} consistently increase in length. In 3D, mesh approximations of each Pareto surface can be shown to increase in surface area: these are included in the \textbf{SI}. This observation underscores the need for future studies to assess Pareto front success by criteria beyond hypervolume alone. As depicted in \autoref{fig:pareto_sets_umap} (left), several alloys did not satisfy design criteria, with a substantial portion in the high SFE experiment. The impact of these invalid designs on the efficacy of the optimization process presents a potential avenue for further investigation. It is generally important to recognize that "improvement from the 1st iteration" cannot be the only evaluation, as such an evaluation would penalize experiments that properly down-select the design space and provide an excellent starting point for optimization.} Other metrics of performance can be used and, in fact, in \autoref{subsec:pareto_set} we present an analysis to test whether the size of the Pareto sets was likely to be arrived at using a baseline random sampling scheme. As shown there, the chances of observing these many points in the Pareto set(s) are extremely low, which implies---with certain caveats---that the proposed Bayesian alloy discovery scheme is performing as intended.

\vspace{-0.4em}\subsection{Overall Trends and Influence of SFE}\vspace{-0.2em}

\begin{figure*}[!h]
    \centering
    \begin{overpic}[width=0.49\textwidth]{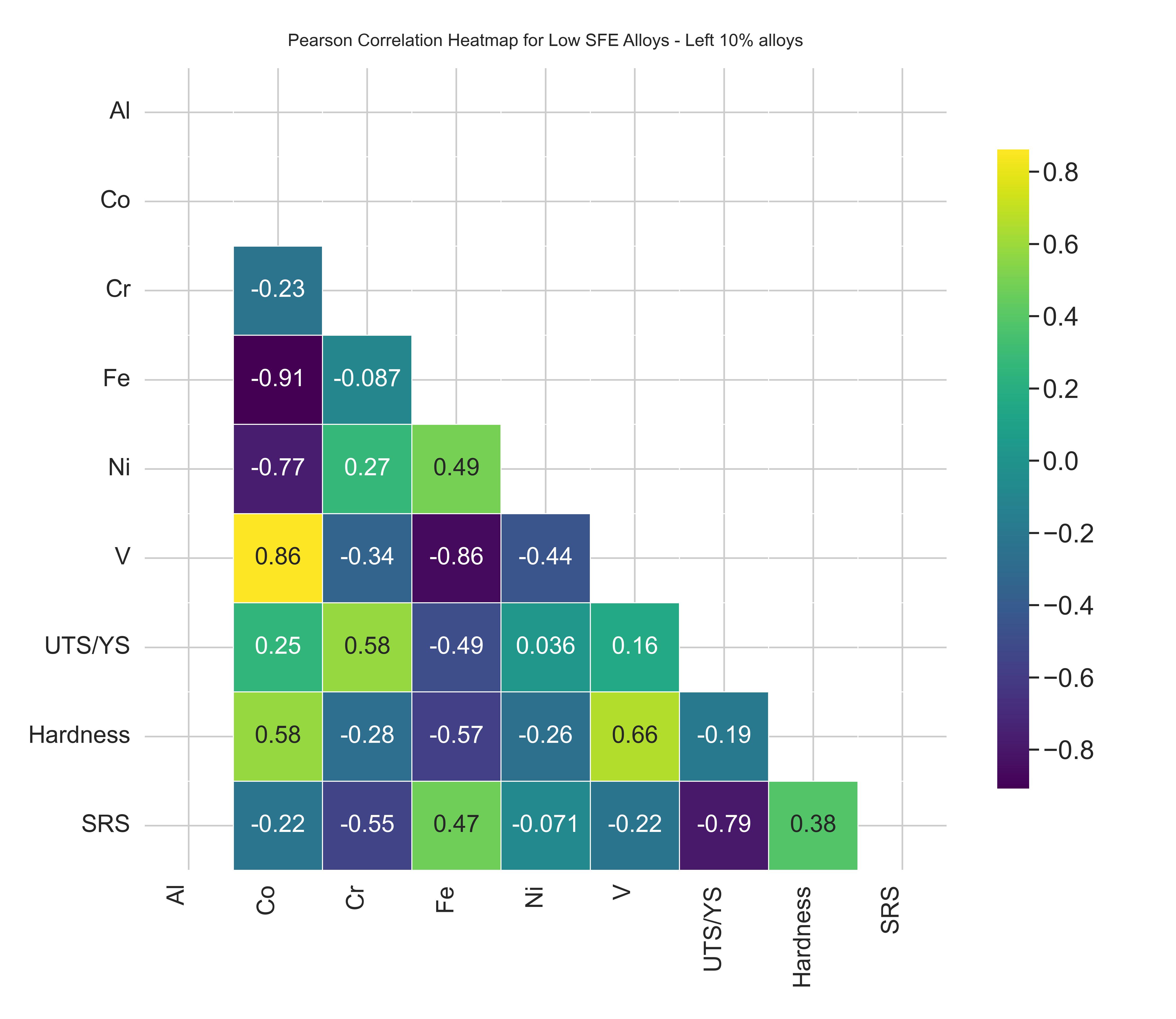}
        \put(5,82){(a)}
        \put(35,50){\includegraphics[width=0.45\columnwidth]{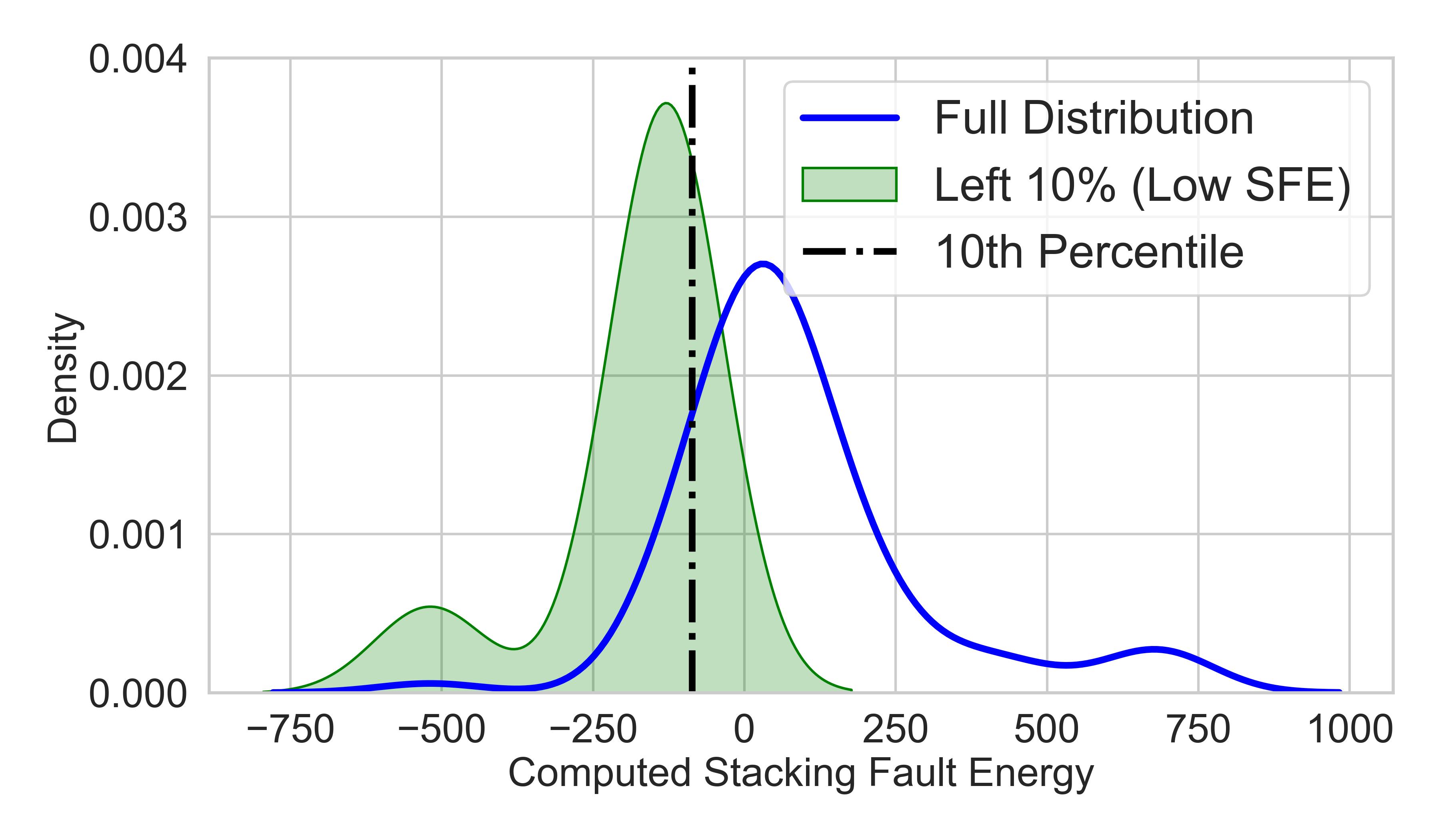}}
        \put(32,76){(c)}        
    \end{overpic}
    \begin{overpic}[width=0.49\textwidth]{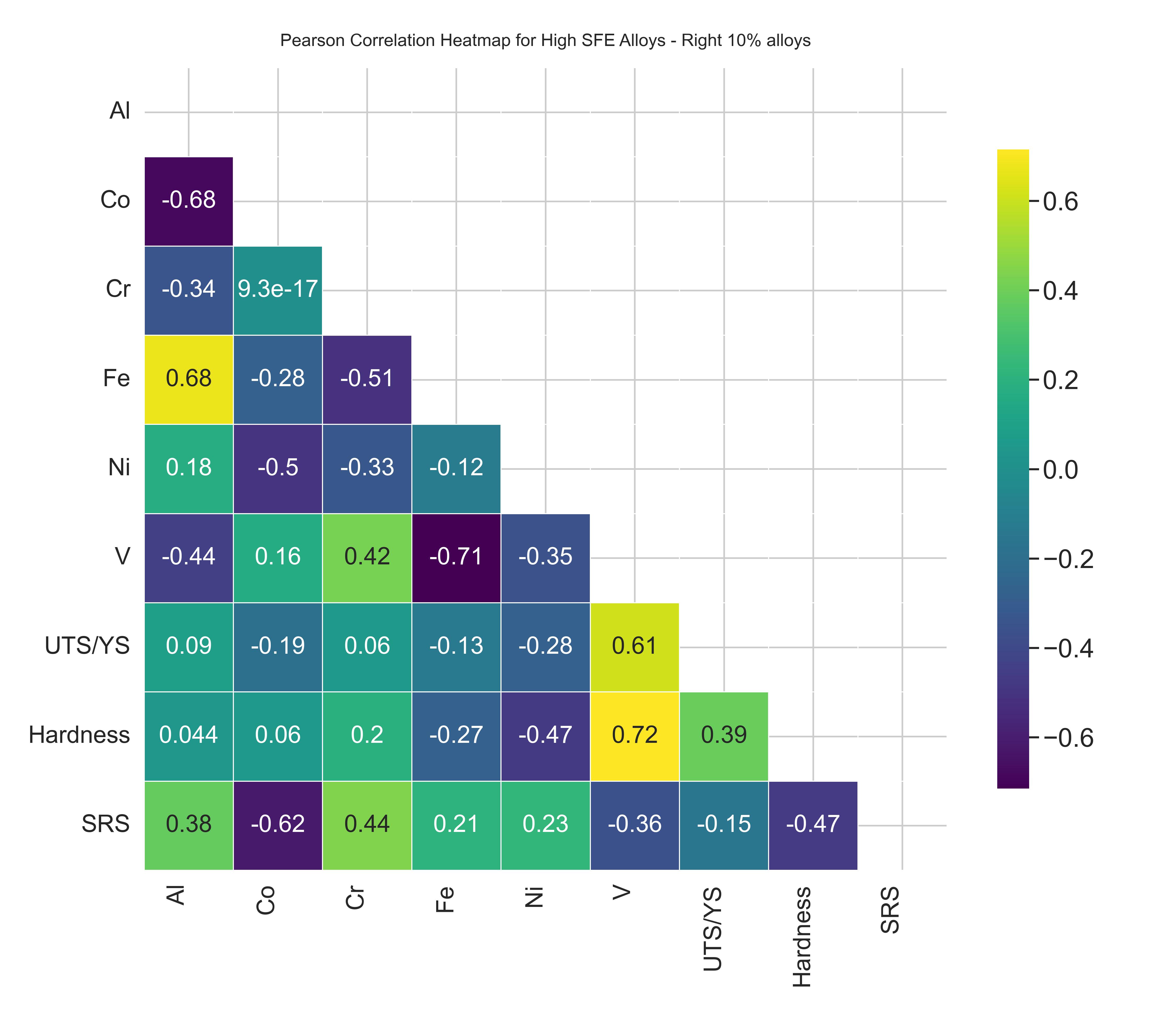}
        \put(5,82){(b)}
        \put(35,50){\includegraphics[width=0.45\columnwidth]{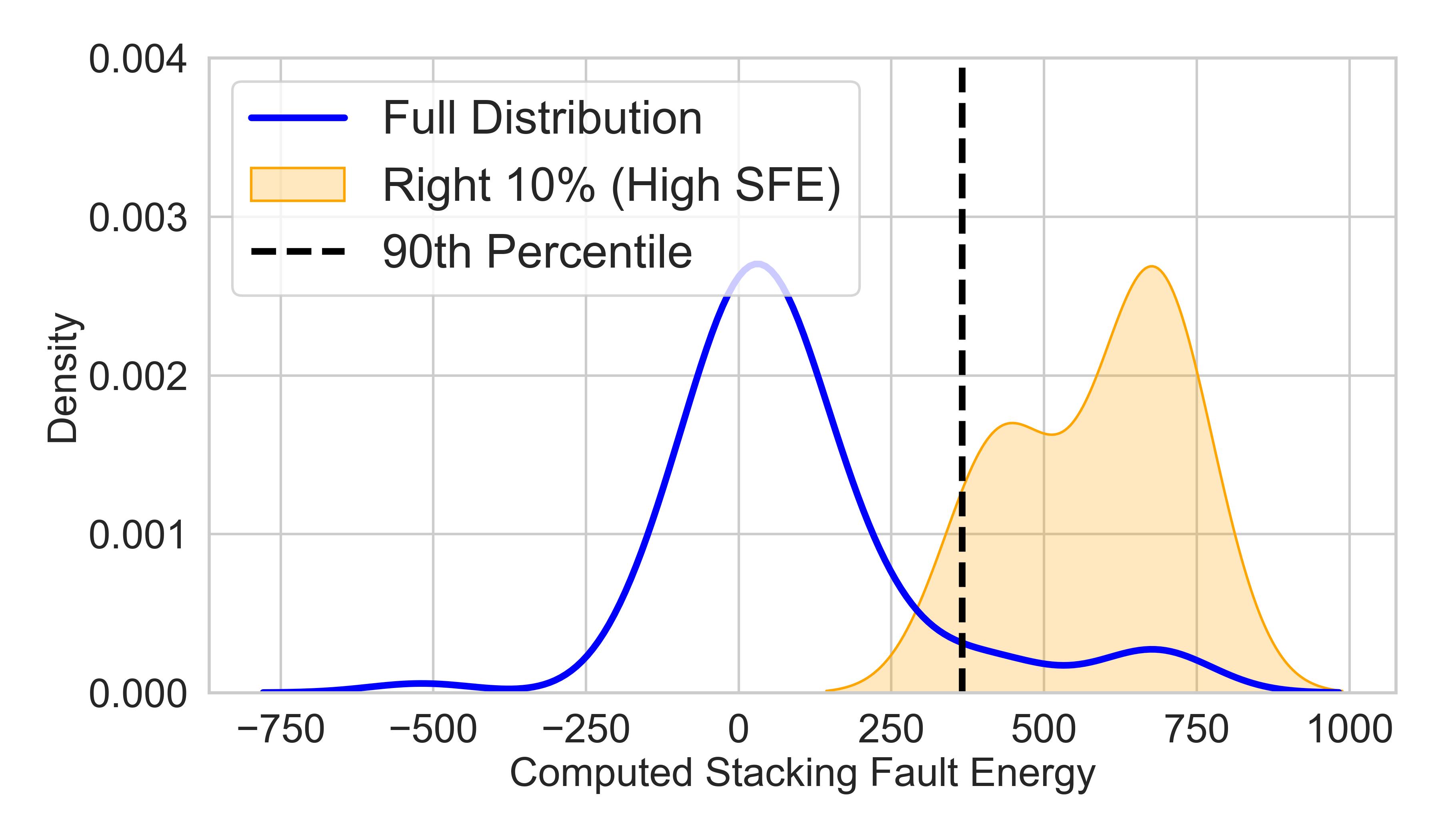}}
        \put(32,76){(d)}        
    \end{overpic}
    \begin{overpic}[width=0.85\textwidth]{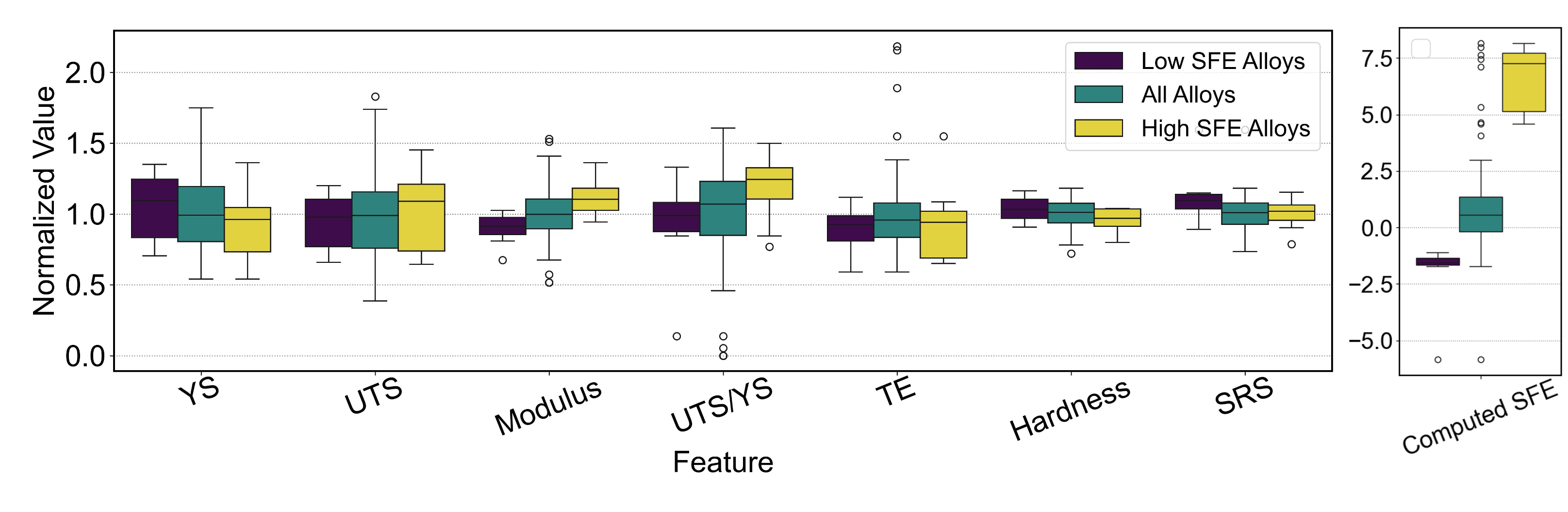}
        \put(5,32){(e)}
        \put(85,32){(f)}
    \end{overpic}
    \vspace{-0.2cm}
    \caption{Pearson correlation heatmaps for alloy properties in different stacking fault energy (SFE) ranges. (a) Correlation heatmap for low SFE alloys representing the left 10\% percentile of the SFE distribution. (b) Correlation heatmap for high SFE alloys representing the right 10\% percentile of the SFE distribution. Each cell shows the Pearson correlation coefficient between alloy elements (e.g., Al, Co, Cr) and key mechanical properties (e.g., UTS/YS, hardness, strain rate sensitivity (SRS)). (c) and (d) Inset plots show the computed SFE distribution with a focus on the left and right 10th percentiles, respectively, highlighting the regions used to generate the heatmaps. The analysis helps to understand the relationship between alloy composition and mechanical properties at extremely low and high SFE levels. (e) Comparison of normalized mechanical properties for alloys with varying stacking fault energy (SFE) levels. (a) Box plots display normalized values (by the average of each property array) for yield strength (YS), ultimate tensile strength (UTS), modulus, UTS/YS ratio, total elongation (TE), hardness, and strain rate sensitivity (SRS) across low SFE alloys (purple), all alloys (green), and high SFE alloys (yellow). The distribution highlights the variation in mechanical properties across different alloy types, indicating trends associated with SFE levels. (f) Inset shows the computed SFE distribution, with low and high SFE ranges corresponding to the data in the main box plot, providing a visual reference for alloy categorization.}
    \label{fig:correlation_heatmaps}
\end{figure*}

\textcolor{black}{This discovery campaign was guided by the hypothesis that differences in stacking fault energy (SFE), as predicted by a machine learning (ML) model developed by Khan \etal~\cite{khan2022towards}, would influence the mechanical properties of the alloys. In face-centered cubic (FCC) materials, SFE plays a key role in determining deformation mechanisms. Alloys with low SFE (below $50\;\text{mJ}/\text{m}^{2}$) are more prone to twinning-induced plasticity (TWIP) and transformation-induced plasticity (TRIP), which enhance ductility and strain hardening. In contrast, alloys with high SFE (above $50\;\text{mJ}/\text{m}^{2}$) primarily deform through dislocation slip, resulting in different mechanical responses.
.} \textcolor{black}{To better understand the impact of computed SFE on the alloys synthesized in this campaign, Pearson correlation heatmaps were calculated for the lowest and highest 10\% of SFE alloys, as shown in \autoref{fig:correlation_heatmaps}. \autoref{fig:correlation_heatmaps} (a) shows the heatmap for low SFE alloys, highlighting key relationships between alloying elements (e.g., Co, Cr, Fe, Ni, V) and material properties. Notably, Aluminum (Al) is absent from low SFE alloys, as it naturally increases stacking fault energy, shifting Al-containing alloys to higher SFE categories. Vanadium (V) and Cobalt (Co) exhibit strong positive correlations with hardness (0.66 and 0.58, respectively), indicating their significant contribution to strengthening. In contrast, Vanadium shows weak correlations with the UTS/YS ratio (0.04) and strain rate sensitivity (SRS), suggesting a limited role in influencing these properties. Chromium (Cr), however, shows a positive correlation with the UTS/YS ratio and a negative correlation with SRS, emphasizing its dual role in balancing strength and deformation behavior.}

\textcolor{black}{Likewise, \autoref{fig:correlation_heatmaps}(b) presents the heatmap for high SFE alloys, revealing trends that differ from those in low SFE alloys. Vanadium shows an even stronger positive correlation with hardness compared to low SFE alloys, further reinforcing its role in improving strength in high SFE systems. Additionally, Vanadium exhibits a strong positive correlation with the UTS/YS ratio, highlighting its importance in enhancing mechanical performance. On the other hand, Cobalt (Co) displays a strong negative correlation with SRS, which may reduce the ability of high SFE alloys to deform effectively under varying strain rates. These differences illustrate the critical need for tailored alloy compositions to optimize specific properties based on the SFE regime, ensuring a balance between strength, ductility, and deformation behavior.}

\textcolor{black}{\autoref{fig:correlation_heatmaps}(e) presents box plots comparing normalized mechanical properties across low SFE alloys, high SFE alloys, and all alloys in the dataset, while \autoref{fig:correlation_heatmaps}(f) displays the computed SFE values for these three groups. Together, these figures provide insights into how stacking fault energy (SFE) trends correlate with mechanical properties. The key properties analyzed include Yield Strength (YS), Ultimate Tensile Strength (UTS), Elastic Modulus, UTS/YS ratio, Total Elongation (TE), Hardness, and Strain Rate Sensitivity (SRS). Low SFE alloys generally exhibit higher Yield Strength (YS), Hardness, and Strain Rate Sensitivity (SRS), indicating superior resistance to deformation and better mechanical stability under dynamic conditions. In contrast, high SFE alloys show higher Ultimate Tensile Strength (UTS), Elastic Modulus, and UTS/YS ratios, suggesting enhanced strength and strain hardening capacity. Notably, there is no apparent correlation between computed SFE and Total Elongation (TE), which may suggest that elongation behavior is influenced by other factors, such as microstructural features like grain size and secondary phases. These findings emphasize the importance of carefully balancing alloy compositions to achieve the optimal trade-offs between strength, hardness, and ductility for specific applications.}

\section{Conclusion}
\label{s5_conclusion}

This study introduced the BIRDSHOT framework, an open-source batch Bayesian optimization approach designed to accelerate materials discovery across diverse design spaces and objectives. By applying BIRDSHOT to a high-dimensional alloy system, we demonstrated its ability to efficiently identify high-performance alloys while balancing multiple objectives. The core insights from this study include:

\begin{itemize}
    \item The combination of batch Bayesian optimization with the squared exponential kernel and EHVI algorithms enabled rapid exploration of a design space with approximately 50,000 candidate alloys. This approach identified a notably improved set of non-dominated alloys within only five experimental iterations.
    \item Unlike sequential optimization methods, BIRDSHOT's batchwise approach aligns with real-world experimental workflows, enabling parallel synthesis and characterization while maintaining optimization efficiency.
    \item The ensemble of Gaussian process regressors with varying length scales ensured robust candidate selection, even when one objective (e.g., strain rate sensitivity) exhibited limited variability, demonstrating the framework's resilience to noisy or weakly correlated objectives.
    \item The use of hypervolume as a primary performance metric effectively captured multi-objective progress. However, future optimization studies should complement hypervolume with additional measures that account for the quality of initial samples and the complexity of design spaces.
    \item The influence of stacking fault energy (SFE) on mechanical properties was evident in both low- and high-SFE alloy groups. Low-SFE alloys exhibited higher yield strength, hardness, and strain rate sensitivity, potentially due to deformation twinning reducing microstructural length scales. In contrast, high-SFE alloys displayed higher ultimate tensile strength, elastic modulus, and UTS/YS ratios, aligning with slip-dominated deformation mechanisms.
\end{itemize}

The BIRDSHOT framework offers significant advantages for both research and industry applications. Its batchwise optimization aligns with the need for parallel experimentation, accelerating discovery while reducing resource consumption. The framework’s generality—supporting arbitrary inputs and objectives—makes it applicable to a wide range of materials, including steels, shape memory alloys, ceramics, and polymers. Moreover, the modular design allows researchers to adapt the approach to different optimization tasks without developing custom frameworks.

This study also highlighted discrepancies between CALPHAD phase predictions and experimental results, emphasizing the need for improved phase stability models. The discovery of alloys that defied conventional solid solution predictions suggests opportunities to refine phase prediction methods using experimental data from compositionally complex systems.

In conclusion, BIRDSHOT provides a scalable, versatile framework capable of accelerating materials discovery across diverse domains. Its combination of batch Bayesian optimization, robust modeling, and practical applicability positions it as a valuable tool for both fundamental research and industry-driven innovation. Future work will focus on expanding the framework to larger design spaces, integrating additional property constraints, and further validating its performance across different material systems.

\section*{CRediT authorship contribution statement}
    \textbf{Trevor Hastings:} Writing – Original Draft, Visualization, Formal Analysis, Data Curation.
    \textbf{Mrinalini Mulukutla:} Writing - Review \& Editing, Formal Analysis, Data Curation.
    \textbf{Danial Khatamsaz:} Software, Visualization, Writing - Review \& Editing.
    \textbf{Daniel Salas:} Investigation, Data Curation.
    \textbf{Wenle Xu:} Investigation.
    \textbf{Daniel Lewis:} Investigation, Data Curation.
    \textbf{Nicole Person:} Investigation.
    \textbf{Matthew Skokan:} Investigation.
    \textbf{Braden Miller:} Investigation.
    \textbf{James Paramore:} Supervision, Resources, Methodology, Writing - Review \& Editing.
    \textbf{Brady Butler:} Supervision, Resources, Methodology, Writing - Review \& Editing.
    \textbf{Douglas Allaire:} Supervision, Resources, Methodology.
    \textbf{Vahid Attari:} Conceptualization, Resources, Supervision.
    \textbf{Ibrahim Karaman:} Conceptualization, Resources, Methodology. Supervision, Funding Acquisition.
    \textbf{George Pharr:} Conceptualization, Resources, Methodology. Supervision.
    \textbf{Ankit Srivastava:} Conceptualization, Resources, Supervision, Funding Acquisition.
    \textbf{Raymundo Arr\'oyave:} Conceptualization, Resources, Supervision, Methodology, Writing - Review \& Editing, Funding Acquisition.

\section*{Declaration of competing interest}

    The authors declare that they have no known competing financial interests or personal relationships that could have appeared to influence the work reported in this paper.

\section*{Data Availability}

    The data generated from this study is included in \underline{Supplementary Information}.

\section*{Code Availability}

    Code used in this project is available on GitHub at https://github.com/trevorhastings/HTMDEC.

\section*{Acknowledgements}

The research was sponsored by the Army Research Laboratory and was accomplished under Cooperative Agreement Number W911NF-22-2-0106. The views and conclusions contained in this document are those of the authors. They should not be interpreted as representing the official policies, either expressed or implied, of the Army Research Laboratory or the US Government. The US Government is authorized to reproduce and distribute reprints for Government Purposes, notwithstanding any copyright notation herein. 
    
We would like to acknowledge Dr. Anup and his group at Texas A\&M University for the use of their XRD Bruker D8 Discover instrument as well as Texas A\&M’s Materials Characterization Facility for the use of their EBSD-capable FIB-SEM. We would like to acknowledge Texas A\&M’s High Performance Research Computing group for access to the computational infrastructure used in this work.

\bibliographystyle{elsarticle-num.bst} 
\bibliography{main.bib} 

\section*{Appendix}
\label{sec:appendix}

\subsection*{Evaluating Bayesian Optimization in Bayesian Materials Discovery}
\label{subsec:pareto_set}

\subsubsection{Theoretical Analysis}

\begin{figure}[hbt]
    \centering
    \includegraphics[width=0.99\columnwidth]{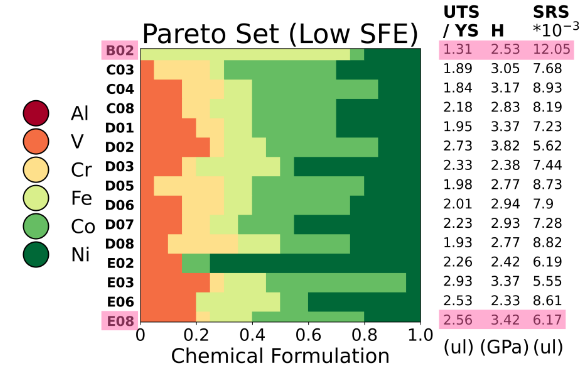}
    \caption{Compositional visualization of the candidate Pareto set in the low SFE regime. Alloys \textbf{B02} and \textbf{E08}, which violate the constraints, are highlighted in pink. Their high strain rate sensitivities would have placed them on the Pareto set had they been single-phase FCC solutions. The probability of identifying at least 14 points in a Pareto set from a design space of \(\sim40\) points is less than 0.3\%.}
\label{fig:pareto_low} 
\end{figure}

\begin{figure}[hbt]
    \centering
    \includegraphics[width=0.99\columnwidth]{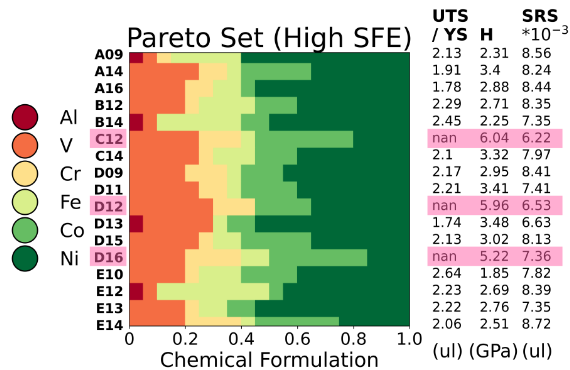}
    \caption{Compositional visualization of the candidate Pareto set in the high SFE regime. Alloys \textbf{C12}, \textbf{D12}, and \textbf{D16}, which violated project constraints, are highlighted in pink. These brittle alloys lack a value for one of the objectives. The probability of identifying at least 15 points in a Pareto set from a design space of \(\sim40\) points is less than 0.08\%.}
\label{fig:pareto_high} 
\end{figure}

In many applications of Bayesian Optimization (BO), researchers benchmark their methods on standard mathematical functions with known optima, allowing for direct performance comparisons against a well-defined ground truth. In contrast, in Bayesian materials discovery we do not have that luxury—the true Pareto set is unknown. Instead, each BO campaign produces a \emph{candidate Pareto set} based solely on the experimental or simulation data available. \autoref{fig:pareto_low} and \autoref{fig:pareto_high} show compositional visualizations\cite{vela2025visualizing} of the candidate Pareto sets in the low and high stacking fault energy (SFE) regimes, respectively. In these figures, the low SFE candidate set contains 14 points while the high SFE candidate set contains 15 points. Naturally, one may ask whether these numbers are statistically significant. 

A useful strategy to evaluate our BO approach is to compare its outcome with what one would expect from random sampling. Although random sampling is not a practical alternative (since even non-BO strategies typically use informed heuristics rather than pure randomness), it nevertheless provides a meaningful baseline for quantifying enrichment. In our experiments, we collect a total of \(n=40\) design points by sampling \(D=8\) points per iteration over \(I=5\) iterations. For a problem with three objectives (\(d=3\)), previous studies have shown that when points are drawn uniformly at random, the expected number of non-dominated (Pareto‐optimal) points in a sample is approximately
\[
E[Y] \approx \frac{(\ln n)^{d-1}}{(d-1)!}.
\]
For \(n=40\) and \(d=3\), this becomes
\[
E[Y] \approx \frac{(\ln 40)^2}{2}.
\]
Since \(\ln(40)\approx 3.69\), we obtain
\[
E[Y] \approx \frac{(3.69)^2}{2} \approx \frac{13.61}{2} \approx 6.81.
\]
Thus, if the 40 points were sampled completely at random, one would expect, on average, about 6.8 non-dominated points.

In contrast, our BO campaign produced candidate Pareto sets with 14 and 15 points in the low and high SFE regimes, respectively. To assess whether these observations represent significant enrichment, we test the null hypothesis that the candidate set arises from random sampling. Although the exact distribution of non-dominated points in a finite sample is complex, a common approximation is to model the count using a Poisson distribution with mean \(\lambda = 6.81\), i.e.,
\[
Y \sim \operatorname{Poisson}(\lambda = 6.81).
\]
Under this model, the probability of obtaining at least \(x\) non-dominated points is given by
\[
P(Y \ge x) = 1 - \sum_{i=0}^{x-1} \frac{e^{-6.81}(6.81)^i}{i!}.
\]

The table below summarizes the tail probabilities for obtaining at least \(x\) non-dominated points (with \(x\) ranging from 10 to 18) when drawing 40 points in a three-objective problem:

\begin{table}[hbt]
\centering
\caption{Tail probabilities for obtaining at least \(x\) non-dominated points by random sampling in 40 draws (3-objective problem).}
\label{tab:pareto_probs}
\begin{tabular}{cc}
\toprule
\textbf{At Least \(x\) Points} & \textbf{Tail Probability \(P(Y \ge x)\)} \\
\midrule
10 & 15.2\% \\
11 & 7.9\% \\
12 & 3.6\% \\
13 & 1.5\% \\
14 & 0.52\% \\
15 & 0.16\% \\
16 & 0.043\% \\
17 & 0.010\% \\
18 & 0.002\% \\
\bottomrule
\end{tabular}
\end{table}

As shown in \autoref{tab:pareto_probs}, the probability of obtaining at least 14 non-dominated points is less than 0.5\%, and for 15 or more points it is only about 0.16\%. These extremely low \(p\)-values indicate that the enrichment observed in our candidate Pareto sets is highly unlikely to have arisen by random chance, providing compelling evidence that our BO method is effectively guiding the search toward high-quality solutions in Bayesian materials discovery.

Finally, we can examine what is the effect of dimensionality of the Pareto set on the probability that a Pareto set is of a given size. The table below summarizes the values of \(E[Y]\) and \(P(Y \ge 10)\) for different numbers of objectives:

\begin{table}[hbt]
\centering
\caption{Expected number of non-dominated points and tail probability \(P(Y \ge 10)\) for random sampling in 40 draws, for various numbers of objectives.}
\label{tab:objective_impact}
\begin{tabular}{ccc}
\toprule
\textbf{Number of Objectives (\(d\))} & \(\mathbf{E[Y]}\) & \(\mathbf{P(Y \ge 10)}\) \\
\midrule
2 & 3.69 & 0.12\% \\
3 & 6.81 & 15.1\% \\
4 & 8.36 & 34.7\% \\
5 & 7.72 & 26.1\% \\
6 & 5.68 & 5.5\% \\
\bottomrule
\end{tabular}
\end{table}

These results reveal that the tail probability \(P(Y \ge 10)\) does not change monotonically with the number of objectives. For a two-objective problem, the chance of obtaining at least 10 non-dominated points is extremely low (around 0.12\%). For three objectives, it increases to about 15.1\%, and peaks at roughly 34.7\% for four objectives. Beyond four objectives, the probability declines to approximately 26.1\% for five objectives and further to about 5.5\% for six objectives. This non-monotonic behavior reflects the interplay between the increased likelihood of non-domination (as additional objectives reduce the chance of one point dominating another) and the combinatorial constraints inherent in high-dimensional objective spaces.

\subsubsection{Empirical Evidence for Bayesian Advantage}

The previous theoretical analysis estimates the likelihood of identifying Pareto-optimal points under the assumption of a uniform random distribution of design points. However, in practical applications the design space is rarely uniform, and the underlying distributions are more complex. This complexity motivates empirical studies to assess the true performance of Bayesian Optimization relative to a baseline random selection process.

\begin{figure}[hbt]
    \centering
    \includegraphics[width=0.99\columnwidth]{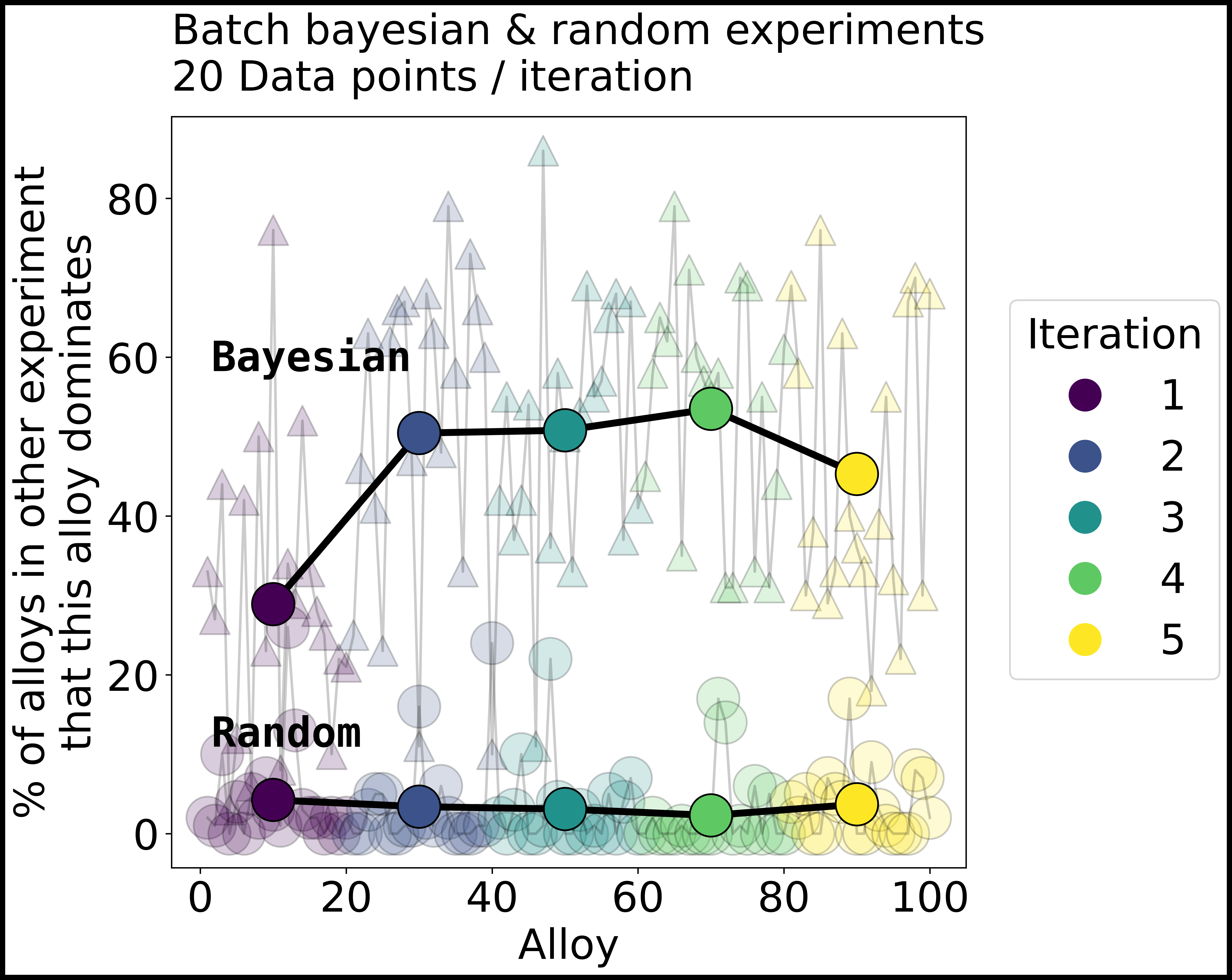}
    \caption{Comparison of performance of a random vs a Bayesian Optimization (BO) policy for the multi-objective toy problem shown in \autoref{fig:toy_optimization}. The figure shows the percentage of alloys discovered in each iteration with a policy that 'dominate' the alloys discovered with the competing policy. The data labeled as 'Random' shows that a random selection policy very seldom dominates the alloys detected using the optimal, 'Bayesian' policy.}
\label{fig:bayesian_advantage} 
\end{figure}

\autoref{fig:bayesian_advantage} compares the probability of dominance between a random policy and an optimal Bayesian policy for the toy multi-objective optimization problem discussed in \autoref{s3_methods} and illustrated in \autoref{fig:toy_optimization}. The figure demonstrates that a random policy very seldom produces solutions that outperform those obtained by the Bayesian approach. Furthermore, as noted in previous work\cite{solomou2018multi}, the advantage of Bayesian Optimization becomes increasingly pronounced with a growing number of objectives.

\end{document}



\title{\Large \textbf{Supplementary Information\footnote{Article: Accelerated Multi-Objective Alloy Discovery through Efficient Bayesian Methods: Application to the FCC Alloy Space}}}

\author[,1]{Trevor Hastings\footnote{Corresponding Author: trevorhastings@tamu.edu}}
\author[1]{Mrinalini Mulukutla}
\author[1]{Danial Khatamsaz}
\author[1]{Daniel Salas}
\author[1]{Wenle Xu}
\author[1]{Daniel Lewis}
\author[1]{Nicole Person}
\author[1]{Matthew Skokan}
\author[1]{Braden Miller}
\author[1]{James Paramore}
\author[1,2]{Brady Butler}
\author[3]{Douglas Allaire}
\author[1]{Ibrahim Karaman}
\author[1]{George Pharr}
\author[1]{Ankit Srivastava}
\author[1]{Raymundo Arroyave}

\affil[1]{\textsl{Texas A\&M University College Station, Department of Materials Science and Engineering, College Station, TX, 77843-3003, USA}}
\affil[2]{\textsl{DEVCOM Army Research Laboratory South at Texas A\&M University, College Station, TX 77843-3003, USA}}
\affil[3]{\textsl{Texas A\&M University College Station, Department of Mechanical Engineering, College Station, TX, 77843, USA}}

\date{\vspace{-2em}}
\maketitle

\addcontentsline{toc}{section}{\textbf{TABLE OF CONTENTS}}
\vspace{-2em}
\renewcommand*\contentsname{}
\tableofcontents

\newpage
\section{\textbf{COMPOSITIONS FOR ALL ALLOYS}}
\label{s1_compositions}
\vspace{-0.25em}

    Some alloys which violated project constraints, which are highlighted in red, were not characterized with EDS.

    \subsection{\textbf{Supplementary \autoref{table_1}} \texorpdfstring{$\mid\;$}{Lg}Iteration 1 alloys}
    \label{s1_1}
    \vspace{-1em}

    \begin{table}[H]
    \caption{}
    \label{table_1}
    \begin{tblr}{
        hlines,
        hline{3,11} = {2pt},
        rowsep=1pt,
        colspec = {
            |X[1.5, c]
            |[2pt]X[1.0, c]|X[1.0, c]|X[1.0, c]
            |X[1.0, c]|X[1.0, c]|X[1.0, c]
            |[2pt]X[1.0, c]|X[1.0, c]|X[1.0, c]
            |X[1.0, c]|X[1.0, c]|X[1.0, c]
            |[2pt]X[3.0, c]
            |X[3.0, c]|
            },
        colsep=0pt,
        }
        \SetCell[r=2,c=1]{c}\textbf{\makecell{Alloy\\Name}}
        & \SetCell[c=6]{c} \textbf{Target Atomic \%}
        &&&&&& \SetCell[c=6]{c} \textbf{EDS Measured Atomic \%}
        &&&&&& \SetCell[r=2,c=1]{c}\textbf{Phase}
        & \SetCell[r=2,c=1]{c}\textbf{Deformation}\\  
        & Al & V & Cr & Fe & Co & Ni 
        & Al & V & Cr & Fe & Co & Ni 
        &  & \\
        A01 & \SetCell{gray!20}0 & 10 & 10 & 20 & 45 & 15 & \SetCell{gray!20}0 & 10.3 & 10.4 & 20.3 & 44.4 & 14.5 & FCC & Ductile\\
        A02 & \SetCell{gray!20}0 & 10 & 10 & 5 & 30 & 45 & \SetCell{gray!20}0 & 10.4 & 10.4 & 5.2 & 30.2 & 43.9 & FCC & Ductile\\
        A03 & \SetCell{gray!20}0 & 15 & 5 & 30 & 30 & 20 & \SetCell{gray!20}0 & 15.3 & 5.3 & 30.1 & 29.8 & 19.5 & FCC & Ductile\\
        A04 & \SetCell{gray!20}0 & 5 & 10 & 20 & 25 & 40 & \SetCell{gray!20}0 & 5.2 & 10.4 & 20.2 & 25.2 & 39.0 & FCC & Ductile\\
        A05 & \SetCell{gray!20}0 & \SetCell{gray!20}0 & 10 & 55 & 10 & 25 & \SetCell{gray!20}0 & \SetCell{gray!20}0 & 10.6 & 54.8 & 10.1 & 24.5 & FCC & Ductile\\
        A06 & \SetCell{gray!20}0 & 5 & 25 & 5 & 35 & 30 & \SetCell{gray!20}0 & 5.2 & 25.6 & 5.0 & 34.9 & 29.3 & FCC & Ductile\\
        A07 & \SetCell{gray!20}0 & 10 & 10 & 10 & 55 & 15 & \SetCell{gray!20}0 & 10.3 & 10.3 & 10.1 & 54.7 & 14.6 & FCC & Ductile\\
        A08 & \SetCell{gray!20}0 & 5 & 10 & 25 & 10 & 50 & \SetCell{gray!20}0 & 5.3 & 10.5 & 25.5 & 10.2 & 48.4 & FCC & Ductile\\
        A09 & 5 & 5 & 5 & 25 & \SetCell{gray!20}0 & 60 & 5.5 & 5.3 & 5.2 & 25.5 & \SetCell{gray!20}0 & 58.6 & FCC & Ductile\\
        A10 & \SetCell{gray!20}0 & 5 & 15 & 5 & 5 & 70 & \SetCell{gray!20}0 & 5.1 & 15.5 & 5.2 & 5.2 & 69.0 & FCC & Ductile\\
        A11 & \SetCell{gray!20}0 & 5 & 5 & 45 & 15 & 30 & \SetCell{gray!20}0 & 5.2 & 5.2 & 45.1 & 15.2 & 29.3 & FCC & Ductile\\
        A12 & 5 & 5 & 5 & 5 & 15 & 65 & 5.3 & 5.1 & 5.3 & 5.2 & 15.2 & 63.8 & FCC & Ductile\\
        \SetCell{red!20}\textbf{A13} & 5 & 5 & 5 & 25 & 10 & 50 & $\cdot$ & $\cdot$ & $\cdot$ & $\cdot$ & $\cdot$ & $\cdot$ & FCC & \SetCell{red!20}\textbf{Brittle}\\
        A14 & \SetCell{gray!20}0 & 25 & 10 & 5 & 25 & 35 & \SetCell{gray!20}0 & 25.2 & 10.1 & 5.0 & 24.9 & 34.8 & FCC & Ductile\\
        A15 & 5 & \SetCell{gray!20}0 & 5 & 45 & \SetCell{gray!20}0 & 45 & 4.9 & \SetCell{gray!20}0 & 5.4 & 45.5 & \SetCell{gray!20}0 & 44.2 & FCC & Ductile\\
        A16 & \SetCell{gray!20}0 & 20 & 10 & 10 & 5 & 55 & \SetCell{gray!20}0 & 20.0 & 10.1 & 10.1 & 5.0 & 54.7 & FCC & Ductile\\
    \end{tblr}
    \end{table}

    \newpage
    \subsection{\textbf{Supplementary \autoref{table_2}} \texorpdfstring{$\mid\;$}{Lg}Iteration 2 alloys}
    \label{s1_2}
    \vspace{-1.25em}

    \begin{table}[H]
    \caption{}
    \label{table_2}
    \begin{tblr}{
        hlines,
        hline{3,11} = {2pt},
        rowsep=1pt,
        colspec = {
            |X[1.5, c]
            |[2pt]X[1.0, c]|X[1.0, c]|X[1.0, c]
            |X[1.0, c]|X[1.0, c]|X[1.0, c]
            |[2pt]X[1.0, c]|X[1.0, c]|X[1.0, c]
            |X[1.0, c]|X[1.0, c]|X[1.0, c]
            |[2pt]X[3.0, c]
            |X[3.0, c]|
            },
        colsep=0pt,
        }
        \SetCell[r=2,c=1]{c}\textbf{\makecell{Alloy\\Name}}
        & \SetCell[c=6]{c} \textbf{Target Atomic \%}
        &&&&&& \SetCell[c=6]{c} \textbf{EDS Measured Atomic \%}
        &&&&&& \SetCell[r=2,c=1]{c}Phase
        & \SetCell[r=2,c=1]{c}Deformation\\  
        & Al & V & Cr & Fe & Co & Ni 
        & Al & V & Cr & Fe & Co & Ni 
        &  & \\
        B01 & \SetCell{gray!20}0 & 10 & 10 & 15 & 50 & 15 & \SetCell{gray!20}0 & 10.2 & 10.2 & 15.3 & 49.6 & 14.7 & FCC & Ductile\\
        \SetCell{red!20}\textbf{B02} & \SetCell{gray!20}0 & \SetCell{gray!20}0 & \SetCell{gray!20}0 & 75 & 5 & 20 & $\cdot$ & $\cdot$ & $\cdot$ & $\cdot$ & $\cdot$ & $\cdot$ & \SetCell{red!20}\textbf{BCC} & Ductile\\
        B03 & \SetCell{gray!20}0 & 5 & 10 & 35 & 25 & 25 & \SetCell{gray!20}0 & 5.1 & 10.2 & 35.2 & 24.8 & 24.7 & FCC & Ductile\\
        B04 & \SetCell{gray!20}0 & 5 & 10 & 25 & 25 & 35 & \SetCell{gray!20}0 & 5.1 & 10.2 & 25.2 & 24.9 & 34.5 & FCC & Ductile\\
        B05 & \SetCell{gray!20}0 & 10 & 5 & 10 & 25 & 50 & \SetCell{gray!20}0 & 10.1 & 5.1 & 10.1 & 25.1 & 49.6 & FCC & Ductile\\
        B06 & \SetCell{gray!20}0 & 5 & 20 & 5 & 30 & 40 & \SetCell{gray!20}0 & 5.1 & 20.5 & 5.1 & 29.9 & 39.3 & FCC & Ductile\\
        B07 & \SetCell{gray!20}0 & 10 & 15 & 5 & 40 & 30 & \SetCell{gray!20}0 & 10.1 & 15.3 & 5.1 & 40.0 & 29.5 & FCC & Ductile\\
        B08 & \SetCell{gray!20}0 & 5 & 10 & 30 & 5 & 50 & \SetCell{gray!20}0 & 5.1 & 10.3 & 30.2 & 5.1 & 49.4 & FCC & Ductile\\
        B09 & \SetCell{gray!20}0 & 5 & \SetCell{gray!20}0 & 30 & 30 & 35 & \SetCell{gray!20}0 & 5.1 & \SetCell{gray!20}0 & 30.5 & 29.9 & 34.4 & FCC & Ductile\\
        B10 & 5 & \SetCell{gray!20}0 & 5 & 25 & 20 & 45 & 5.1 & \SetCell{gray!20}0 & 5.1 & 25.3 & 20.1 & 44.3 & FCC & Ductile\\
        B11 & \SetCell{gray!20}0 & 10 & 20 & 10 & 5 & 55 & \SetCell{gray!20}0 & 10.2 & 20.3 & 10.2 & 5.0 & 54.3 & FCC & Ductile\\
        B12 & \SetCell{gray!20}0 & 20 & 5 & 20 & 15 & 40 & \SetCell{gray!20}0 & 20.4 & 5.1 & 20.2 & 15.0 & 39.4 & FCC & Ductile\\
        B13 & 5 & \SetCell{gray!20}0 & 10 & 35 & 5 & 45 & 5.2 & \SetCell{gray!20}0 & 10.2 & 35.2 & 5.1 & 44.2 & FCC & Ductile\\
        B14 & 5 & 5 & \SetCell{gray!20}0 & 30 & 10 & 50 & 5.3 & 5.1 & \SetCell{gray!20}0 & 30.3 & 10.2 & 49.0 & FCC & Ductile\\
        B15 & 5 & 5 & 10 & 20 & 5 & 55 & 4.6 & 4.9 & 10.1 & 20.3 & 5.1 & 55.0 & FCC & Ductile\\
        \SetCell{red!20}\textbf{B16} & 15 & \SetCell{gray!20}0 & \SetCell{gray!20}0 & 15 & 5 & 65 & $\cdot$ & $\cdot$ & $\cdot$ & $\cdot$ & $\cdot$ & $\cdot$ & \SetCell{red!20}$\mathbf{FCC + L1_{2}}$ & Ductile\\
    \end{tblr}
    \end{table}

    \subsection{\textbf{Supplementary \autoref{table_3}} \texorpdfstring{$\mid\;$}{Lg}Iteration 3 alloys}
    \label{s1_3}
    \vspace{-1.25em}

    \begin{table}[H]
    \caption{}
    \label{table_3}
    \begin{tblr}{
        hlines,
        hline{3,11} = {2pt},
        rowsep=1pt,
        colspec = {
            |X[1.5, c]
            |[2pt]X[1.0, c]|X[1.0, c]|X[1.0, c]
            |X[1.0, c]|X[1.0, c]|X[1.0, c]
            |[2pt]X[1.0, c]|X[1.0, c]|X[1.0, c]
            |X[1.0, c]|X[1.0, c]|X[1.0, c]
            |[2pt]X[3.0, c]
            |X[3.0, c]|
            },
        colsep=0pt,
        }
        \SetCell[r=2,c=1]{c}\textbf{\makecell{Alloy\\Name}}
        & \SetCell[c=6]{c} \textbf{Target Atomic \%}
        &&&&&& \SetCell[c=6]{c} \textbf{EDS Measured Atomic \%}
        &&&&&& \SetCell[r=2,c=1]{c}Phase
        & \SetCell[r=2,c=1]{c}Deformation\\  
        & Al & V & Cr & Fe & Co & Ni 
        & Al & V & Cr & Fe & Co & Ni 
        &  & \\
        C01 & \SetCell{gray!20}0 & 5 & 25 & 5 & 15 & 50 & \SetCell{gray!20}0 & 5.1 & 25.2 & 5.1 & 15.1 & 49.6 & FCC & Ductile\\
        C02 & \SetCell{gray!20}0 & 15 & 5 & 35 & 10 & 35 & \SetCell{gray!20}0 & 15.0 & 5.0 & 35.1 & 10.0 & 34.8 & FCC & Ductile\\
        C03 & \SetCell{gray!20}0 & 5 & 20 & 5 & 40 & 30 & \SetCell{gray!20}0 & 5.1 & 20.2 & 5.1 & 40.1 & 29.5 & FCC & Ductile\\
        C04 & \SetCell{gray!20}0 & 15 & 15 & 10 & 45 & 15 & \SetCell{gray!20}0 & 15.1 & 15.1 & 10.1 & 44.8 & 14.8 & FCC & Ductile\\
        C05 & \SetCell{gray!20}0 & 10 & \SetCell{gray!20}0 & 15 & 75 & \SetCell{gray!20}0 & \SetCell{gray!20}0 & 10.2 & \SetCell{gray!20}0 & 15.4 & 74.4 & \SetCell{gray!20}0 & FCC & Ductile\\
        C06 & \SetCell{gray!20}0 & 10 & 20 & 10 & 25 & 35 & \SetCell{gray!20}0 & 10.2 & 20.3 & 10.1 & 24.9 & 34.4 & FCC & Ductile\\
        C07 & \SetCell{gray!20}0 & 10 & 5 & 15 & 45 & 25 & \SetCell{gray!20}0 & 9.9 & 5.1 & 15.2 & 45.0 & 24.7 & FCC & Ductile\\
        C08 & \SetCell{gray!20}0 & 15 & 10 & 15 & 30 & 30 & \SetCell{gray!20}0 & 15.2 & 10.1 & 15.2 & 29.8 & 29.6 & FCC & Ductile\\
        C09 & 5 & 5 & 5 & 20 & 10 & 55 & 5.2 & 5.1 & 5.1 & 20.3 & 10.0 & 54.2 & FCC & Ductile\\
        C10 & \SetCell{gray!20}0 & 5 & 20 & 5 & 10 & 60 & \SetCell{gray!20}0 & 5.1 & 20.4 & 5.1 & 10.3 & 59.2 & FCC & Ductile\\
        C11 & \SetCell{gray!20}0 & 30 & 5 & 5 & 10 & 50 & \SetCell{gray!20}0 & 30.5 & 5.0 & 5.0 & 10.0 & 49.5 & FCC & Ductile\\
        \SetCell{red!20}\textbf{C12} & \SetCell{gray!20}0 & 25 & 15 & 5 & 35 & 20 & $\cdot$ & $\cdot$ & $\cdot$ & $\cdot$ & $\cdot$ & $\cdot$ & \SetCell{red!20}$\mathbf{FCC + Sigma}$ & \SetCell{red!20}\textbf{Brittle}\\
        C13 & \SetCell{gray!20}0 & 15 & 20 & 5 & 15 & 45 & \SetCell{gray!20}0 & 15.3 & 20.2 & 5.1 & 15.1 & 44.4 & FCC & Ductile\\
        C14 & \SetCell{gray!20}0 & 25 & 5 & 15 & 15 & 40 & \SetCell{gray!20}0 & 25.4 & 5.0 & 15.1 & 15.0 & 39.5 & FCC & Ductile\\
        C15 & 5 & 15 & 5 & 20 & 5 & 50 & 5.5 & 15.1 & 5.0 & 20.2 & 4.9 & 49.4 & FCC & Ductile\\
        C16 & \SetCell{gray!20}0 & 20 & 15 & 5 & 25 & 35 & \SetCell{gray!20}0 & 20.4 & 15.2 & 5.2 & 24.9 & 34.3 & FCC & Ductile\\
    \end{tblr}
    \end{table}
    \vspace{-2em}

    \newpage
    \subsection{\textbf{Supplementary \autoref{table_4}} \texorpdfstring{$\mid\;$}{Lg}Iteration 4 alloys}
    \label{s1_4}
    \vspace{-1.25em}

    \begin{table}[H]
    \caption{}
    \label{table_4}
    \begin{tblr}{
        hlines,
        hline{3,11} = {2pt},
        rowsep=1pt,
        colspec = {
            |X[1.5, c]
            |[2pt]X[1.0, c]|X[1.0, c]|X[1.0, c]
            |X[1.0, c]|X[1.0, c]|X[1.0, c]
            |[2pt]X[1.0, c]|X[1.0, c]|X[1.0, c]
            |X[1.0, c]|X[1.0, c]|X[1.0, c]
            |[2pt]X[3.0, c]
            |X[3.0, c]|
            },
        colsep=0pt,
        }
        \SetCell[r=2,c=1]{c}\textbf{\makecell{Alloy\\Name}}
        & \SetCell[c=6]{c} \textbf{Target Atomic \%}
        &&&&&& \SetCell[c=6]{c} \textbf{EDS Measured Atomic \%}
        &&&&&& \SetCell[r=2,c=1]{c}Phase
        & \SetCell[r=2,c=1]{c}Deformation\\  
        & Al & V & Cr & Fe & Co & Ni 
        & Al & V & Cr & Fe & Co & Ni 
        &  & \\
        D01 & \SetCell{gray!20}0 & 20 & 10 & 10 & 30 & 30 & \SetCell{gray!20}0 & 21.2 & 10.6 & 9.9 & 29.6 & 28.6 & FCC & Ductile\\
        D02 & \SetCell{gray!20}0 & 25 & 5 & 15 & 40 & 15 & \SetCell{gray!20}0 & 26.1 & 5.3 & 14.9 & 39.4 & 14.3 & FCC & Ductile\\
        D03 & \SetCell{gray!20}0 & 15 & 5 & 30 & 5 & 45 & \SetCell{gray!20}0 & 15.8 & 5.3 & 30.4 & 5.3 & 43.2 & FCC & Ductile\\
        D04 & \SetCell{gray!20}0 & 15 & 5 & 25 & 20 & 35 & \SetCell{gray!20}0 & 15.8 & 5.3 & 25.2 & 20.2 & 33.5 & FCC & Ductile\\
        D05 & \SetCell{gray!20}0 & 5 & 25 & 10 & 40 & 20 & \SetCell{gray!20}0 & 5.3 & 26.2 & 10.0 & 39.2 & 19.2 & FCC & Ductile\\
        D06 & \SetCell{gray!20}0 & 15 & 15 & 10 & 35 & 25 & \SetCell{gray!20}0 & 15.8 & 15.7 & 10.0 & 34.5 & 24.0 & FCC & Ductile\\
        D07 & \SetCell{gray!20}0 & 15 & 10 & 10 & 20 & 45 & \SetCell{gray!20}0 & 15.9 & 10.6 & 10.1 & 20.1 & 43.2 & FCC & Ductile\\
        D08 & \SetCell{gray!20}0 & 10 & 25 & 15 & 25 & 25 & \SetCell{gray!20}0 & 10.6 & 26.2 & 14.7 & 24.7 & 23.9 & FCC & Ductile\\
        D09 & \SetCell{gray!20}0 & 20 & 15 & 5 & 15 & 45 & \SetCell{gray!20}0 & 21.0 & 15.8 & 4.9 & 15.0 & 43.3 & FCC & Ductile\\
        D10 & \SetCell{gray!20}0 & 5 & 5 & 30 & 5 & 55 & \SetCell{gray!20}0 & 5.3 & 5.4 & 30.9 & 5.4 & 53.1 & FCC & Ductile\\
        D11 & \SetCell{gray!20}0 & 25 & 10 & 5 & 20 & 40 & \SetCell{gray!20}0 & 26.3 & 10.5 & 4.9 & 19.8 & 38.5 & FCC & Ductile\\
        \SetCell{red!20}\textbf{D12} & \SetCell{gray!20}0 & 30 & 15 & \SetCell{gray!20}0 & 20 & 35 & $\cdot$ & $\cdot$ & $\cdot$ & $\cdot$ & $\cdot$ & $\cdot$ & \SetCell{red!20}\textbf{FCC + Sigma} & \SetCell{red!20}\textbf{Brittle}\\
        D13 & 5 & 25 & \SetCell{gray!20}0 & 5 & 10 & 55 & 3.1 & 26.6 & \SetCell{gray!20}0 & 5.2 & 10.4 & 54.7 & FCC & Ductile\\
        D14 & \SetCell{gray!20}0 & 15 & 15 & 15 & 15 & 40 & \SetCell{gray!20}0 & 15.8 & 15.8 & 14.9 & 15.1 & 38.4 & FCC & Ductile\\
        D15 & \SetCell{gray!20}0 & 25 & 5 & 10 & 25 & 35 & \SetCell{gray!20}0 & 26.1 & 5.3 & 10.0 & 24.9 & 33.7 & FCC & Ductile\\
        \SetCell{red!20}\textbf{D16} & \SetCell{gray!20}0 & 20 & 20 & 10 & 35 & 15 & $\cdot$ & $\cdot$ & $\cdot$ & $\cdot$ & $\cdot$ & $\cdot$ & \SetCell{red!20}\textbf{FCC + Sigma} & \SetCell{red!20}\textbf{Brittle}\\
    \end{tblr}
    \end{table}

    \subsection{\textbf{Supplementary \autoref{table_5}} \texorpdfstring{$\mid\;$}{Lg}Iteration 5 alloys}
    \label{s1_5}
    \vspace{-1.25em}

    \begin{table}[H]
    \caption{}
    \label{table_5}
    \begin{tblr}{
        hlines,
        hline{3,11} = {2pt},
        rowsep=1pt,
        colspec = {
            |X[1.5, c]
            |[2pt]X[1.0, c]|X[1.0, c]|X[1.0, c]
            |X[1.0, c]|X[1.0, c]|X[1.0, c]
            |[2pt]X[1.0, c]|X[1.0, c]|X[1.0, c]
            |X[1.0, c]|X[1.0, c]|X[1.0, c]
            |[2pt]X[3.0, c]
            |X[3.0, c]|
            },
        colsep=0pt,
        }
        \SetCell[r=2,c=1]{c}\textbf{\makecell{Alloy\\Name}}
        & \SetCell[c=6]{c} \textbf{Target Atomic \%}
        &&&&&& \SetCell[c=6]{c} \textbf{EDS Measured Atomic \%}
        &&&&&& \SetCell[r=2,c=1]{c}Phase
        & \SetCell[r=2,c=1]{c}Deformation\\  
        & Al & V & Cr & Fe & Co & Ni 
        & Al & V & Cr & Fe & Co & Ni 
        &  & \\
        E01 & \SetCell{gray!20}0 & \SetCell{gray!20}0 & 20 & 55 & \SetCell{gray!20}0 & 25 & \SetCell{gray!20}0 & \SetCell{gray!20}0 & 21.5 & 55.3 & \SetCell{gray!20}0 & 23.2 & FCC & Ductile\\
        E02 & \SetCell{gray!20}0 & 15 & \SetCell{gray!20}0 & \SetCell{gray!20}0 & 10 & 75 & \SetCell{gray!20}0 & 15.4 & \SetCell{gray!20}0 & \SetCell{gray!20}0 & 10.6 & 73.9 & BCC & Ductile\\
        E03 & \SetCell{gray!20}0 & 25 & 5 & 15 & 50 & 5 & \SetCell{gray!20}0 & 26.0 & 5.3 & 15.0 & 48.8 & 4.9 & FCC & Ductile\\
        E04 & \SetCell{gray!20}0 & \SetCell{gray!20}0 & 20 & 30 & 35 & 15 & \SetCell{gray!20}0 & \SetCell{gray!20}0 & 20.6 & 29.9 & 35.0 & 14.5 & FCC & Ductile\\
        E05 & \SetCell{gray!20}0 & 10 & 15 & \SetCell{gray!20}0 & 65 & 10 & \SetCell{gray!20}0 & 10.5 & 15.8 & 0.0 & 64.0 & 9.8 & FCC & Ductile\\
        E06 & \SetCell{gray!20}0 & 20 & \SetCell{gray!20}0 & 30 & 10 & 40 & \SetCell{gray!20}0 & 21.0 & \SetCell{gray!20}0 & 30.6 & 10.4 & 38.0 & FCC & Ductile\\
        E07 & \SetCell{gray!20}0 & 10 & 20 & 15 & 15 & 40 & \SetCell{gray!20}0 & 10.7 & 21.2 & 14.9 & 15.1 & 38.1 & FCC & Ductile\\
        \SetCell{red!20}\textbf{E08} & \SetCell{gray!20}0 & 20 & 5 & 15 & 40 & 20 & $\cdot$ & $\cdot$ & $\cdot$ & $\cdot$ & $\cdot$ & $\cdot$ & \SetCell{red!20}$\mathbf{FCC + L1_{2}}$ & Ductile\\
        E09 & \SetCell{gray!20}0 & 5 & \SetCell{gray!20}0 & 35 & 25 & 35 & \SetCell{gray!20}0 & 5.2 & \SetCell{gray!20}0 & 36.0 & 25.6 & 33.3 & FCC & Ductile\\
        E10 & \SetCell{gray!20}0 & 20 & 5 & 15 & 20 & 40 & \SetCell{gray!20}0 & 20.9 & 5.3 & 15.1 & 20.2 & 38.4 & FCC & Ductile\\
        E11 & \SetCell{gray!20}0 & 25 & 10 & \SetCell{gray!20}0 & 20 & 45 & \SetCell{gray!20}0 & 26.2 & 10.4 & 0.0 & 20.0 & 43.5 & FCC & Ductile\\
        E12 & 5 & 5 & \SetCell{gray!20}0 & 40 & 5 & 45 & 3.9 & 5.5 & \SetCell{gray!20}0 & 42.2 & 5.6 & 42.8 & FCC & Ductile\\
        E13 & \SetCell{gray!20}0 & 20 & 5 & 10 & 10 & 55 & \SetCell{gray!20}0 & 20.9 & 5.4 & 10.1 & 10.3 & 53.3 & FCC & Ductile\\
        E14 & \SetCell{gray!20}0 & 15 & 20 & 15 & 20 & 30 & \SetCell{gray!20}0 & 15.9 & 21.1 & 14.8 & 19.8 & 28.5 & FCC & Ductile\\
        \SetCell{red!20}\textbf{E15} & \SetCell{gray!20}0 & 30 & 5 & 15 & 30 & 20 & $\cdot$ & $\cdot$ & $\cdot$ & $\cdot$ & $\cdot$ & $\cdot$ & \SetCell{red!20}\textbf{FCC + Sigma} & \SetCell{red!20}\textbf{Brittle}\\
        E16 & \SetCell{gray!20}0 & 5 & \SetCell{gray!20}0 & 15 & 30 & 50 & \SetCell{gray!20}0 & 5.2 & \SetCell{gray!20}0 & 15.6 & 30.9 & 48.4 & FCC & Ductile\\
    \end{tblr}
    \end{table}

\begin{figure}[H]
        \centering
        \includegraphics[width=0.9\linewidth]{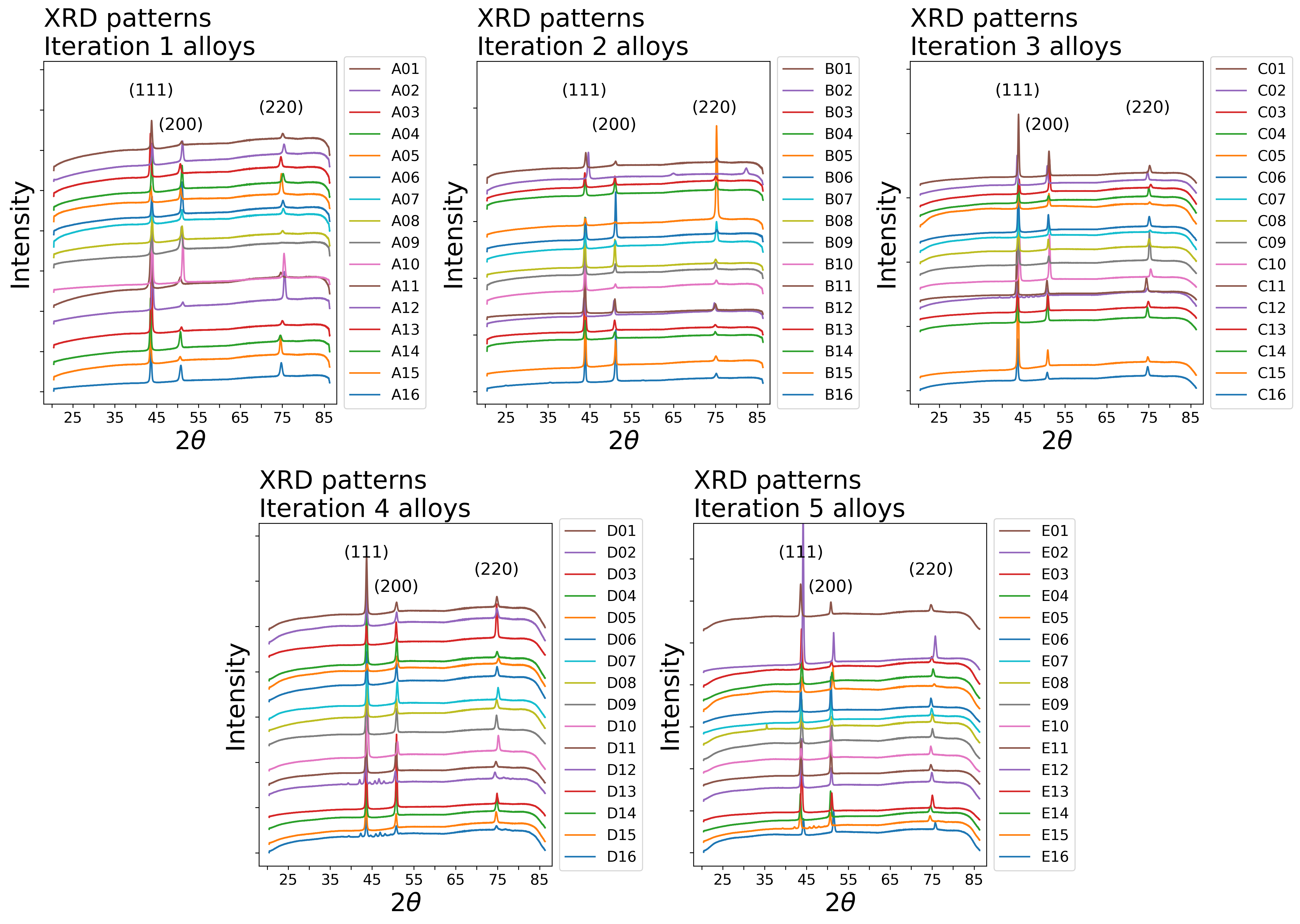}
        \caption{X-ray diffraction patterns for all alloys. Intensities within each iteration are shifted relative to each other to provide a visual of several samples at once. The expected positions for FCC (111), (200), and (220) peaks are labeled. Nearly all alloys have these confirmed peaks; phase exceptions, such as the clearly visible splitting in D12 and D16, are noted in \autoref{table_1} through \autoref{table_5}.}
        \label{ff1_xrd} 
    \end{figure}

    \newpage
    \subsection{\textbf{Supplementary \autoref{ff1_xrd}} \texorpdfstring{$\mid\;$}{Lg}XRD data for all alloys}
    \label{s1_6}

\begin{figure}[H]
        \centering
        \includegraphics[width=0.9\linewidth]{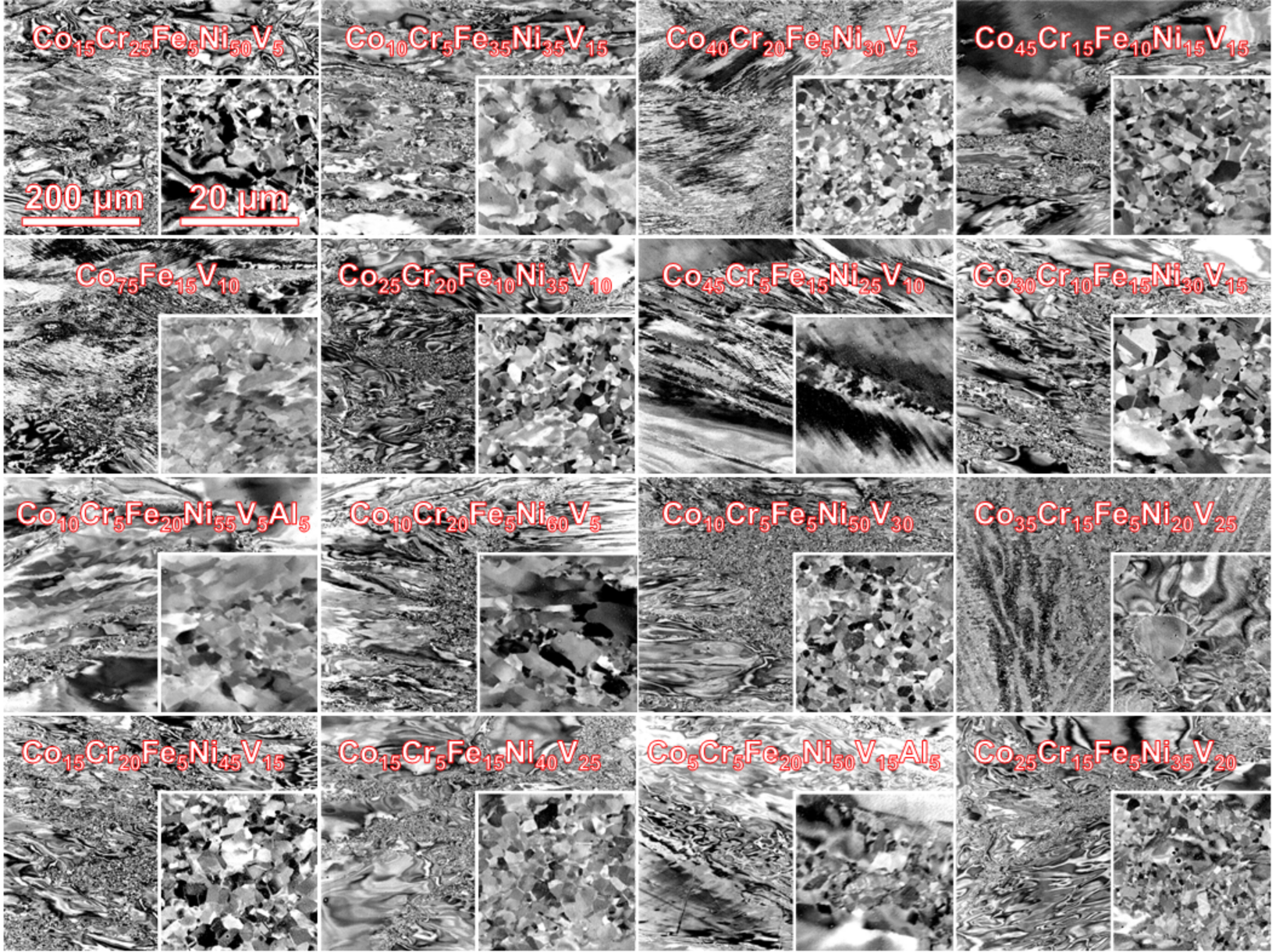}
        \caption{Back scattered electron micrographs for the 16 alloys of iteration 3. For each case, an inset at the bottom right of the image presents a section with a magnification 10 times larger than the base micrograph. Typically, the inset presents a region recrystallized during the high-temperature thermomechanical processing of the ingots. Samples in this iteration were single phase with the exception of the alloy C12 (FCC+Sigma phase) with a nominal composition of \ce{V_{25}Cr_{15}Fe_{5}Co_{35}Ni_{20}} (3rd row, 4th column), which has white second phase particles.}
        \label{ff2_sem}
    \end{figure}

    \newpage
    \subsection{\textbf{Supplementary \autoref{ff2_sem}} \texorpdfstring{$\mid\;$}{Lg}Example SEM micrographs}
    \label{s1_7}

    \autoref{ff2_sem} shows a bimodal grain structure for most alloys, with no visible porosity.

\begin{figure}[H]
        \centering
        \includegraphics[width=0.9\linewidth]{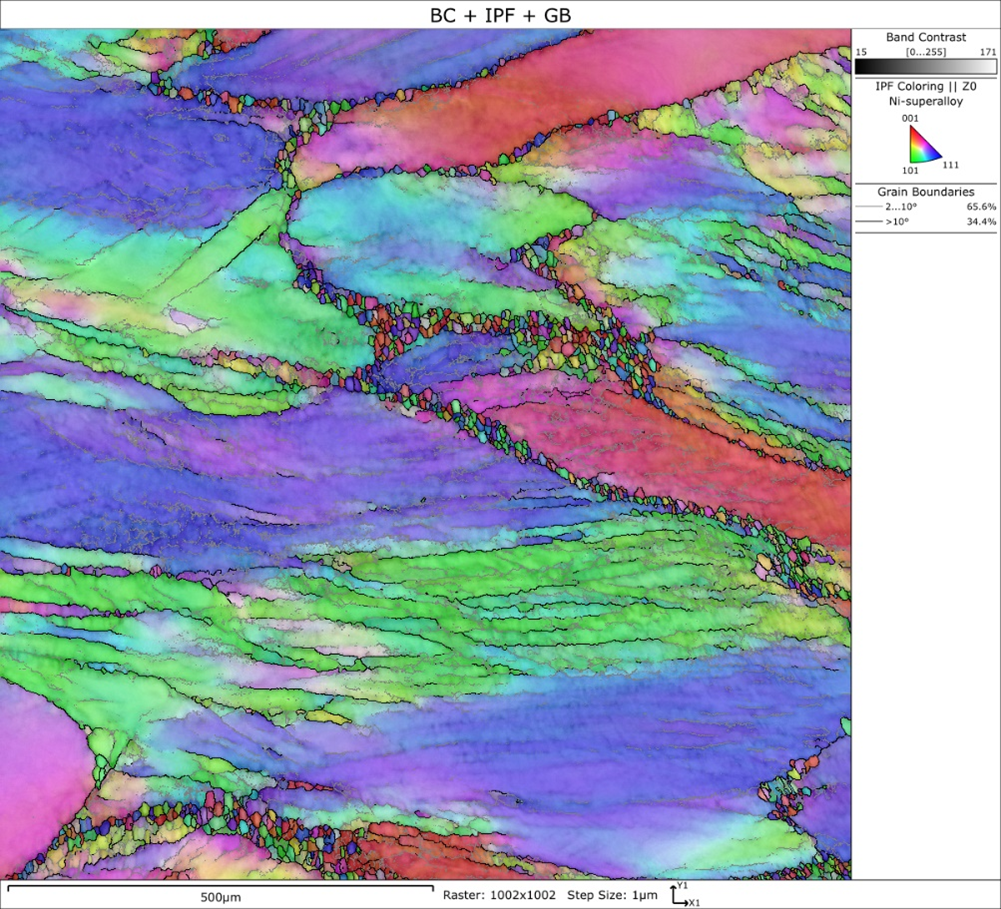}
        \caption{Electron backscatter diffraction micrograph for the fourth alloy in the first iteration. The bimodal grain structure resultant from vacuum arc meleting is clearly visible.}
        \label{ff3_ebsd}
    \end{figure}

    \newpage
    \subsection{\textbf{Supplementary \autoref{ff3_ebsd}} \texorpdfstring{$\mid\;$}{Lg}Example EBSD}
    \label{s1_8}

\newpage
\section{\textbf{PROPERTIES FOR ALL ALLOYS}}
\label{s2_properties}
\vspace{-0.25em}

    Some alloys which violated project constraints, which are highlighted in red, were not recorded with a tension experiment. Alloys for the final iteration were not recorded for their modulus of elasticity.

    \subsection{\textbf{Supplementary \autoref{table_6}} \texorpdfstring{$\mid\;$}{Lg}Iteration 1 alloy properties}
    \label{s2_1}
    \vspace{-1em}

    \begin{table}[H]
    \caption{}
    \label{table_6}
    \begin{tblr}{
        hlines,
        hline{3,11} = {2pt},
        rowsep=1pt,
        colspec = {
            |X[1.5, c]
            |[2pt]X[1.0, c]|X[1.0, c]|X[1.0, c]|X[1.0, c]
            |[2pt]X[1.0, c]|X[1.0, c]|X[1.0, c]
            |[2pt]X[1.5, c]|
            },
        colsep=0pt,
        }
        \SetCell[r=2,c=1]{c}\textbf{\makecell{Alloy\\Name}}
        & \SetCell[c=4]{c} \textbf{Tension Test Properties}
        &&&& \SetCell[c=3]{c} \textbf{Objective Measurements}
        &&& \SetCell[r=2,c=1]{c}\textbf{Pareto}\\  
        & \makecell{E\\(GPa)} & \makecell{YS\\(MPa)} & \makecell{UTS\\(MPa)} & \makecell{EL\\(\%)}
        & \makecell{UTS/YS\\(ul.)} & \makecell{H\\(GPa)} & \makecell{SRS\\$(ul. *10^{-3})$}
        &\\
        A01 & 229 & 487 & 839 & 25.7 & 1.67 & 2.62 & 8.17 & No\\
        A02 & 220 & 537 & 829 & 20.8 & 1.66 & 2.85 & 7.13 & No\\
        A03 & 206 & 427 & 721 & 22.1 & 1.59 & 2.40 & 8.83 & No\\
        A04 & 195 & 383 & 621 & 19.5 & 1.61 & 2.31 & 7.01 & No\\
        A05 & 162 & 310 & 446 & 18.3 & 1.38 & 1.82 & 8.05 & No\\
        A06 & 230 & 650 & 974 & 20.5 & 1.66 & 2.90 & 8.24 & No\\
        A07 & 191 & 606 & 952 & 24.9 & 1.68 & 2.88 & 8.57 & No\\
        A08 & 189 & 447 & 660 & 14.2 & 1.57 & 2.15 & 7.17 & No\\
        \SetCell{green!20}\textbf{A09} & 185 & 331 & 782 & 40.1 & 2.13 & 2.31 & 8.56 & \SetCell{green!20}\textbf{Yes}\\
        A10 & 205 & 376 & 764 & 28.1 & 1.90 & 2.24 & 6.32 & No\\
        A11 & 165 & 364 & 547 & 11.9 & 1.45 & 1.91 & 7.97 & No\\
        A12 & 244 & 391 & 622 & 16.7 & 1.59 & 2.47 & 7.46 & No\\
        \SetCell{red!20}\textbf{A13} & 185 & 275 & 816 & \SetCell{red!20}\textbf{0} & \SetCell{red!20}\textbf{N/A} & 2.32 & 7.90 & \SetCell{red!20}\textbf{N/A}\\
        \SetCell{green!20}\textbf{A14} & 192 & 601 & 1144 & 29.9 & 1.91 & 3.40 & 8.24 & \SetCell{green!20}\textbf{Yes}\\
        A15 & 148 & 253 & 492 & 23.3 & 1.93 & 1.96 & 7.73 & No\\
        \SetCell{green!20}\textbf{A16} & 199 & 517 & 977 & 26.7 & 1.78 & 2.88 & 8.44 & \SetCell{green!20}\textbf{Yes}\\
    \end{tblr}
    \end{table}

    \newpage
    \subsection{\textbf{Supplementary \autoref{table_7}} \texorpdfstring{$\mid\;$}{Lg}Iteration 2 alloy properties}
    \label{s2_2}
    \vspace{-1.25em}

    \begin{table}[H]
    \caption{}
    \label{table_7}
    \begin{tblr}{
        hlines,
        hline{3,11} = {2pt},
        rowsep=1pt,
        colspec = {
            |X[1.5, c]
            |[2pt]X[1.0, c]|X[1.0, c]|X[1.0, c]|X[1.0, c]
            |[2pt]X[1.0, c]|X[1.0, c]|X[1.0, c]
            |[2pt]X[1.5, c]|
            },
        colsep=0pt,
        }
        \SetCell[r=2,c=1]{c}\textbf{\makecell{Alloy\\Name}}
        & \SetCell[c=4]{c} \textbf{Tension Test Properties}
        &&&& \SetCell[c=3]{c} \textbf{Objective Measurements}
        &&& \SetCell[r=2,c=1]{c}\textbf{Pareto}\\  
        & \makecell{E\\(GPa)} & \makecell{YS\\(MPa)} & \makecell{UTS\\(MPa)} & \makecell{EL\\(\%)}
        & \makecell{UTS/YS\\(ul.)} & \makecell{H\\(GPa)} & \makecell{SRS\\$(ul. *10^{-3})$}
        &\\
        B01 & 230 & 539 & 1038 & 34.6 & 1.94 & 2.68 & 8.36 & No\\
        \SetCell{red!20}\textbf{B02} & 185 & 610 & 798 & \SetCell{red!20}\textbf{3.6} & 1.31 & 2.53 & 12.05 & \SetCell{red!20}\textbf{N/A}\\
        B03 & 175 & 363 & 576 & 17.2 & 1.59 & 2.13 & 7.02 & No\\
        B04 & 193 & 413 & 676 & 20.1 & 1.64 & 2.29 & 6.78 & No\\
        B05 & 214 & 332 & 634 & 28.7 & 1.97 & 2.35 & 6.98 & No\\
        B06 & 241 & 508 & 857 & 24.4 & 1.69 & 2.60 & 7.52 & No\\
        B07 & 221 & 448 & 863 & 28.2 & 1.93 & 2.78 & 7.78 & No\\
        B08 & 172 & 324 & 604 & 23.4 & 1.87 & 2.11 & 6.42 & No\\
        B09 & 159 & 347 & 610 & 20.0 & 1.76 & 1.98 & 6.97 & No\\
        B10 & 167 & 313 & 615 & 24.6 & 1.98 & 2.11 & 6.48 & No\\
        B11 & 220 & 427 & 886 & 31.5 & 2.09 & 2.62 & 7.21 & No\\
        \SetCell{green!20}\textbf{B12} & 189 & 430 & 986 & 37 & 2.29 & 2.71 & 8.35 & \SetCell{green!20}\textbf{Yes}\\
        B13 & 179 & 250 & 549 & 27.5 & 2.19 & 1.87 & 7.43 & No\\
        \SetCell{green!20}\textbf{B14} & 191 & 282 & 691 & 41.8 & 2.45 & 2.25 & 7.35 & \SetCell{green!20}\textbf{Yes}\\
        B15 & 221 & 367 & 759 & 33.8 & 2.10 & 2.57 & 7.30 & No\\
        \SetCell{red!20}\textbf{B16} & 160 & 299 & 333 & \SetCell{red!20}\textbf{1.4} & 1.12 & 2.97 & 5.86 & \SetCell{red!20}\textbf{N/A}\\
    \end{tblr}
    \end{table}

    \subsection{\textbf{Supplementary \autoref{table_8}} \texorpdfstring{$\mid\;$}{Lg}Iteration 3 alloy properties}
    \label{s2_3}
    \vspace{-1.25em}

    \begin{table}[H]
    \caption{}
    \label{table_8}
    \begin{tblr}{
        hlines,
        hline{3,11} = {2pt},
        rowsep=1pt,
        colspec = {
            |X[1.5, c]
            |[2pt]X[1.0, c]|X[1.0, c]|X[1.0, c]|X[1.0, c]
            |[2pt]X[1.0, c]|X[1.0, c]|X[1.0, c]
            |[2pt]X[1.5, c]|
            },
        colsep=0pt,
        }
        \SetCell[r=2,c=1]{c}\textbf{\makecell{Alloy\\Name}}
        & \SetCell[c=4]{c} \textbf{Tension Test Properties}
        &&&& \SetCell[c=3]{c} \textbf{Objective Measurements}
        &&& \SetCell[r=2,c=1]{c}\textbf{Pareto}\\  
        & \makecell{E\\(GPa)} & \makecell{YS\\(MPa)} & \makecell{UTS\\(MPa)} & \makecell{EL\\(\%)}
        & \makecell{UTS/YS\\(ul.)} & \makecell{H\\(GPa)} & \makecell{SRS\\$(ul. *10^{-3})$}
        &\\
        C01 & 201 & 503 & 854 & 22.5 & 1.70 & 2.82 & 7.07 & No\\
        C02 & 191 & 391 & 832 & 29.7 & 2.13 & 2.52 & 7.86 & No\\
        \SetCell{green!20}\textbf{C03} & 216 & 461 & 863 & 26.2 & 1.89 & 3.05 & 7.68 & \SetCell{green!20}\textbf{Yes}\\
        \SetCell{green!20}\textbf{C04} & 225 & 569 & 1044 & 29.1 & 1.84 & 3.17 & 8.93 & \SetCell{green!20}\textbf{Yes}\\
        C05 & 199 & 382 & 646 & 27.2 & 1.70 & 2.25 & 8.16 & No\\
        C06 & 193 & 460 & 932 & 32.6 & 2.03 & 2.73 & 7.90 & No\\
        C07 & 223 & 357 & 671 & 26.6 & 1.88 & 2.58 & 7.42 & No\\
        \SetCell{green!20}\textbf{C08} & 213 & 444 & 966 & 33.8 & 2.18 & 2.83 & 8.19 &  \SetCell{green!20}\textbf{Yes}\\
        C09 & 222 & 366 & 804 & 38.4 & 2.20 & 2.67 & 7.77 & No\\
        C10 & 240 & 397 & 734 & 22.4 & 1.85 & 2.58 & 6.78 & No\\
        C11 & 191 & 637 & 1115 & 25.0 & 1.75 & 3.32 & 7.76 & No\\
         \SetCell{red!20}\textbf{C12} & 219 & 790 & 790 &  \SetCell{red!20}\textbf{0} &  \SetCell{red!20}\textbf{N/A} & 6.04 & 6.22 &  \SetCell{red!20}\textbf{N/A}\\
        C13 & 205 & 551 & 1060 & 31.0 & 1.93 & 2.96 & 7.84 & No\\
        \SetCell{green!20}\textbf{C14} & 207 & 541 & 1136 & 33.7 & 2.10 & 3.32 & 7.97 & \SetCell{green!20}\textbf{Yes}\\
        C15 & 232 & 542 & 1006 & 29.1 & 1.86 & 3.35 & 5.90 & No\\
        C16 & 217 & 613 & 1197 & 31.8 & 1.95 & 3.29 & 7.76 & No\\
    \end{tblr}
    \end{table}

    \newpage
    \subsection{\textbf{Supplementary \autoref{table_9}} \texorpdfstring{$\mid\;$}{Lg}Iteration 4 alloy properties}
    \label{s2_4}
    \vspace{-1.25em}

    \begin{table}[H]
    \caption{}
    \label{table_9}
    \begin{tblr}{
        hlines,
        hline{3,11} = {2pt},
        rowsep=1pt,
        colspec = {
            |X[1.5, c]
            |[2pt]X[1.0, c]|X[1.0, c]|X[1.0, c]|X[1.0, c]
            |[2pt]X[1.0, c]|X[1.0, c]|X[1.0, c]
            |[2pt]X[1.5, c]|
            },
        colsep=0pt,
        }
        \SetCell[r=2,c=1]{c}\textbf{\makecell{Alloy\\Name}}
        & \SetCell[c=4]{c} \textbf{Tension Test Properties}
        &&&& \SetCell[c=3]{c} \textbf{Objective Measurements}
        &&& \SetCell[r=2,c=1]{c}\textbf{Pareto}\\  
        & \makecell{E\\(GPa)} & \makecell{YS\\(MPa)} & \makecell{UTS\\(MPa)} & \makecell{EL\\(\%)}
        & \makecell{UTS/YS\\(ul.)} & \makecell{H\\(GPa)} & \makecell{SRS\\$(ul. *10^{-3})$}
        &\\
        \SetCell{green!20}\textbf{D01} & 226 & 625 & 1221 & 31 & 1.95 & 3.37 & 7.23 & \SetCell{green!20}\textbf{Yes}\\
        \SetCell{green!20}\textbf{D02} & 236 & 551 & 1504 & 24 & 2.73 & 3.82 & 5.62 & \SetCell{green!20}\textbf{Yes}\\
        \SetCell{green!20}\textbf{D03} & 189 & 361 & 842 & 34 & 2.33 & 2.38 & 7.44 & \SetCell{green!20}\textbf{Yes}\\
        D04 & 187 & 391 & 822 & 30 & 2.10 & 2.30 & 8.03 & No\\
        \SetCell{green!20}\textbf{D05} & 211 & 502 & 996 & 32 & 1.98 & 2.77 & 8.73 & \SetCell{green!20}\textbf{Yes}\\
        \SetCell{green!20}\textbf{D06} & 191 & 521 & 1042 & 30 & 2.01 & 2.94 & 7.90 & \SetCell{green!20}\textbf{Yes}\\
        \SetCell{green!20}\textbf{D07} & 210 & 443 & 986 & 34 & 2.23 & 2.93 & 7.28 & \SetCell{green!20}\textbf{Yes}\\
        \SetCell{green!20}\textbf{D08} & 197 & 491 & 945 & 29 & 1.93 & 2.77 & 8.82 & \SetCell{green!20}\textbf{Yes}\\
        \SetCell{green!20}\textbf{D09} & 194 & 525 & 1142 & 35 & 2.17 & 2.95 & 8.41 & \SetCell{green!20}\textbf{Yes}\\
        D10 & 176 & 308 & 620 & 23 & 2.02 & 2.15 & 6.72 & No\\
        \SetCell{green!20}\textbf{D11} & 206 & 590 & 1274 & 32 & 2.21 & 3.41 & 7.41 & \SetCell{green!20}\textbf{Yes}\\
        \SetCell{red!20}\textbf{D12} & 209 & 559 & 561 & \SetCell{red!20}\textbf{0} & \SetCell{red!20}\textbf{N/A} & 5.96 & 6.53 & \SetCell{red!20}\textbf{N/A}\\
        \SetCell{green!20}\textbf{D13} & 229 & 669 & 1164 & 26 & 1.74 & 3.48 & 6.63 & \SetCell{green!20}\textbf{Yes}\\
        D14 & 187 & 459 & 931 & 30 & 2.03 & 2.57 & 7.52 & No\\
        \SetCell{green!20}\textbf{D15} & 216 & 614 & 1311 & 37 & 2.13 & 3.02 & 8.13 & \SetCell{green!20}\textbf{Yes}\\
        \SetCell{red!20}\textbf{D16} & 207 & 460 & 462 & \SetCell{red!20}\textbf{0} & \SetCell{red!20}\textbf{N/A} & 5.22 & 7.36 & \SetCell{red!20}\textbf{N/A}\\
    \end{tblr}
    \end{table}

    \subsection{\textbf{Supplementary \autoref{table_10}} \texorpdfstring{$\mid\;$}{Lg}Iteration 5 alloy properties}
    \label{s2_5}
    \vspace{-1.25em}

    \begin{table}[H]
    \caption{}
    \label{table_10}
    \begin{tblr}{
        hlines,
        hline{3,11} = {2pt},
        rowsep=1pt,
        colspec = {
            |X[1.5, c]
            |[2pt]X[1.0, c]|X[1.0, c]|X[1.0, c]|X[1.0, c]
            |[2pt]X[1.0, c]|X[1.0, c]|X[1.0, c]
            |[2pt]X[1.5, c]|
            },
        colsep=0pt,
        }
        \SetCell[r=2,c=1]{c}\textbf{\makecell{Alloy\\Name}}
        & \SetCell[c=4]{c} \textbf{Tension Test Properties}
        &&&& \SetCell[c=3]{c} \textbf{Objective Measurements}
        &&& \SetCell[r=2,c=1]{c}\textbf{Pareto}\\  
        & \makecell{E\\(GPa)} & \makecell{YS\\(MPa)} & \makecell{UTS\\(MPa)} & \makecell{EL\\(\%)}
        & \makecell{UTS/YS\\(ul.)} & \makecell{H\\(GPa)} & \makecell{SRS\\$(ul. *10^{-3})$}
        &\\
        E01 & 188 & 318 & 569 & 23 & 1.79 & 1.63 & 6.73 & No\\
        \SetCell{green!20}\textbf{E02} & 204 & 398 & 899 & 36 & 2.26 & 2.42 & 6.19 & \SetCell{green!20}\textbf{Yes}\\
        \SetCell{green!20}\textbf{E03} & 220 & 539 & 1581 & 27 & 2.93 & 3.37 & 5.55 & \SetCell{green!20}\textbf{Yes}\\
        E04 & 241 & 448 & 892 & 31 & 1.99 & 2.19 & 7.97 & No\\
        E05 & 235 & 548 & 960 & 22 & 1.75 & 3.09 & 8.67 & No\\
        \SetCell{green!20}\textbf{E06} & 178 & 336 & 850 & 36 & 2.53 & 2.33 & 8.61 & \SetCell{green!20}\textbf{Yes}\\
        E07 & 193 & 431 & 895 & 32 & 2.08 & 2.43 & 7.70 & No\\
        \SetCell{red!20}\textbf{E08} & 191 & 518 & 1325 & 21 & 2.56 & 3.42 & 6.17 & \SetCell{red!20}\textbf{N/A}\\
        E09 & 142 & 305 & 557 & 20 & 1.83 & 1.80 & 7.59 & No\\
        \SetCell{green!20}\textbf{E10} & 142 & 244 & 644 & 39 & 2.64 & 1.85 & 7.82 & \SetCell{green!20}\textbf{Yes}\\
        E11 & 183 & 615 & 1256 & 36 & 2.04 & 3.00 & 7.92 & No\\
        \SetCell{green!20}\textbf{E12} & 194 & 477 & 1062 & 34 & 2.23 & 2.69 & 8.39 & \SetCell{green!20}\textbf{Yes}\\
        \SetCell{green!20}\textbf{E13} & 194 & 469 & 1040 & 33 & 2.22 & 2.76 & 7.35 & \SetCell{green!20}\textbf{Yes}\\
        \SetCell{green!20}\textbf{E14} & 205 & 470 & 967 & 31 & 2.06 & 2.51 & 8.72 & \SetCell{green!20}\textbf{Yes}\\
        \SetCell{red!20}\textbf{E15} & $\cdot$ & $\cdot$ & $\cdot$ & $\cdot$ & $\cdot$ & 4.28 & 5.95 & \SetCell{red!20}\textbf{N/A}\\
        E16 & 205 & 339 & 620 & 22 & 1.83 & 1.92 & 6.83 & No\\
    \end{tblr}
    \end{table}

\begin{figure}[H]
        \centering
        \includegraphics[width=0.6\linewidth]{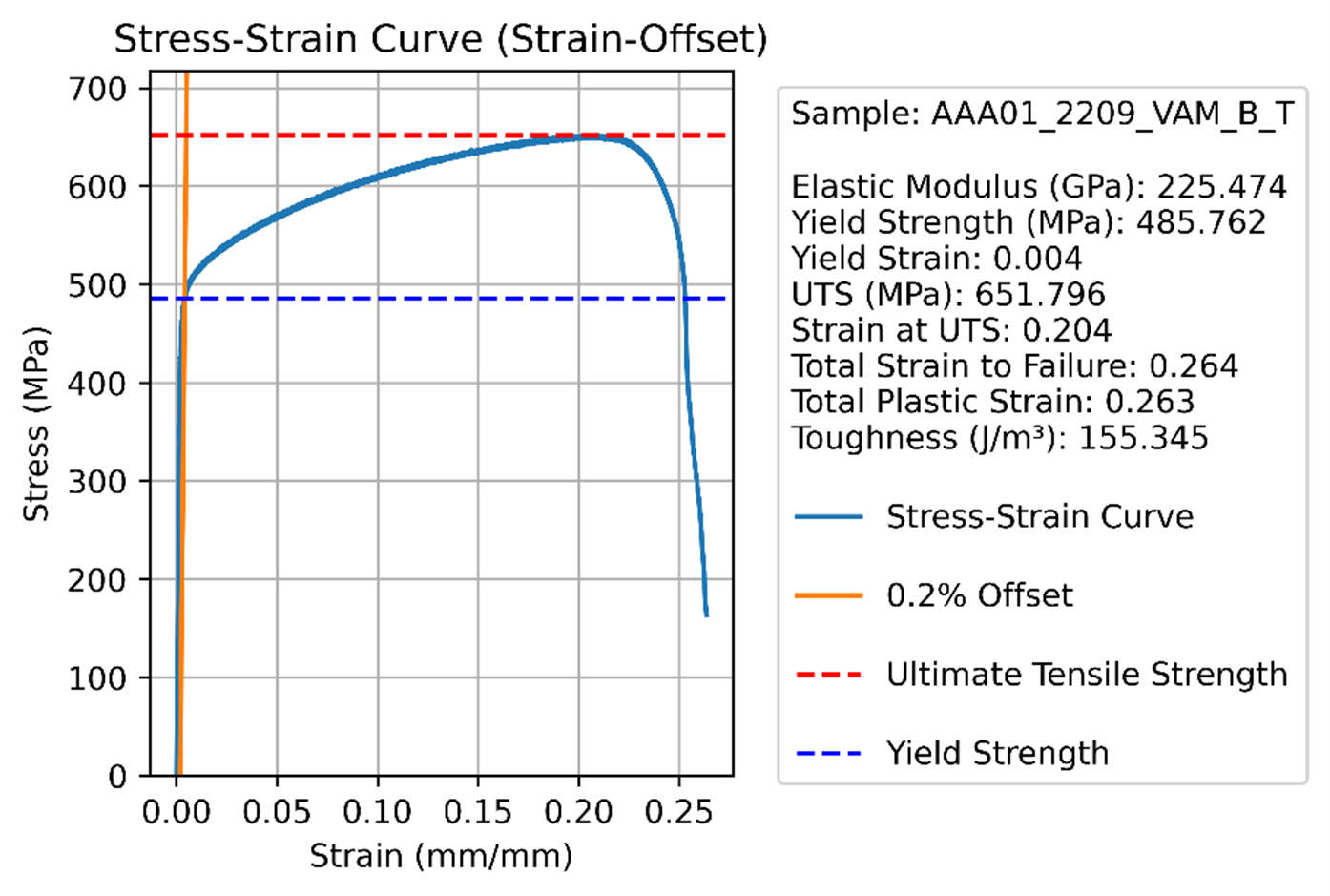}
        \caption{Example tension test data extraction.}
        \label{ff4_tension}
    \end{figure}

    \newpage
    \subsection{\textbf{Supplementary \autoref{ff4_tension}} \texorpdfstring{$\mid\;$}{Lg}Example Tension Test}
    \label{s2_6}
    \vspace{-1.25em}

Axial force and axial displacement, read from tab-delimited text files, were used to determine stress and strain, with data points after fracture removed. The stress-strain curve was initially searched for the most linear region with a fixed window of 200 strain data points  to find the elastic region of interest. This method worked for most (but not all) examples of material deformation: a more robust method was developed by finding the largest linear region that is above a set linear-fit threshold (e.g., $R^2=0.995$). The selected elastic region was used to calculate the elastic modulus of the sample, which can identify the strain offset so that the data passes through (0,0). This offset is minimal but necessary for accurately determining the yield point when the stress-strain curve deviates from ideal behavior. The yield strength, ultimate tensile strength, toughness, and strain to failure were calculated after the strain-offset correction was applied to the engineering stress-strain curve. True-stress and true-strain was calculated throughout the region of uniform elongation. Strain rates are assessed throughout the test with numerical methods (e.g., moving average, Savitzky-Golay filtering, spline interpolation) to verify that prescribed strain rates (of $5*10^{-4} s^{-1}$) were achieved during the tensile test.

\begin{figure}[H]
        \centering
        \includegraphics[width=0.6\linewidth]{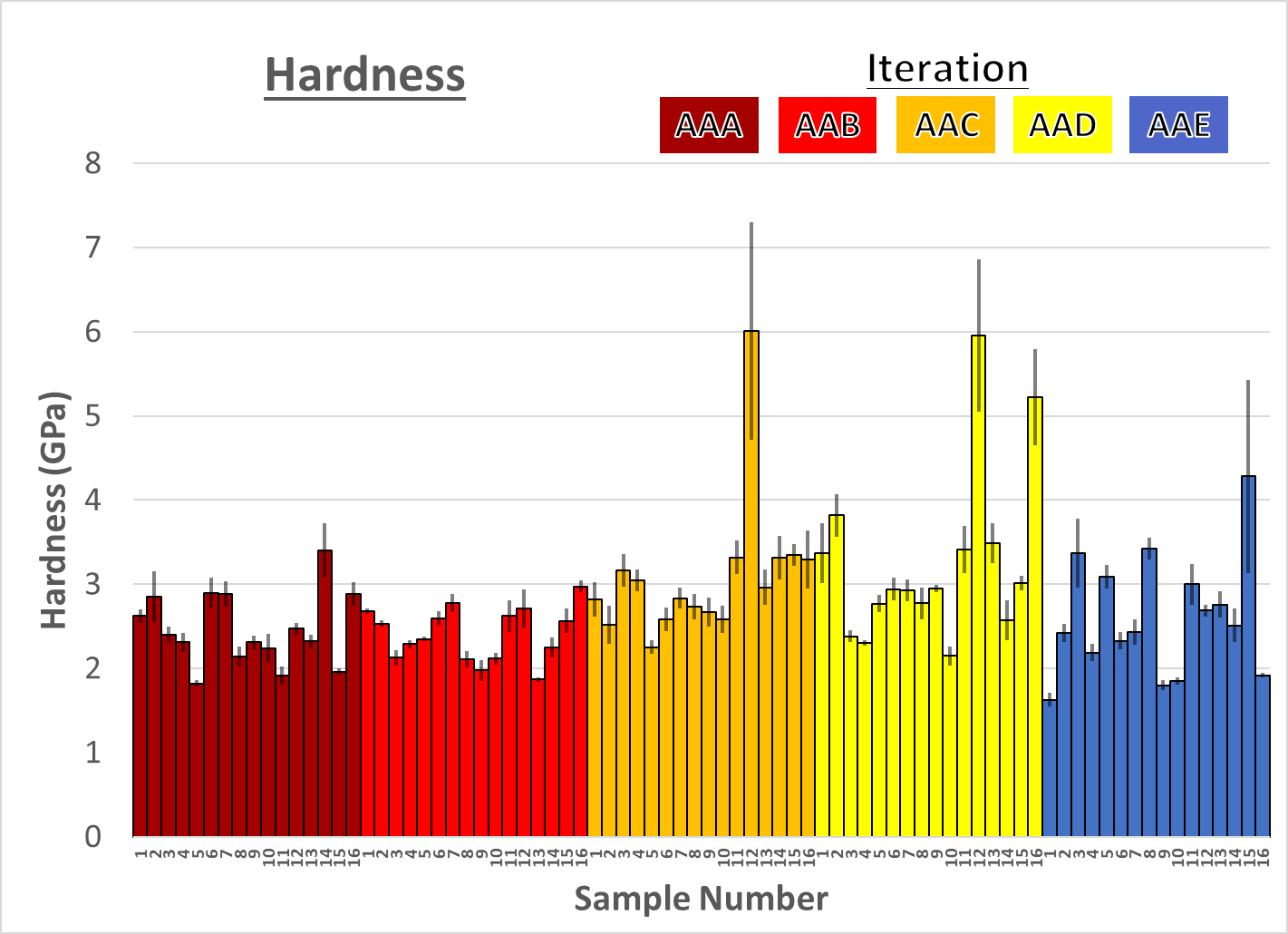}
        \caption{Hardness values of each alloy, separated by iteration. Large values that correspond to alloys with constraint violations, as seen in \autoref{s1_1} through \autoref{s1_5}, were excluded from the optimization loop.}
        \label{ff5_hardness}
    \end{figure}

    \newpage
    \subsection{\textbf{Supplementary \autoref{ff5_hardness}} \texorpdfstring{$\mid\;$}{Lg}Summary of Nanoindentation Hardness}
    \label{s2_7}
    \vspace{-1.25em}

\begin{figure}[H]
        \centering
        \includegraphics[width=0.6\linewidth]{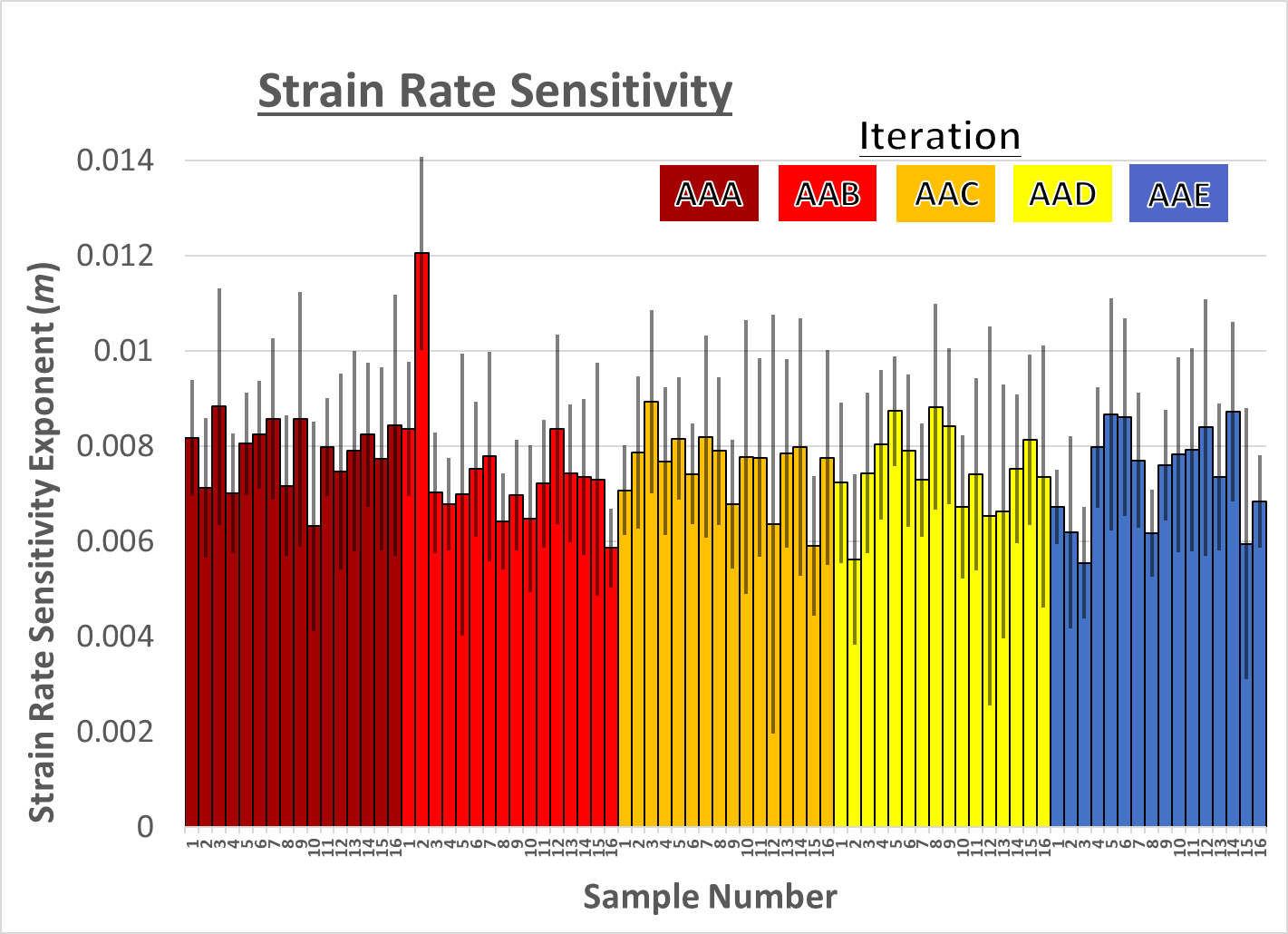}
        \caption{Strain rate sensitivity exponents for each alloy. The scale is multiplied in the optimization loop as well as ease of viewing; these materials are largely strain rate insensitive.}
        \label{ff6_values}
    \end{figure}

    \subsection{\textbf{Supplementary \autoref{ff6_values}} \texorpdfstring{$\mid\;$}{Lg}Summary of Nanoindentation SRS}

\begin{figure}[H]
        \centering
        \includegraphics[width=1.0\linewidth]{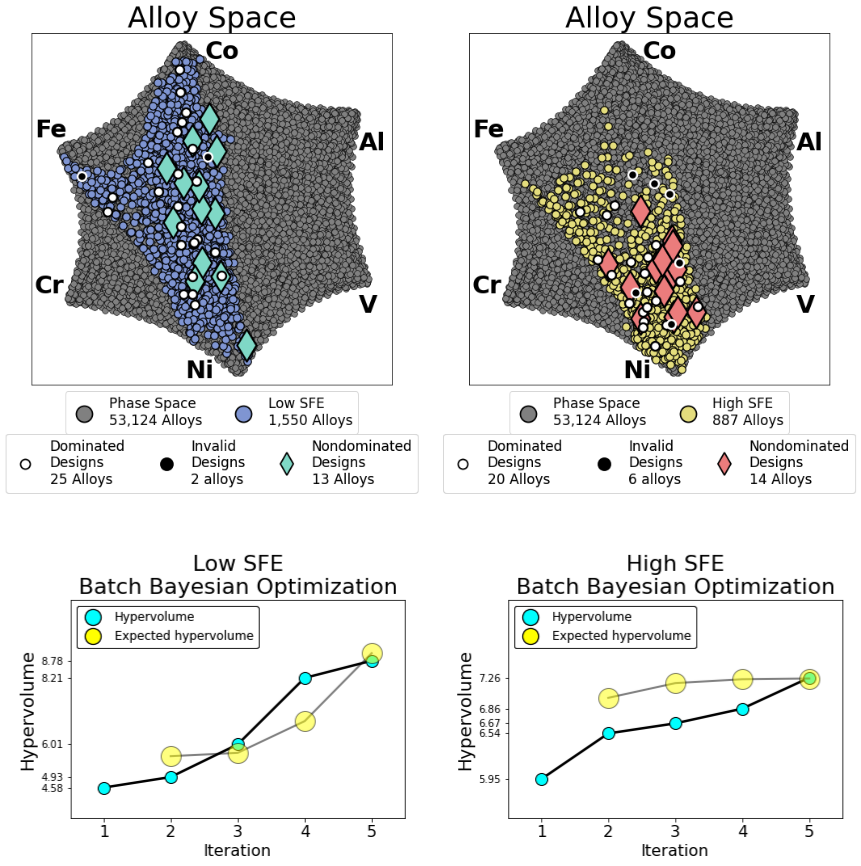}
        \caption{EHVI estimates for each of the iterations along with their experimental hypervolumes.}
        \label{EHVI}
    \end{figure}

    \newpage
    \subsection{\textbf{Supplementary \autoref{EHVI}} \texorpdfstring{$\mid\;$}{Lg}EHVI Estimates}
    \label{s3_2}
    \vspace{-1.25em}

\autoref{EHVI} Since EVHI was calculated with an ensemble, no one value describes the "EHVI" of the system. However, an estimate for EHVI can be provided by taking the improvement predicted by each GP and averaging the improvements predicted for identical candidates. Then, the improvements associated with the candidates selected for batch production can be de-normalized and compared to the HV found after experimental testing. \autoref{EHVI} shows the expected HV in this manner along with the actual.

\newpage
\section{\textbf{ALLOY PARETO SETS}}
\label{s3_pareto_alloys}
\vspace{-0.25em}

    \subsection{\textbf{Supplementary \autoref{table_11}} \texorpdfstring{$\mid\;$}{Lg}Alloy Pareto Sets}
    \label{s2_5}
    \vspace{-1.25em}

    \begin{table}[H]
    \caption{}
    \label{table_11}
    \begin{tblr}{
        hlines,
        hline{3,16,30} = {2pt},
        rowsep=1pt,
        colspec = {
            |X[1.5, c]
            |[2pt]X[1.0, c]|X[1.0, c]|X[1.0, c]
            |X[1.0, c]|X[1.0, c]|X[1.0, c]
            |[2pt]X[1.5, c]|X[1.5, c]|X[1.5, c]|
            },
        colsep=0pt,
        }
        \SetCell[r=2,c=1]{c}\textbf{\makecell{Alloy\\Name}}
        & \SetCell[c=6]{c} \textbf{EDS Measured Atomic \%}
        &&&&&& \SetCell[c=3]{c} \textbf{Objectives}
        &&\\
        & Al & V & Cr & Fe & Co & Ni 
        & \makecell{UTS/YS\\(ul.)} & \makecell{H\\(GPa)} & \makecell{SRS\\$(ul. *10^{-3})$}\\
        C03 & \SetCell{gray!20}0 & 5.1 & 20.2 & 5.1 & 40.1 & 29.5 & 1.89 & 3.05 & 7.68\\
        C04 & \SetCell{gray!20}0 & 15.1 & 15.1 & 10.1 & 44.8 & 14.8 & 1.84 & 3.17 & \SetCell{green!20}\textbf{8.93}\\
        C08 & \SetCell{gray!20}0 & 15.2 & 10.1 & 15.2 & 29.8 & 29.6 & 2.18 & 2.83 & 8.19\\
        D01 & \SetCell{gray!20}0 & 21.2 & 10.6 & 9.9 & 29.6 & 28.6 & 1.95 & 3.37 & 7.23\\
        D02 & \SetCell{gray!20}0 & 26.1 & 5.3 & 14.9 & 39.4 & 14.3 & 2.73 & \SetCell{green!20}\textbf{3.82} & 5.62\\
        D03 & \SetCell{gray!20}0 & 15.8 & 5.3 & 30.4 & 5.3 & 43.2 & 2.33 & 2.38 & 7.44\\
        D05 & \SetCell{gray!20}0 & 5.3 & 26.2 & 10.0 & 39.2 & 19.2 & 1.98 & 2.77 & 8.73\\
        D06 & \SetCell{gray!20}0 & 15.8 & 15.7 & 10.0 & 34.5 & 24.0 & 2.01 & 2.94 & 7.9\\
        D07 & \SetCell{gray!20}0 & 15.9 & 10.6 & 10.1 & 20.1 & 43.2 & 2.23 & 2.93 & 7.28\\
        D08 & \SetCell{gray!20}0 & 10.6 & 26.2 & 14.7 & 24.7 & 23.9 & 1.93 & 2.77 & 8.82\\
        E02 & \SetCell{gray!20}0 & 15.4 & \SetCell{gray!20}0 & \SetCell{gray!20}0 & 10.6 & 73.9 & 2.26 & 2.42 & 6.19\\
        E03 & \SetCell{gray!20}0 & 26.0 & 5.3 & 15.0 & 48.8 & 4.9 & \SetCell{green!20}\textbf{2.93} & 3.37 & 5.55\\
        E06 & \SetCell{gray!20}0 & 21.0 & \SetCell{gray!20}0 & 30.6 & 10.4 & 38.0 & 2.53 & 2.33 & 8.61\\
        A09 & 5.5 & 5.3 & 5.2 & 25.5 & \SetCell{gray!20}0 & 58.6 & 2.13 & 2.31 & 8.56\\
        A14 & \SetCell{gray!20}0 & 25.2 & 10.1 & 5.0 & 24.9 & 34.8 & 1.91 & 3.4 & 8.24\\
        A16 & \SetCell{gray!20}0 & 20.0 & 10.1 & 10.1 & 5.0 & 54.7 & 1.78 & 2.88 & 8.44\\
        B12 & \SetCell{gray!20}0 & 20.4 & 5.1 & 20.2 & 15.0 & 39.4 & 2.29 & 2.71 & 8.35\\
        B14 & 5.3 & 5.1 & \SetCell{gray!20}0 & 30.3 & 10.2 & 49.0 & 2.45 & 2.25 & 7.35\\
        C14 & \SetCell{gray!20}0 & 25.4 & 5.0 & 15.1 & 15.0 & 39.5 & 2.1 & 3.32 & 7.97\\
        D09 & \SetCell{gray!20}0 & 21.0 & 15.8 & 4.9 & 15.0 & 43.3 & 2.17 & 2.95 & 8.41\\
        D11 & \SetCell{gray!20}0 & 26.3 & 10.5 & 4.9 & 19.8 & 38.5 & 2.21 & 3.41 & 7.41\\
        D13 & 3.1 & 26.6 & \SetCell{gray!20}0 & 5.2 & 10.4 & 54.7 & 1.74 & \SetCell{green!20}\textbf{3.48} & 6.63\\
        D15 & \SetCell{gray!20}0 & 26.1 & 5.3 & 10.0 & 24.9 & 33.7 & 2.13 & 3.02 & 8.13\\
        E10 & \SetCell{gray!20}0 & 20.9 & 5.3 & 15.1 & 20.2 & 38.4 & \SetCell{green!20}\textbf{2.64} & 1.85 & 7.82\\
        E12 & 3.9 & 5.5 & \SetCell{gray!20}0 & 42.2 & 5.6 & 42.8 & 2.23 & 2.69 & 8.39\\
        E13 & \SetCell{gray!20}0 & 20.9 & 5.4 & 10.1 & 10.3 & 53.3 & 2.22 & 2.76 & 7.35\\
        E14 & \SetCell{gray!20}0 & 15.9 & 21.1 & 14.8 & 19.8 & 28.5 & 2.06 & 2.51 & \SetCell{green!20}\textbf{8.72}\\
    \end{tblr}
    \end{table}

    Pareto sets for each parallel experiment. Elements not present in the composition are grayed out for ease of viewing common alloy systems. The edges of the Pareto surface, marked by the materials with the highest value in a single property, are highlighted in green.

\begin{figure}[H]
        \centering
        \includegraphics[width=1.0\linewidth]{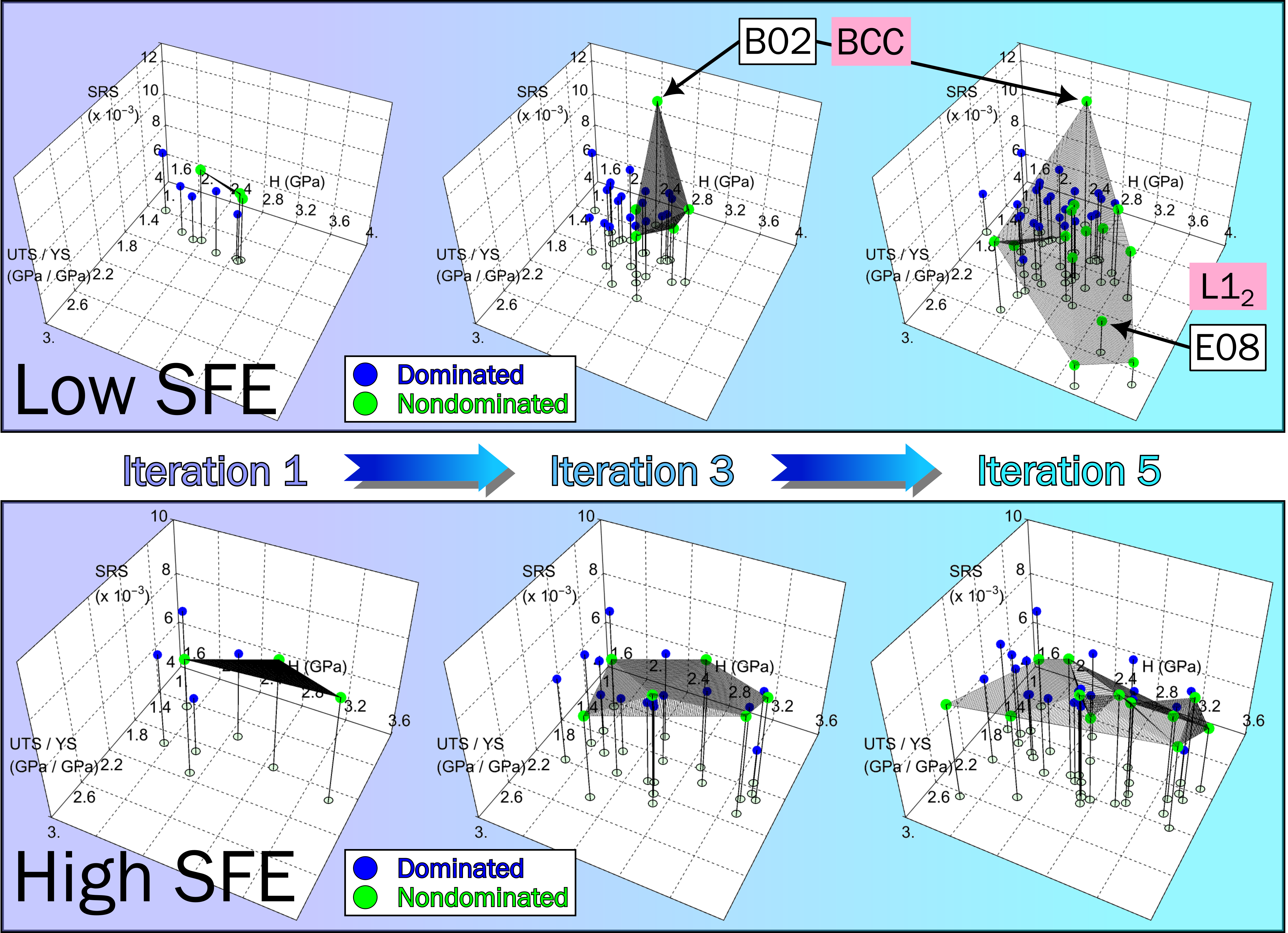}
        \caption{3D scatter plot of the nondominated and dominated points for each iteration after iterations 1, 3, and 5. Nondominated points are green; dominated points are blue. Nondominated surfaces are approximated with a connected mesh. A surface overlays the points forming the known Pareto set at that time. Note that this surface is only a visual approximation of the discretized points.}
        \label{ff_surfaces}
    \end{figure}

    \newpage
    \subsection{\textbf{Supplementary \autoref{ff_surfaces}} \texorpdfstring{$\mid\;$}{Lg}3D Surfaces}
    \label{s3_2}
    \vspace{-1.25em}

\autoref{ff_surfaces} shows the iterative expansion of the Pareto surface, reflecting how successive cycles refined alloy compositions and increased the hypervolume (HV). As part of this effort, points within the Pareto set---shown as green data points in \autoref{ff_surfaces}---are currently being scaled-up for further studies in order to elucidate in a more complete manner the underpinnings of their superior performance. The 3D scatter plot effectively captures how each iteration advanced the alloy designs toward improved mechanical properties across all properties of interest, as observed by the significant expansion in the Pareto surface as a function of iteration count.

\begin{figure}[H]
        \centering
        \includegraphics[width=1.0\linewidth]{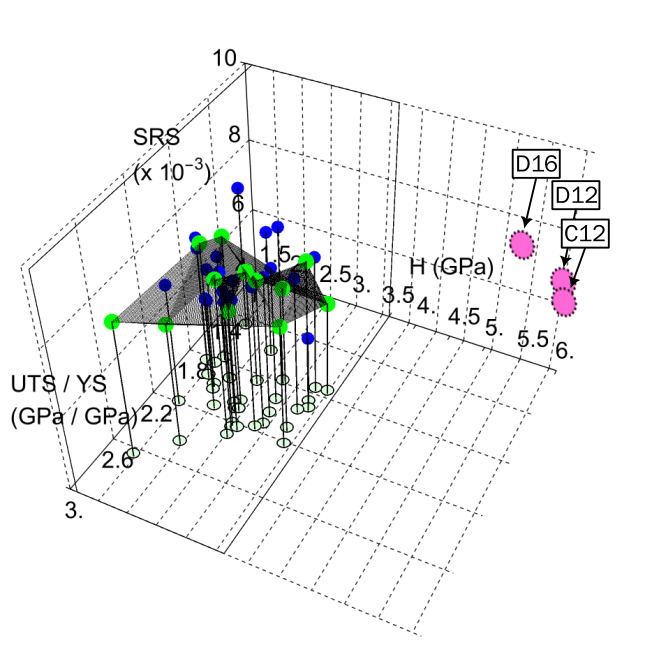}
        \caption{3D scatter plot of the nondominated and dominated points for the high SFE experiment after 5. The location of brittle violators have been included on the Hardness-SRS plane; since they have no yield value, they cannot be added to an approximate mesh.}
        \label{ff_surfaces2}
    \end{figure}

    \newpage
    \subsection{\textbf{Supplementary \autoref{ff_surfaces2}} \texorpdfstring{$\mid\;$}{Lg}Brittle Violators}
    \label{s3_2}
    \vspace{-1.25em}

\begin{figure}[H]
        \centering
        \includegraphics[width=0.8\linewidth]{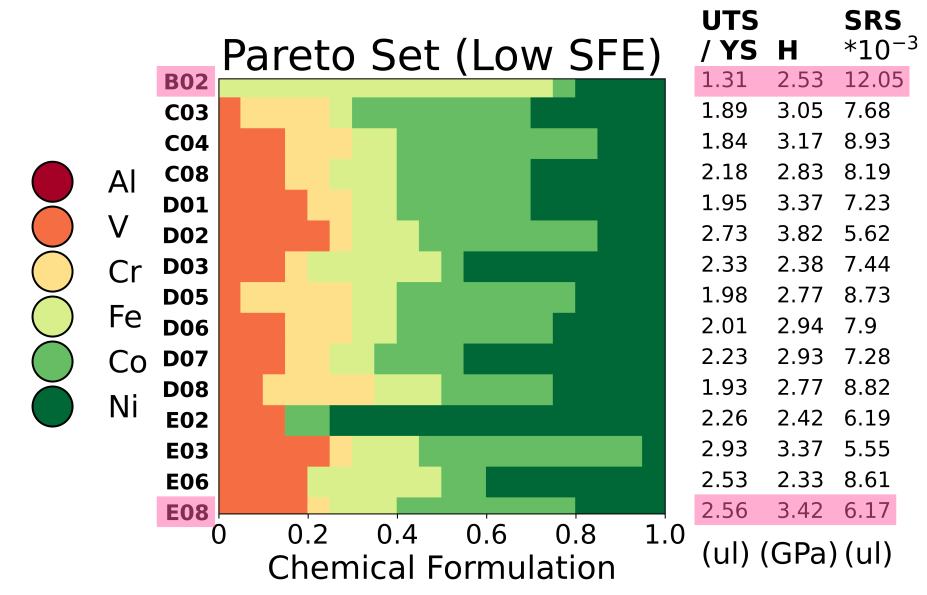}
        \caption{Compositional visual of low SFE Pareto set. Two of the alloys, B02 and E08, which violated project constraints, are highlighted in pink. Their high strain rate sensitivities would have put them on the Pareto set had they been single phase solid solution FCC.}
        \label{ff_pareto1}
    \end{figure}

    \newpage
    \subsection{\textbf{Supplementary \autoref{ff_pareto1}} \texorpdfstring{$\mid\;$}{Lg}Compositional Pareto Chart low SFE}
    \label{s3_3}
    \vspace{-1.25em}

\begin{figure}[H]
        \centering
        \includegraphics[width=0.8\linewidth]{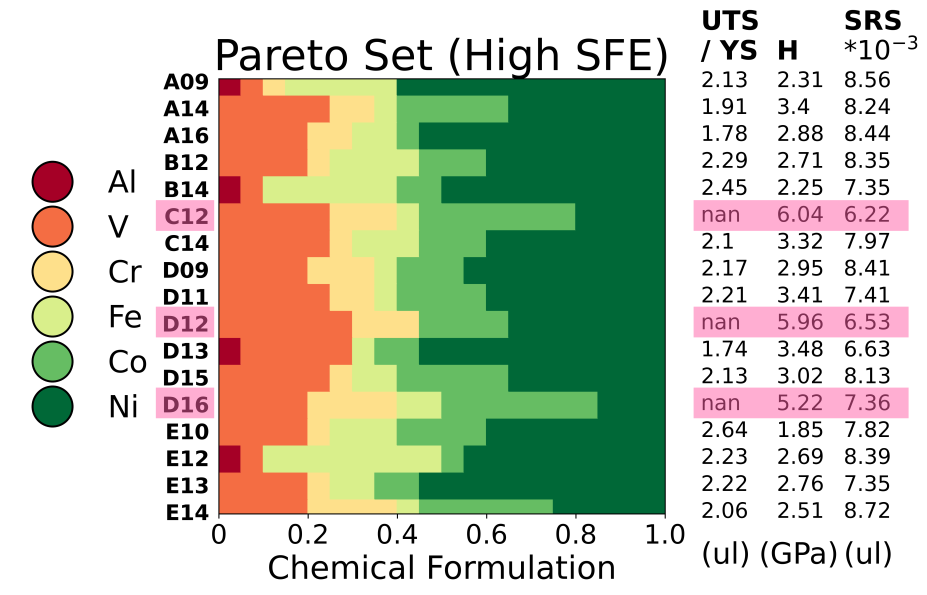}
        \caption{Compositional visual of high SFE Pareto set. C12, D12, and D16, which violated project constraints, are highlighted in pink. These alloys were brittle and thus do not have a value for one of the objectives.}
        \label{ff_pareto2}
    \end{figure}

    \subsection{\textbf{Supplementary \autoref{ff_pareto2}} \texorpdfstring{$\mid\;$}{Lg}Compositional Pareto Chart high SFE}
    \label{s3_3}
    \vspace{-1.25em}

\newpage
\section{\textbf{SIZE OF THE COMPLEX ALLOY SPACE}}
\label{s4_size}

    Determining the number of compositions possible given a number of elements and a maximum divisor is identical to a combinatorics problem wherein it is asked "How many ways \textbf{T} are there of arranging a fractionated quantity into \textbf{n} bins, where the quantity is divided into \textbf{d} equally sized identical partitions, using any number of the partitions for each bin, including zero?" The solution to this problem, excluding the cases where 1 bin receives all of the quantity, is listed in \autoref{eq_stirling}.

    \subsection{\textbf{Supplementary \autoref{eq_stirling}} \texorpdfstring{$\mid\;$}{Lg}Partition function T for n elements and d resolution}
    \label{s4_1}

    \vspace{-0.75em}
    {\small
    \begin{equation}
        \begin{aligned}
        T(n,d) = \left(\frac{1}{(n-1)!}\right)\left(\sum_{i=1}^{n-1} (|StirlingS1[n, i+1]|*d^{i}) - (n! - (n - 1)!)\right)
        \label{eq_stirling}
        \end{aligned}
    \end{equation}
    }

    \autoref{eq_stirling} uses the Stirling Numbers of the First Kind, which are evaluated recursively. When n is set to a constant, \autoref{eq_stirling} reduces to a polynomial, shown for n = 6 (this work) in \autoref{eq_stirling_constant}.

    \subsection{\textbf{Supplementary \autoref{eq_stirling_constant}} \texorpdfstring{$\mid\;$}{Lg}Partition function T for 6 elements and d resolution}
    \label{s4_2}

    \vspace{-0.75em}
    {\small
    \begin{equation}
        \begin{aligned}
        T(6,d) = \left(\frac{1}{5!}\right)(d^5+15d^4+85d^3+225d^2+274d-600)
        \label{eq_stirling_constant}
        \end{aligned}
    \end{equation}
    }

\begin{figure}[H]
        \centering
        \includegraphics[width=0.9\linewidth]{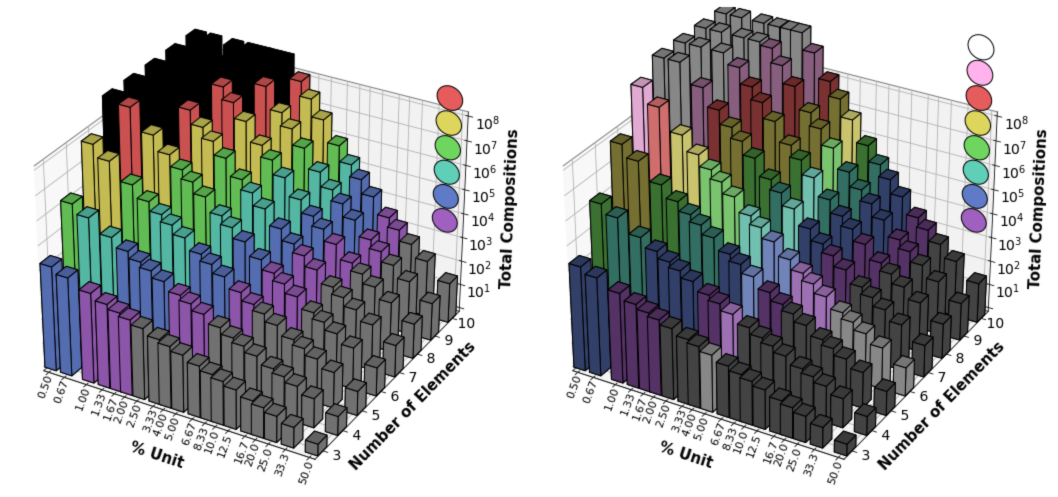}
        \caption{(left) The number of compositions contained in a list, to the nearest order of magnitude, given a number of elements and an elemental resolution, for typical percentages. (right) This work's 6 elements and 5\% resolution highlighted, resulting in a total alloy space of 53,124 possible compositions.}
        \label{ff5_space}
    \end{figure}

    \subsection{\textbf{Supplementary \autoref{ff5_space}} \texorpdfstring{$\mid\;$}{Lg}Sizes of typical composition lists}
    \label{s4_3}
    \vspace{-1.25em}

\newpage
\section{\textbf{VISUALIZATION OF HIGHER DIMENSIONAL SPACES}}
\label{s5_visual}

    UMAP (Uniform Manifold Approximation and Projection) is a dimensionality reduction technique particularly suitable for complex alloy spaces, as the graphs it can produce can be easily interpreted \cite{mcinnes_umap_2018}. Notably, when provided a complete list of compositions, it produces shapes whose vertices correspond to 100\% of an element, with alloys having higher entropy in the center. Additional information on producing these graphs can be found in \cite{vela_visualizing_2024}.
    
    \autoref{ff6_umap} (left) is an example of a UMAP of the 6-dimensional space with 53,124 compositions, sorted by cobalt fraction. This system, despite having over $10^{4}$ data points stacked on top of each other, provides an intuitive understanding of relative compositions: alloys nearly 1.00 mole fraction of an element will lie near that element’s vertex while high entropy alloys are moved toward the center. \autoref{ff6_umap} (middle, right) show the same space with entropy and density visualizations. As expected, higher entropy compositions are in the center, and the lowest densities are near the aluminum vertex. The "ridges" that can be seen are the result of UMAP mapping binary compositions directly from one vertex to another with regular spacing.

    While UMAP, like other dimensionality reduction algorithms, does not allow you to backtrack the higher dimensional data, it does give a general sense of comparison for nearby data points. For example, a data point closer to the vanadium vertex will have higher vanadium content.

\begin{figure}[H]
        \centering
        \includegraphics[width=0.9\linewidth]{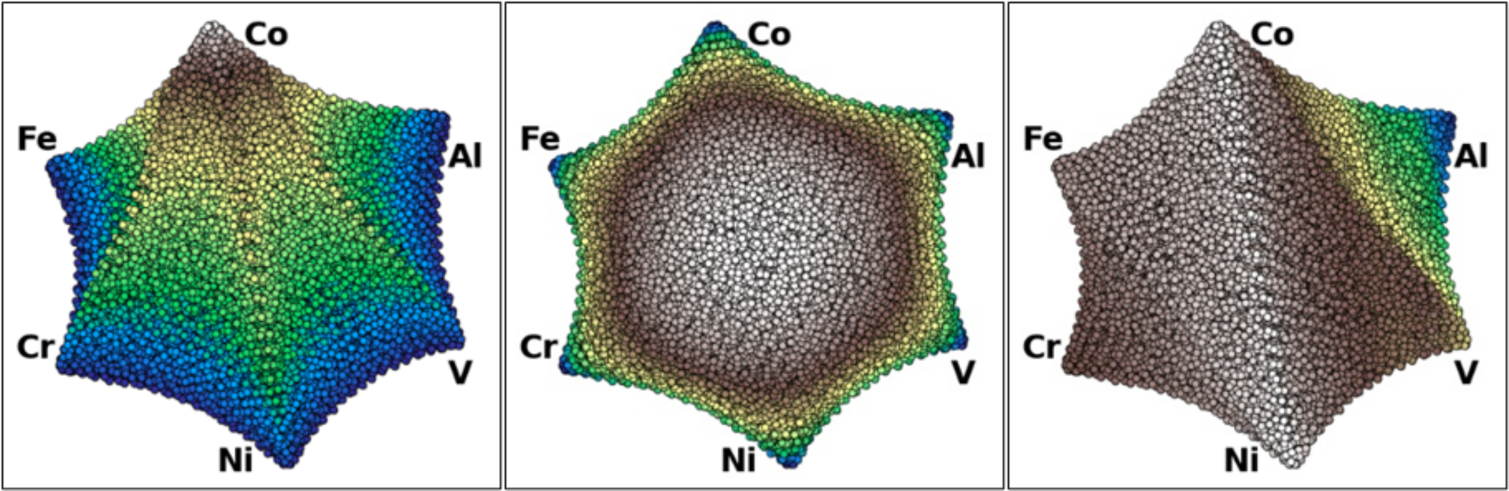}
        \caption{UMAPs of the \ce{Al-V-Cr-Fe-Co-Ni} space sorted by a property from high, white, to low, dark blue. (left) Sorted by cobalt fraction. (center) Sorted by configurational entropy (right) Sorted by density.}
        \label{ff6_umap}
    \end{figure}

    \subsection{\textbf{Supplementary \autoref{ff6_umap}} \texorpdfstring{$\mid\;$}{Lg}UMAP of alloy space}
    \label{s5_1}
    \vspace{-1.25em}

\begin{figure*}[htb]
        \centering
        \includegraphics[width=0.9\textwidth]{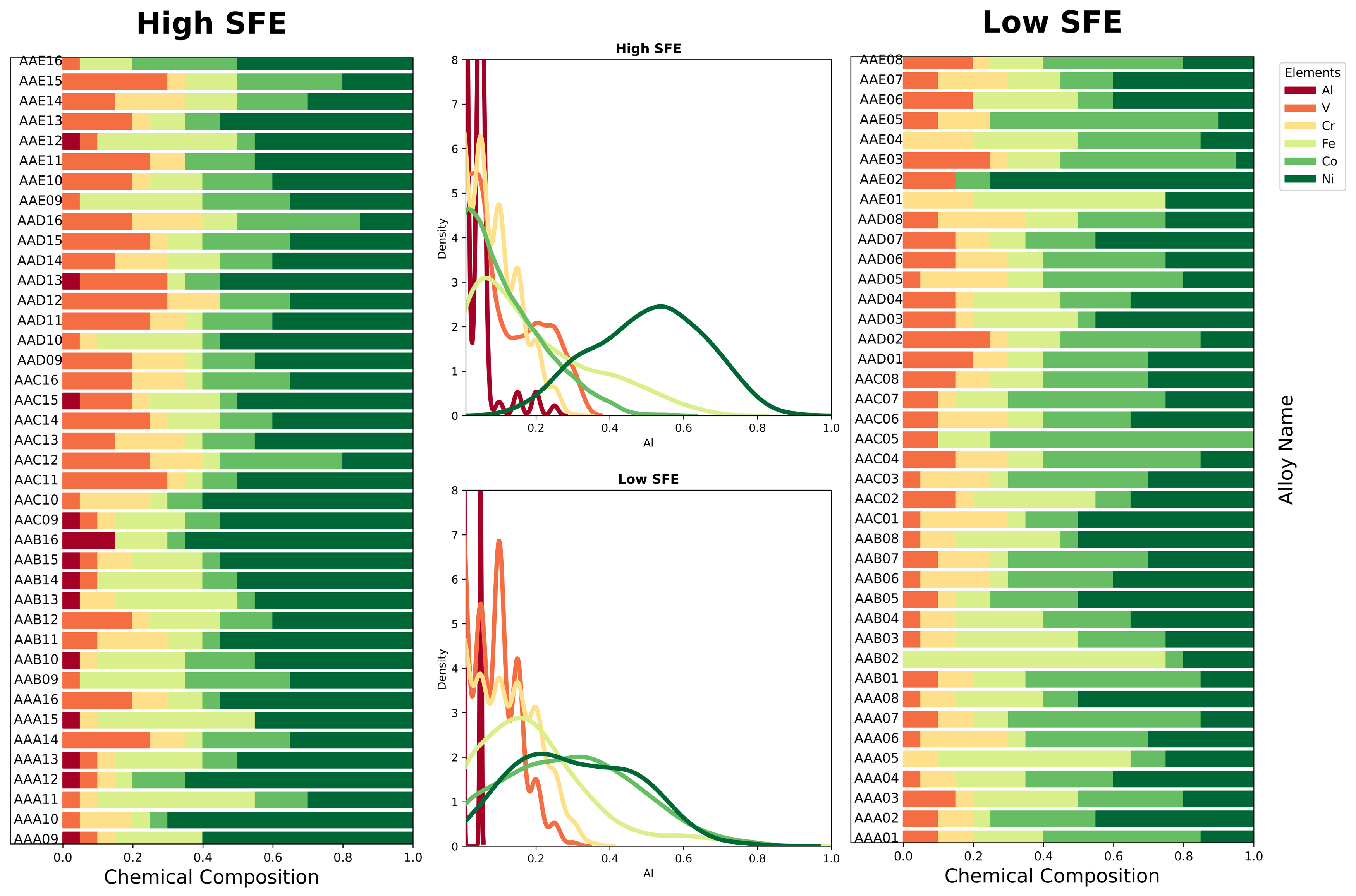}
        \caption{Composition profile of the alloys iteratively selected in this campaign for the "low" and "high" SFE subsets. The "low" SFE alloy set is (on average) richer in Co (an hcp stabilizer), while the "high" SFE set is (on average) richer in Ni.}
        \label{fig:composition_profile} 
    \end{figure*}

    \newpage
    \subsection{\textbf{Supplementary \autoref{fig:composition_profile}} \texorpdfstring{$\mid\;$}{Lg}Alloy composition profile}
    \label{s5_1}

    \autoref{fig:composition_profile} Shows a compositional profile of each of the alloys. The stacking fault energy definition precluded any low-SFE alloys from being designed with aluminum, while it ensured that high-SFE alloys would maintain a large nickel balance. Additionally, the low-SFE alloys are more rich in cobalt.

\newpage
\section{\textbf{METHODS OF PREDICTING SINGLE PHASE SOLID SOLUTION ALLOYS}}
\label{s6_methods}

    This work exclusively used CALPHAD (ThermoCalc) to predict the stability of single phase solid solutions. Three other methods of note include: building off of the Hume-Rothery Rules \cite{callister_materials_2018} \cite{takeuchi_classification_2005} \cite{zhang_miedema_2016}, inspecting binary phase diagram constituents \cite{gao_searching_2013} \cite{liaw_high-entropy_2016} \cite{zhang_microstructures_2014} \cite{zhao_hexagonal_2016}, and computing first order approximations as an extension of known databases \cite{curtarolo_aflow_2012} \cite{curtarolo_high-throughput_2013}. This section details how these methods appear to be insufficient for the purposes of this work.
    
    \bigskip

    A Hume-Rothery variant can be described by interactions between atomic fraction and empirical radii, and regular solution model interaction parameters from literature, shown in \autoref{eq_hume} \cite{yang_prediction_2012}. Above a certain $\Omega$ value and below a certain $\delta$ value indicates single phase stability.

    \subsection{\textbf{Supplementary \autoref{eq_hume}} \texorpdfstring{$\mid\;$}{Lg}Modified Hume-Rothery Rules}
    \label{s6_1}

    \vspace{-0.75em}
    {\small
    \begin{equation}
        \begin{aligned}
        \delta = \sqrt{\sum_{i=1}^{n}c_i\left(1-\frac{r_{i}}{\sum_{i=1}^{n}(c_{i}r_{i})}\right)^{2}} ; \Omega = \frac{T_{m}\Delta S_{mix}}{\Delta H_{mix}}
        \label{eq_hume}
        \end{aligned}
    \end{equation}
    }

    The modified Hume-Rothery parameters have the problem of leniency: this method estimates that $\approx$40,000 of the 53,124 alloys modeled in this work would turn out to be single phase solid solution, and that over 15,000 alloys would still be single phase at $700\degree C$. While ThermoCalc was used to evaluate FCC specifically, this is a stark comparison to the $<$2,500 alloys it predicted.

    \newpage
    The logic behind the inspection of binary phase diagrams method would proceed as follows: if every possible binary alloy within the subsystems of an arbitrary alloy space all exhibit either (1) isomorphous solid solutions of a phase, (2) a nontrivial solubility of that phase within one element, or (3) a homogeneous range (e.g. $>$5 at.\%) of that phase anywhere within the phase diagram, then one can speculate that this phase is a salient facet of the higher order system. \autoref{ff7_pseudo} shows a pseudobinary of the system investigated in this work, varying nickel. From this perspective alone, it is difficult to predict that low-nickel content alloys could result in a stable FCC phase (in \autoref{s1_compositions}, there are multiple FCC examples with 15\% nickel).

\begin{figure}[H]
        \centering
        \includegraphics[width=0.3\linewidth]{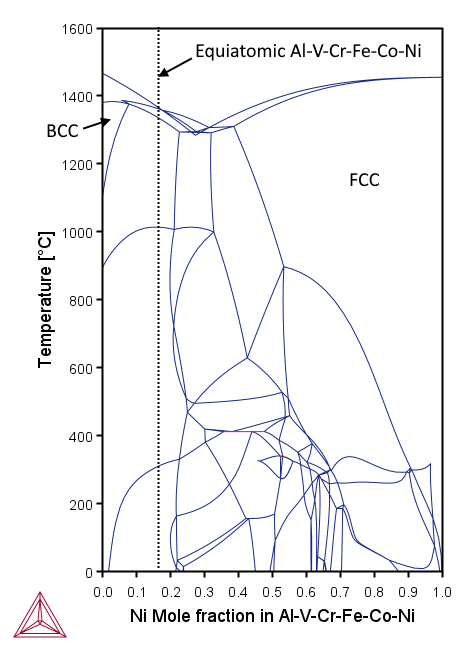}
        \caption{A pseudobinary phase diagram of \ce{Ni} within the \ce{Al-V-Cr-Fe-Co-Ni} space. A small single phase BCC region and large single phase FCC region are labeled.}
        \label{ff7_pseudo}
    \end{figure}

    \vspace{-1.0em}
    \subsection{\textbf{Supplementary \autoref{ff7_pseudo}} \texorpdfstring{$\mid\;$}{Lg}Pseudobinary phase diagram}
    \label{s6_2}
    \vspace{-1.25em}

    \vspace{-1.0em}

    \autoref{ff8_diagram} shows three of the constituent binaries within the \ce{Al-V-Cr-Fe-Co-Ni} space that violate the previous supposition.

\begin{figure}[H]
        \centering
        \includegraphics[width=0.8\linewidth]{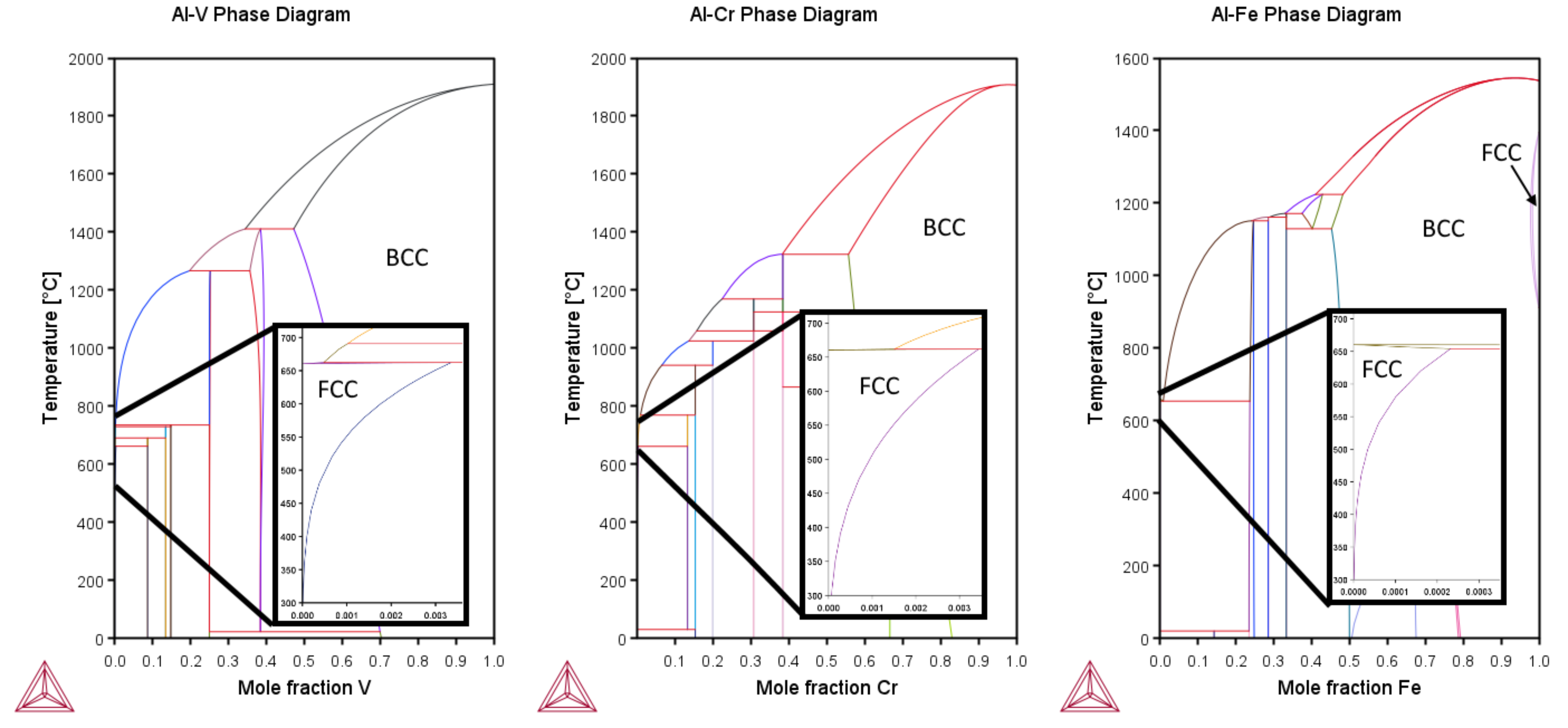}
        \caption{Binary phase diagrams. (left) \ce{Al-V} (center) \ce{Al-Cr} (right) \ce{Al-Fe} (all) Each of these binaries has a vanishingly small FCC region.}
        \label{ff8_diagram}
    \end{figure}

    \vspace{-1.0em}
    \subsection{\textbf{Supplementary \autoref{ff8_diagram}} \texorpdfstring{$\mid\;$}{Lg}Binary phase diagrams}
    \label{s6_3}
    \vspace{-1.25em}

    \newpage
    The method involving first order approximations comes from a DFT (density functional theory) model of binary enthalpies \cite{troparevsky_criteria_2015}. As long as the configurational contribution of a system is lower than its lowest DFT evaluated binary subsystem, then it will have a single phase present (and conversely, the reverse of this indicates no single phases can be present). \autoref{eq_enthalpy}, enthalpy of formation, is then used to determine phase stability. As long as the configurational contribution of a system is lower than its lowest evaluated binary subsystem using DFT, then a single homogenous phase will be present. \autoref{ff9_dft} shows the formation enthalpies for the binary compounds subsisting within the space tested in this work.

    \vspace{-1.0em}
    \subsection{\textbf{Supplementary \autoref{eq_enthalpy}} \texorpdfstring{$\mid\;$}{Lg}Enthalpy of formation}
    \label{s6_4}

    \vspace{-0.75em}
    {\small
    \begin{equation}
        \begin{aligned}
        \Delta H_{formation} >= -T\Delta S_{mix ideal} = -RT\sum_{i=1}^{n}X_{i}ln(X_{i})
        \label{eq_enthalpy}
        \end{aligned}
    \end{equation}
    }

\begin{figure}[H]
        \centering
        \includegraphics[width=0.4\linewidth]{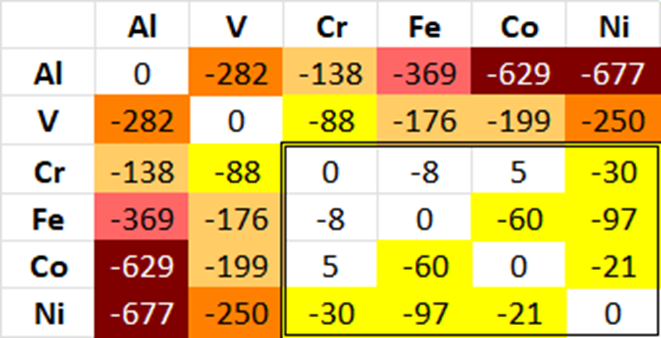}
        \caption{Minimum $\delta H_{f}$ in meV/atom, relative to pure elements required for stable single phase solid solution, adapted from \cite{troparevsky_criteria_2015}}
        \label{ff9_dft}
    \end{figure}

    \subsection{\textbf{Supplementary \autoref{ff9_dft}} \texorpdfstring{$\mid\;$}{Lg}Minimum binary compound enthalpy relative to pure elements}
    \label{s6_5}
    \vspace{-1.25em}

    For example, an equiatomic \ce{Cr-Fe-Co-Ni} alloy annealed at $700\degree C$ results in an enthalpy of -116 meV/atom, while the lowest DFT binary enthalpy formation for this system is -97 meV/atom, supporting the conclusion that such an HEA would be single phase. Adding aluminum to this alloy suggests phase separation, as its configuration term is -135 eV/atom while the \ce{Al-Ni} DFT calculation is -677 meV/atom. This phase separation has been experimentally verified for some compounds \cite{manzoni_phase_2013}.

    In the DFT matrix, the \ce{Al-Ni} binary enthalpy of -677 meV/atom would require an annealing temperature to exceed $4100\degree C$ (a nonsensical answer), and the \ce{V-Ni} binary would require such a temperature to be above the alloys’ solidus in most cases. Even for the more approachable \ce{V-Cr-Fe-Co-Ni} system (whose enthalpy need only be less than -250 meV/atom), an equiatomic set requires an annealing temp in excess of $1525\degree C$, which is above the solidus temperature of most combinations of these elements in delineations of 5\%. Many of the alloys with vanadium and nickel were predicted to be FCC by ThermoCalc.
    
    Examples in \autoref{s1_compositions} include single phase solid solutions with V and Ni content that are both above 25\%. The \ce{Al-V-Cr-Fe-Co-Ni} and \ce{V-Cr-Fe-Co-Ni} single phase alloys produced by this work directly violate the predictions made from the above calculations.

    \bigskip
    
    Regardless of predictive method, various combinations of these “impossible” alloys have been verified experimentally as single phase solid solutions by other authors \cite{li_superior_2022} \cite{wang_effects_2012} \cite{zhang_solidsolution_2008}. While these approaches might seem sensible in common HEA contexts (like that of \ce{Cr-Fe-Co-Ni}), they incontrovertibly ignore higher dimensional phase complexities. A notable shortcoming likely comes from elements like aluminum acting as a stabilizer of crystal structures distinct from its primary unary structure, \cite{liu_formation_2022} and also from the lack of any consideration of higher order mixing parameters like that of the Redlich-Kirsten model on which modern CALPHAD methods are based \cite{redlich_algebraic_1948}.

\newpage
\setstretch{1.0}
\printbibliography